\documentclass[journal]{IEEEtran}

\ifCLASSINFOpdf
\else
\fi

\usepackage{pifont}
\usepackage{mathrsfs}
\usepackage{bbm}
\usepackage{amsmath, amsthm}
\usepackage{tabularx}
\usepackage{cite}
\usepackage{listings}
\usepackage{graphicx}
\usepackage{float}
\usepackage{amssymb}
\usepackage{array}
\usepackage{subfigure}
\usepackage{fancyhdr}
\usepackage{cases}
 \usepackage{supertabular}
\usepackage{url}
\usepackage{psfrag}
\usepackage{color}
\usepackage{bbding}
\newcommand{\qeda}{\hfill\ensuremath{\blacksquare}}
\usepackage{relsize}
\newcommand{\f}{it follows that }
\newcommand{\h}{it holds that }
\newcommand{\bcap} {\hspace{2pt} \mathlarger{\cap}
\hspace{2pt}}
\newcommand{\cl}{it is clear that }
\newcommand{\m}[1]{\mbox{$#1$}}
\newcommand {\n} {for any sufficiently large $n$}

\newcommand{\junzhao}[1]{\ding{110}\ding{43}\textcolor{red}{Jun Zhao: #1}}

\newcommand {\C} {{\rm I\kern-5.5pt C}}

\usepackage{enumitem}

\setlist{leftmargin=2pt}

\newtheorem{lem}{Lemma}
\newtheorem{thm}{Theorem}

\newtheorem{rem}{Remark}
\newtheorem{cor}{Corollary}

\newtheorem*{proposition1.1}{Proposition 1.1}
\newtheorem*{proposition1.2}{Proposition 1.2}
\newtheorem*{proposition1.3}{Proposition 1.3}
\newtheorem*{proposition2.1}{Proposition 2.1}
\newtheorem*{proposition2.2}{Proposition 2.2}
\begin{document}
%
\title{On resilience and connectivity of secure wireless sensor networks under node capture attacks}


%

%
%
%

\author{Jun Zhao,~\IEEEmembership{Member,~IEEE}
\IEEEcompsocitemizethanks{\IEEEcompsocthanksitem Copyright (c) 2016 IEEE. Personal use of this material is permitted. However, permission to use this material for any other purposes must be obtained from the IEEE by sending a request to pubs-permissions@ieee.org

The author was with the Cybersecurity Lab (CyLab) at Carnegie Mellon University, Pittsburgh, PA 15213, USA. He is now with Arizona State University, Tempe, AZ 85281, USA (Email: junzhao@alumni.cmu.edu).

This research was supported in part by CyLab and Department of Electrical \& Computer Engineering
at Carnegie Mellon University, by School of Electrical, Computer and Energy Engineering at Arizona State University, and also by
  Grants CCF-1617934 and CNS-1422277 from the U.S. National Science Foundation.}}

\maketitle

%
%

\markboth{published in IEEE Transactions on Information Forensics and Security}{published in IEEE Transactions on Information Forensics and Security}
%



  \begin{abstract}
Despite much research on probabilistic key predistribution schemes
for wireless sensor networks over the past decade, few formal
analyses exist that define schemes' resilience to node-capture
attacks precisely and under realistic conditions. In this paper, we
analyze the resilience of the $q$-composite key predistribution
scheme, which mitigates the node capture vulnerability of the
Eschenauer--Gligor scheme in the neighbor discovery phase. We derive scheme parameters to have a desired level of resiliency, and   obtain optimal parameters that defend against different adversaries as much as possible. We also
show that this scheme can be easily enhanced to achieve the same
``perfect resilience" property as in the random pairwise key
predistribution for attacks launched \emph{after} neighbor
discovery. Despite considerable attention to this scheme, much prior
work explicitly or implicitly uses an \emph{incorrect} computation
for the probability of link compromise under node-capture attacks
and ignores real-world transmission constraints of sensor nodes. Moreover, we
derive the critical network parameters to ensure connectivity in
both the absence and presence of node-capture attacks. We also investigate node replication attacks by analyzing the adversary's optimal strategy.
\end{abstract}


\section{Introduction}\label{introduction}

\subsection{Background}

\IEEEPARstart{W}{ireless} sensor networks (WSNs) deployed in hostile environments are
subject to adversary attacks that can lead to sensor capture
\cite{virgil,adrian,5717499,Conti:2008:EPD:1352533.1352568}. Sensors
are particularly vulnerable to such attacks because their physical
protection is limited by low-cost considerations and their operation
is unattended
\cite{Conti:2008:EPD:1352533.1352568,Vu:2010:SWS:1755688.1755703,5717499}.
As a result of such attacks, all secret keys of a captured
node are
discovered by an adversary. Although random key predistribution
schemes can be successfully used to secure communications in WSNs,
they are often not explicitly designed to address node-capture
vulnerabilities, and hence their resilience to such attacks is
seldom analyzed formally. The main goal of this paper is to provide
precise analytical results for resilience to node capture of one of
the best known random key predistribution schemes, namely the {\em
$q$-composite} extension of the basic Eschenauer and Gligor
scheme~\cite{virgil} (hereafter referred to as the EG scheme), which
was proposed by Chan \emph{et al.} \cite{adrian}.

{\bf Background.} The EG scheme \cite{virgil}, which is widely
considered to be the basic random key predistribution mechanism,
works as follows. For a WSN with $n$ nodes, in the key
predistribution phase, a large \emph{key pool} consisting of $P_n$
cryptographic keys is used to select \emph{uniformly at random}
$K_n$ distinct keys for each sensor node. These $K_n$ keys
constitute the \emph{key ring} of a sensor, and are installed in the
sensor's memory. After deployment, two sensors establish secure
communication over an existing link if and only if their key rings
have at least one key in common. Common keys are found in the
neighbor discovery phase whereby a random constant is enciphered in
all keys of a node and broadcast along with the resulting ciphertext
block in a given area limited by the transmission power/range; i.e.,
in a local neighborhood. ${P}_n$ and $K_n$ are both functions of $n$
for generality, with the natural condition $1 \leq K_n \leq P_n$.

According to the $q$-composite extension of the EG scheme \cite{adrian},
a secure link between two sensors is established if they share at least $q$ key(s) in
their key rings, where $1 \leq q \leq
K_n$. Clearly, the $q$-composite scheme with $q=1$ reduces to the EG
scheme. The $q$-composite scheme with $q\geq 2$ outperforms the EG
scheme in terms of resilience (defined below) to small-scale sensor
capture attacks while trading off increased vulnerability in the
presence of large-scale attacks.

In addressing the resilience to node-capture attacks, we adopt the
metric proposed by Chan \emph{et al.} \cite{adrian} where an
adversary captures some random set of $m$ nodes. Communication
between two arbitrary nodes, which are not among these $m$ nodes and
have a secure communication link in between, may still be
compromised (in particular be decrypted) by the adversary under the
EG scheme and its $q$-composite version due to key collisions that
can be generated during the predistribution phase. Thus, capturing
nodes enables an adversary to compromise communications between
non-captured nodes which happen to use keys that are also shared by
captured nodes. We denote the probability of such compromise by
$p_{\textnormal{compromised}}$ and, following Chan \emph{et al.}
\cite{adrian}, we say that a random key predistribution scheme is
``perfectly resilient" to node-capture attacks if
$p_{\textnormal{compromised}} = 0$. This implies that only
communications between a captured node and its direct neighbors are
compromised in a perfectly resilient scheme. Hence, analysis of
prefect resilience of a scheme must account for the number of
neighbors of any captured node (e.g., minimum, average) to assess
inherent vulnerability to node capture. Specifically, resilience
analysis must explicitly address the connectivity property of a key
graph arising from a key predistribution scheme under the classic
disk model of
communication~\cite{ISIT_RKGRGG,Krzywdzi,YaganThesis,pietro2004connectivity},
which enables the computation of the number of neighbors of a
captured sensor under realistic transmission constraints.

\subsection{Motivation}

\textbf{Why the $q$-composite scheme matters?}

Although many key predistribution schemes have been proposed in the literature  \cite{Rybarczyk,bloznelis2013,DiPietroTissec,Liu2003CCS,Du:2005:PKP:1065545.1065548}, we explain that  the $q$-composite scheme matters as follows.

\begin{itemize}
\item[\ding{172}] Although there are other key predistribution schemes that  by design are perfectly resilient after the neighbor discovery phase, the basic EG scheme and $q$-composite scheme can also be easily extended to be perfectly resilient after the neighbor discovery phase.
\item[\ding{173}]  Schemes are still vulnerable \emph{during} the neighbor discovery phase, even being perfectly resilient after the neighbor discovery phase. In terms of resilience against node capture during the neighbor discovery stage, the $q$-composite scheme often outperforms the EG scheme and many other schemes.
\end{itemize}

 We now discuss the above points \ding{172} and \ding{173} respectively. We look at \ding{172} first. Although the basic EG scheme and the $q$-composite scheme are  {\em not} perfectly resilient, they can be easily extended to
become perfectly resilient after neighbor discovery. For example, after the neighbor discovery phase of the EG
scheme ends and two neighbor nodes identified by $ID_i$ and $ID_j$
discover that they share key $k_{ij}$, they can each compute a new
shared key $K_{ij} = hash(ID_i|| ID_j||k_{ij})$, where $i < j$ and
$hash(\cdot)$ is a cryptographic hash function with pseudo-random
output, and erase the old key $k_{ij}$.
 $K_{ij}$ is statistically unique (i.e., up to the birthday bounds)
and hence the EG scheme becomes perfectly resilient by the above
definition. Clearly, its $q$-composite version also becomes
perfectly resilient if we include the uniquely ordered $q$ keys in
the hash operation instead of the single key $k_{ij}$.

We now explain the above point \ding{173}. The $q$-composite scheme increases \cite{adrian} the
resilience of the EG scheme in the neighbor discovery phase where
perfect resilience {\em cannot} hold; i.e., when
$p_{\textnormal{compromised}} > 0$. This is important since many
random key predistribution schemes are also vulnerable to node
capture in their neighbor-discovery phase even when
they are perfectly resilient after neighbor discovery, and yet their vulnerabilities
cannot be easily mitigated. For example, the random {\em
pairwise-key predistribution} scheme of Chan \emph{et al.}
\cite{adrian} is perfectly resilient and yet vulnerable to non-local
communication compromise. In this scheme, prior to deployment, each
sensor is matched up with a certain number of other randomly
selected sensors. Then for each pair of sensors that are matched
together, a {\em pairwise} key is generated and loaded in both
sensors' key rings. After deployment, any two nodes sharing a common
pairwise key establish secure communication in the actual
neighbor-discovery phase. Following neighbor discovery, each node
erases its locally unused keys; i.e., keys that are shared with
remote sensors located outside the transmission radius of a sensor.
However, before these keys are erased, an adversary can capture and
use them in his/her nodes strategically placed at various locations
in the network to establish connections to sensors that are
unreachable to remote legitimate neighbors; e.g., the adversary's
nodes can broadcast encrypted messages at these locations, find and
connect to legitimate local sensor nodes.

Other pairwise predistribution schemes
such as the ones by Liu and Ning \cite{Liu2003CCS} are constructed
on a polynomial-based key predistribution protocol, while the scheme
by Du \emph{et al.} \cite{Du:2005:PKP:1065545.1065548} is
built on a threshold key exchange protocol. Both schemes exhibit the
following \textit{threshold resilience} property: after the adversary has
captured some $m$ number of nodes, the fraction of compromised
communications among non-captured nodes is close to $0$ for $m$ less
than a certain threshold and sharply grows toward $1$ when $m$
surpasses the threshold. The larger $m$, the more sensor energy
these schemes use for shared key generation. Thus, these schemes are
not applicable in environments where $m$ is on the order of tens and
hundreds of nodes. For this reason, threshold key predistribution
schemes are less relevant to our analysis, despite their other
interesting properties.


\textbf{What are the limitations of    prior work on the $q$-composite scheme in the literature?}

Since the introduction of the $q$-composite scheme, it has received tremendous interest in the literature over the past decade
\cite{Rybarczyk,bloznelis2013,Perfectmatchings,Liu2003CCS,Du:2005:PKP:1065545.1065548}. However, prior researches have a few limitations which can be possibly addressed only via very recent advances. We discuss these limitations in terms of resiliency against node capture, network connectivity and replication attacks, respectively.

\begin{itemize}
\item \textbf{Resilience against node capture in much prior studies: Incorrect analysis of link compromise.} It was only
recently discovered \cite{6170857} that the computation of the
probability of link compromise, $p_{\textnormal{compromised}}$, was
incorrectly computed more than a decade ago by Chan \emph{et al.}
\cite{adrian}. Hence, most prior work
\cite{Liu2003CCS,Chan2008134,YaganThesis,Vu:2010:SWS:1755688.1755703,Yang:2005:TRS:1062689.1062696}
used the incorrect computation either explicitly or implicitly,
yielding inaccurate or even misleading results.
\item \textbf{Connectivity analysis in much prior studies: No consideration of transmission constraints, or relatively weak results when there is consideration.}
  The analysis of
 resilience (even perfect resilience) of a scheme must account for the number of
neighbors of any captured node to assess
\textit{inherent} vulnerability to node capture. Hence, resilience
analysis must explicitly address   connectivity   of a key
graph arising from key predistribution. This builds the connection between resilience and connectivity. In fact, connectivity analysis of the secure sensor network under key predistribution is also of significant virtue in its own because network-wide communication requires connectivity. Given the importance of network connectivity, there have been considerable studies on this issue \cite{pietro2004connectivity,r1,ISIT_RKGRGG,Krzywdzi,MobiCom14}. However, few results address physical transmission constraints because considering both the security aspect (i.e., the key predistribution scheme) and transmission constraints will render the study much more challenging. Transmission constraints reflect real-world implementations of WSNs in
which sensors have limited transmission capabilities so two remote sensors may not be able to communicate directly. Even under the classic
disk model of
communication where two sensors have to be within a certain distance from each
other to communicate, connectivity analysis of secure sensor networks is difficult. Because of the difficulty, limited studies that consider secure connectivity under the disk model all present quite weak results. Note that the topology induced by the $q$-composite scheme is a \textit{key graph} in which each of $n$ nodes selects $K_n$ number of keys uniformly at random from a common pool of $P_n$ keys, and two nodes establish an edge upon sharing at least $q$ keys (we denote the above key graph by $G_q(n,K_n,P_n)$). The disk model induces the so-called \textit{random geometric graph} in which any two of $n$ uniformly distributed nodes on an area $\mathcal{A}$ have an edge in between if and only if their distance is at most some value, say $r_n$ (we denote the above random geometric graph $G_{RGG}(n, r_n,
\mathcal{A})$). Then the topology of a secure sensor network deploying the $q$-composite scheme under the disk model is given by the intersection of the key graph and the random geometric graph; i.e., a secure link exists only when the two nodes have an edge in the key graph and also have an edge in the random geometric graph. The reason is that a secure link between two sensors require them to share at least $q$ keys and also to be within distance $r_n$. For this graph intersection $G_q(n,K_n,P_n) \bcap G_{RGG}(n, r_n,
\mathcal{A})$, due to the intertwining
of the key graph and the random geometric graph, it is extremely difficult to obtain strong connectivity result. To illustrate the difficultly, we discuss a related yet simper problem that had been open for 18 years. Let us replace the key graph with a much simpler graph, the renowned Erd\H{o}s--R\'{e}nyi
graph \cite{citeulike:4012374} $G_{ER}(n,p_n)$, which is defined on $n$ nodes such
that any two nodes establish an edge in between \emph{independently} with
the same probability $p_n$. This graph is much simpler than the key graph because it removes the edge dependencies \cite{5383986}. Then we consider the graph intersection $G_{ER}(n,p_n) \bcap G_{RGG}(n, r_n,
\mathcal{A})$. After establishing connectivity result of $G_{RGG}(n, r_n,
\mathcal{A})$, Gupta
and Kumar \cite{Gupta98criticalpower} in 1998 proposed a conjecture for connectivity of $G_{ER}(n,p_n) \bcap G_{RGG}(n, r_n,
\mathcal{A})$ (for the details of this conjecture, see Page \pageref{Penroseground} where we discuss related work). Despite many attempts to address this conjecture, it was resolved by Penrose \cite{penrose2016connectivity} not until 2016. The difficulty is to analyze the connection structure when two distinct types of random graphs intersect: even individual graphs are highly connected, the resulting topology can become disconnected after intersection. We resolve an analog of the above conjecture for connectivity in $G_q(n,K_n,P_n) \bcap G_{RGG}(n, r_n,
\mathcal{A})$, the intersection of the key graph and the random geometric graph. In particular, we show the connectivity has a sharp transition when the probability of a secure link is   $\frac{\ln n}{n}$. To summarize, our strong connectivity result of the studied secure sensor network $G_q(n,K_n,P_n) \bcap G_{RGG}(n, r_n,
\mathcal{A})$ is not obtained in the literature due to the difficulty of analyzing graph intersection. Prior results \emph{either} ignore transmission constraints to analyze just the key graph, \emph{or} consider the graph intersection but obtain relatively weak results (more details in Section \ref{secompare}).
\item \textbf{Node-replication attacks in prior studies: Lack of formal or tractable analyses to determine the adversary's optimal strategy.} Although our main focus is to analyze node-capture resiliency and connectivity of the $q$-composite scheme, we also discuss node-replication attacks on the scheme (following Review 1's suggestion). After node capture, the adversary can deploy   replica nodes of   compromised nodes by extracting keys from   compromised nodes and then inserting some keys into the memory of replica nodes. Prior studies have investigated node-replication attacks, but most researches lacked formal analyses \cite{parno2005distributed,conti2007randomized,zhu2007efficient}. Limited studies have formally addressed node replication, but existing expressions to quantify node-replication attack are complex \cite{fu2008replication}, making it intractable to determine the adversary's optimal strategy to maximize node-replication attack given limited resource.
\end{itemize}

  \subsection{\textbf{Summary of Our Results}} \label{sec:result-summary}

Our main results can be summarized as follows:
\begin{itemize}
  \item We analyze the
$q$-composite scheme's resilience against node-capture attacks. We derive a tractable expression for the fraction of compromised communications under node capture, identify scheme parameters to have a desired level of resiliency, and   obtain optimal parameters that defend against different adversaries as much as possible.
  \item We study  connectivity of
sensor networks with the $q$-composite scheme under real-world
transmission constraints. We derive the critical
conditions to guarantee secure network connectivity with high
probability (i.e., with a probability converging to $1$ as the network size goes to infinity).
  \item For node-replication attacks, we investigate the adversary's optimal strategy.
\end{itemize}

\subsection{\textbf{Comparison with Related Work}} \label{secompare}

We will first explain the improvements of this paper over related studies, and then discuss additional related work.

%


\textbf{Improvement over prior work in terms of resiliency analysis against node capture.} For a secure sensor network using the EG scheme \cite{virgil} (i.e.,
the $q$-composite scheme with $q=1$), Di Pietro \emph{et al.}
\cite{4198829,DiPietroTissec} prove that if $\frac{P_n}{K_n} =
\Omega(n)$, the network is resilient in the sense that an adversary
capturing sensors at random has to obtain at least a constant
fraction of nodes of the network in order to compromise a constant
fraction of secure links. We obtain the corresponding
result for general $q$. Chan \emph{et al.} \cite{adrian} show that for the $q$-composite scheme, larger $q$ outperforms smaller $q$ in terms of resilience against small-scale sensor
capture attacks while trading off increased vulnerability in the
presence of large-scale attacks, but no analytical result that identifies the optimal $q$ is given. Furthermore, their computation is shown to be incorrect by Yum and
Lee \cite{6170857} recently. In contrast, we formally obtain optimal $q$ that defends against different adversaries as much as possible. Although Yum and
Lee \cite{6170857} have provided the correct expression to quantify node capture, but the expression is complex and intractable to analyze. We significantly simplify the expression via an asymptotic analysis, and this enables us to identify optimal scheme parameters.

\textbf{Improvement over prior work  in terms of connectivity analysis.} Many studies on connectivity in secure sensor networks ignore real-world transmission constraints \cite{pietro2004connectivity,r1,adrian} because considering both the security aspect (i.e., the key predistribution scheme) and transmission constraints will render the analysis much more challenging. Because of the challenges, limited studies incorporating transmission constraints often present quite weak results \cite{ISIT_RKGRGG,Krzywdzi}. In the
special case of $q=1$, Krzywdzi\'{n}ski and Rybarczyk
\cite{Krzywdzi}, and Krishnan \emph{et al.} \cite{ISIT_RKGRGG} recently show upper bounds $\frac{8\ln n}{n}$ and $\frac{2\pi \ln n}{n}$ for the critical threshold that the edge probability (i.e., the probability of a secure link between two sensors) takes for connectivity, while we prove   the \emph{exact} threshold as $\frac{\ln n}{n}$ for \emph{general} $q$. The threshold is identified because we derive a sharp zero-one law for connectivity. Intuitively, a sharp zero--one law means that as the edge probability surpasses
(resp., falls below) the critical value and grows (resp., declines)
further, the network immediately becomes asymptotically connected
(resp., disconnected). Chan and Fekri \cite{Chan2008134} approximate the key
 graph of the $q$-composite scheme by an Erd\H{o}s--R\'{e}nyi graph. However, there is
 a lack of rigorous argument for this approximation. A formal argument
 is needed because an Erd\H{o}s--R\'{e}nyi graph and a key
 graph used to represent the $q$-composite scheme are quite different; e.g., edges are all independent in the
 former but are not in the latter \cite{5383986,Assortativity,QcompTech14}. In
 this paper, we rigorously bridge these two graphs (we provide the details in Section VI of the full version \cite{full} due to space limitation).

\textbf{Improvement over prior work  in terms of replication attack analysis.} For
 node-replication attacks, prior researches either lack formal analyses \cite{parno2005distributed,conti2007randomized,zhu2007efficient}, or provide intractable analyses  \cite{fu2008replication}. We close this gap by presenting a simple asymptotic analysis, which further enables us to determine the adversary's optimal strategy.

%
%
%
%

%
%
%

{\bf Other related studies.} Key predistribution schemes by Liu and Ning
\cite{Liu2003CCS} and Du \emph{et al.}
\cite{Du:2005:PKP:1065545.1065548} have the threshold
behavior: the probability of link compromise is close to $0$ under
small-scale attacks but becomes close to $1$ after more than a
threshold number of nodes are captured. Alarifi and Du
\cite{Alarifi:2006:DSN:1180345.1180359} investigate data and code
obfuscation techniques to improve resilience against node capture.
Yang \emph{et al.} \cite{Yang:2005:TRS:1062689.1062696} propose an
approach of binding keys to geographic locations to improve the
resilience against node capture. Defense mechanisms are also
discussed by Vu \emph{et al.} \cite{Vu:2010:SWS:1755688.1755703}.
Conti \emph{et al.} \cite{Conti:2008:EPD:1352533.1352568} introduce
techniques to detect node-capture attacks. Bonaci \emph{et al.}
\cite{5717499} apply control theory to model network behavior under
node capture. Tague and Poovendran \cite{tague2008modeling} evaluate
node-capture attacks in sensor networks by decomposing them into
collections of primitive events.

Erd\H{o}s and R\'{e}nyi
  \cite{citeulike:4012374} analyze connectivity in the classical Erd\H{o}s--R\'{e}nyi
graph \cite{citeulike:4012374} $G_{ER}(n,p_n)$ (the graph is named after them), while Gupta and Kumar \cite{Gupta98criticalpower} address a random geometric graph $G_{RGG}(n, r_n,
\mathcal{A})$. Both studies show that a critical threshold of connectivity in either graph is that the edge probability  equals $\frac{\ln n}{n}$; i.e., $G_{ER}(n,p_n)$ is connected (resp., disconnected) if $p_n\sim \frac{a\ln n}{n}$ for some constant $a >1$ (resp., some constant $a <1$), while $G_{RGG}(n, r_n,
\mathcal{A})$ is connected (resp., disconnected) if $\pi {r_n}^2\sim \frac{a\ln n}{n}$ for some constant $a >1$ (resp., some constant $a <1$). Gupta and Kumar \cite{Gupta98criticalpower}
  conjecture that the intersection $G_{ER}(n, p_n) \bcap G_{RGG}(n, r_n,
\mathcal{D})$ will also have a similar connectivity result where a critical threshold of connectivity is that the edge probability $\pi  {r_n}^2 p_n $ equals $\frac{\ln n}{n}$. Despite the seemingly simple extension, the Gupta--Kumar
  conjecture had been open for 18 years (1998--2016) before being confirmed by Penrose's recent
 ground-breaking work \cite{penrose2016connectivity}. \label{Penroseground}

\subsection{Organization and Notation}

\textbf{Roadmap.} The rest of the paper is organized as follows. We
introduce useful notation below. In Section \ref{sec:resi}, we
analyze the resilience of the $q$-composite scheme against node
capture. Section \ref{sec:main:res} presents connectivity results of
a sensor network under the $q$-composite scheme, in
consideration of sensors' limited transmission ranges. We quantify node-replication attacks in Section \ref{node-quan}. Section  \ref{sec-derive-pcCapture} provides analytical details for resilience. We conclude the paper in Section \ref{sec:Conclusion}.


\textbf{Notation}. Throughout the paper, $q$ is an arbitrary positive integer and does
not scale with $n$, the number of nodes in the sensor network. We define $p_s$ as the probability that the key
rings of two sensors share at least $q$ keys in the $q$-composite scheme, where the
subscript ``s'' means ``secure''. $p_s$ is a function of the
parameters $K_n, P_n$ and $q$.
All
limits are taken with $n\to\infty$. We use the standard   asymptotic
notation $o(\cdot), O(\cdot), \omega(\cdot),
\Omega(\cdot),\Theta(\cdot)$ and $ \sim$; see \cite[Page 2-Footnote 1]{ZhaoYaganGligor}.
In particular, for two positive sequences $x_n$ and $y_n$, the
relation $x_n \sim y_n$ signifies $\lim_{n \to
  \infty} (x_n/y_n)=1$.

\section{Resilience of $q$-Composite Scheme\\against Node Capture} \label{sec:resi}

We now rigorously evaluate the resilience against node capture in
sensor networks employing the $q$-composite scheme.
We consider the case where the adversary has captured some random
set of $m$ nodes. Throughout the rest of the paper, we let $p_{\textnormal{compromised}}$ denote the probability
that secure communication
between two non-captured nodes 
 is compromised by the adversary (we often refer to this probability as probability of link compromise). Due to node homogeneity, $p_{\textnormal{compromised}}$ is also the fraction of compromised
communications among non-captured nodes. In the work proposing the $q$-composite scheme, Chan
\emph{et al.} \cite{adrian} have already computed $p_{\textnormal{compromised}}$. Unfortunately, this computation, cited in numerous following studies \cite{Chan2008134,Liu2003CCS,YaganThesis,Vu:2010:SWS:1755688.1755703,Yang:2005:TRS:1062689.1062696}, is shown to be incorrect by Yum and
Lee \cite{6170857} recently. Both the incorrect and correct formulas are quite involved, so we defer presenting them to Section \ref{sec-derive-pc}. Given the correct yet complex formula, it is still challenging to interpret how $p_{\textnormal{compromised}}$ varies with respect to the parameters. To this end, we significantly simplify $p_{\textnormal{compromised}}$ via an asymptotic analysis, and this enables us to identify scheme parameters to have a desired level of resiliency, and to choose optimal parameters that defend against different adversaries as much as possible. The intuition of our analysis is that although $p_{\textnormal{compromised}}$ is complex for finite $n$, we manage to derive simple asymptotic result of $p_{\textnormal{compromised}}$. Despite simple asymptotics obtained, the analysis to derive the result turns out to be non-trivial since it involves significant simplification of a complex expression.

\subsection{Summary of Results}

We now summarize the main results formally. First, we identify the condition for negligible probability of link compromise, where a probability is negligible if it converges to $0$ as $n \to \infty$.

\begin{thm}[\textbf{Conditions for negligible probability of link compromise}] \label{thm:pcomp1}

The probability $p_{\textnormal{compromised}}$ of link compromise is
negligible if $m\Big/\big(\frac{P_n}{K_n}\big)$ is negligible. In
other words,
 if
\begin{align}
m = o\bigg(\frac{P_n}{K_n}\bigg), \label{moPnKn}
\end{align}
then $
  p_{\textnormal{compromised}}   = o(1). $

\end{thm}

Intuitively, Theorem \ref{thm:pcomp1} exhibits the following finding: to
compromise a constant fraction of communication links between
non-captured nodes in the network, the adversary has to capture at
least some number of nodes such that these nodes combined have a
number of keys at least a constant fraction of the key pool.
An immediate implication of Theorem \ref{thm:pcomp1} gives the condition of scheme parameters to have a desired level of resiliency: the adversary has to capture a constant fraction of nodes in
order to compromise a constant fraction of communication links
in the network. Specifically, we obtain Corollary \ref{cor:pcomp1} below.

\begin{cor}[\textbf{Scheme condition to achieve desired resiliency}]  \label{cor:pcomp1}

If
\begin{align}
\frac{P_n}{K_n} = \Omega(n), \label{PnKnn}
\end{align}
then the network has a desired level of resilience against node capture in the sense
that the adversary has to capture a constant fraction of nodes in
order to compromise a constant fraction of communication links
in the network. Put formally, we have  $
  p_{\textnormal{compromised}}   = o(1)$ if $m=o(n)$.

\end{cor}

Theorem \ref{thm:pcomp1} above provides condition for $
  p_{\textnormal{compromised}}   = o(1)$, but there is no detailed expression of $p_{\textnormal{compromised}}$. This is given by
Theorem \ref{thm:pcomp2} below, which establishes the asymptotically
exact probability of link compromise. Theorem \ref{thm:pcomp2} will enable to us to optimize scheme parameters to defend against different attackers.

\begin{thm}[\textbf{Asymptotic probability of link compromise}] \label{thm:pcomp2}
Let the notation ``$\sim$'' denote asymptotic equivalence. Specifically, for two positive sequences $x_n$ and $y_n$, the
relation $x_n \sim y_n$ means $\lim_{n \to
  \infty} (x_n/y_n)=1$. We also define $p_s$ as the probability that the key
rings of two sensors share at least $q$ keys. If
\begin{align}
m = o\bigg(\sqrt{\frac{P_n}{{K_n}^2}} \hspace{2pt}\bigg),
\label{moPnKn-sqr}
\end{align}
and
\begin{align}
K_n   = \omega(1) , \label{Knomega1}
\end{align}
 then
\begin{align}
  p_{\textnormal{compromised}}   & \sim
\bigg(\frac{mK_n}{P_n} \bigg)^q, \label{ptnc}
\end{align}
%
and
\begin{align}
 \frac{p_{\textnormal{compromised}}}{p_s}   & \sim q!
  \bigg(\frac{m}{K_n}\bigg)^q. \label{fq}
\end{align}
\end{thm}

Now we use Theorem \ref{thm:pcomp2} for optimizing scheme parameters to defend against different adversaries. In particular, we will consider two different adversaries and show how to choose the key-sharing requirement $q$ (two sensors should share at least $q$ keys in order to establish a secure link in between).

To see the impact of $q$ on node-capture attacks, we will fix the key ring size $K_n$ and the key-setup probability $p_s$. This means that the key pool size $P_n$ needs to
decrease as $q$ increases. We first consider an adversary who has captured a certain number of nodes, $m$. In other words, given $m$, we study how $p_{\textnormal{compromised}}$ varies with respect to $q$. The result is that there exists an optimal $q$ to minimize $p_{\textnormal{compromised}}$ for an adversary that has captured $m$ nodes. More specifically, we have the following Corollary \ref{cor:capture_vary_q_with_m}.

\begin{cor}[\textbf{Minimizing link compromise with respect to $q$ given the number of captured nodes}]  \label{cor:capture_vary_q_with_m}

To minimize the fraction of  communications compromised by an adversary who has captured $m$ nodes, the optimal $q$ is
$q^*=\max\{\big\lfloor\frac{K_n}{m}\big\rfloor, 1\}$; i.e., $q^* =\big\lfloor\frac{K_n}{m}\big\rfloor$ if $\frac{K_n}{m} > 1$, and $q^* =1$ if $\frac{K_n}{m} \leq 1$, where the floor function $\lfloor x \rfloor$ maps a real number $x$ to the largest integer that is no greater than $x$.

\end{cor}

We then consider an adversary whose goal is to compromise a certain fraction of communications. We show that there exists an optimal $q$ to maximize the number of nodes needed to be captured by such an adversary (again, the key ring size $K_n$ and the key-setup probability $p_s$ are both fixed). More specifically, we present  Corollary \ref{cor:capture_vary_q_with_pc} below.

\begin{figure}[H]
\begin{center}
\vspace{-20pt}  \includegraphics[scale=0.4]{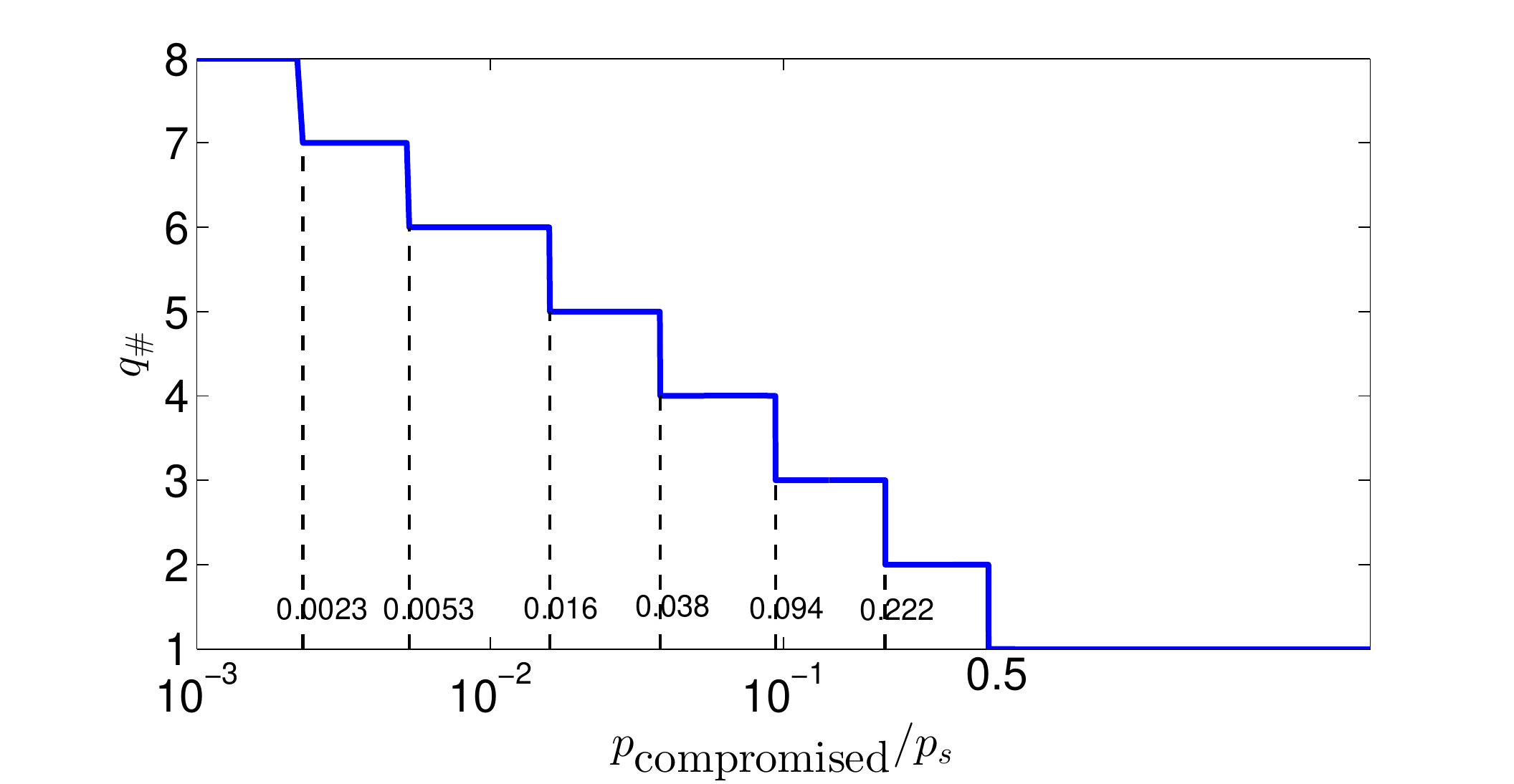}
 \vspace{-9pt} \caption{The plot of $q^{\#}$ with respect to ${p_{\textnormal{compromised}}/p_s}$, where $q^{\#}$ maximizes $ (\frac{p_{\textnormal{compromised}}/p_s}{q!})^{1/q}$.\vspace{-20.0pt}} \label{figure-optimal-q-pound}
  \end{center}
  \end{figure}

\begin{cor}[\textbf{Maximizing needed node capture with respect to $q$ given the fraction of link compromise}]  \label{cor:capture_vary_q_with_pc}

To maximize the resource (specifically, to maximize the number of captured nodes) needed for an adversary whose goal is to compromise $p_{\textnormal{compromised}}$ fraction of communications, the optimal $q$ is the solution $q^{\#}$ to maximize $ (\frac{p_{\textnormal{compromised}}/p_s}{q!})^{1/q}$. The solution $q^{\#}$ is difficult to derive formally, but can given empirically by Table \ref{table-optimal-q-pound} and illustrated by Figure \ref{figure-optimal-q-pound}. The above result is obtained by proving that the number of captured nodes needed to compromise $p_{\textnormal{compromised}}$ fraction of communications equals
$K_n \times (\frac{p_{\textnormal{compromised}}/p_s}{q!})^{1/q}$ asymptotically.\vspace{-15pt}

\end{cor}

\begin{table}[H]
\caption{This table presents the relationship between $\frac{p_{\textnormal{compromised}}}{p_s}$ and $q^{\#}$ which maximizes $ (\frac{p_{\textnormal{compromised}}/p_s}{q!})^{1/q}$.}\vspace{-10.0pt}
\label{table-optimal-q-pound}
\setlength\tabcolsep{.5pt}{\scriptsize\begin{tabular}{|c|c|c|c|c|c|}
\hline \rule{0pt}{3ex}
$\frac{p_{\textnormal{compromised}}}{p_s}$ & $[0.5, \infty)$ & $[0.222,0.5]$ & $[0.094,0.222]$ & $[0.038,0.094]$ & $[0.016,0.038]$   \\[2pt] \hline \rule{0pt}{2ex}
$q^{\#}$ (i.e., optimal $q$)                                & 1               & 2             & 3              & 4               & 5                 \\[2pt] \hline
\end{tabular}}
\\[2pt]
\setlength\tabcolsep{1pt}{\scriptsize\begin{tabular}{|c|c|c|c|c|}
\hline \rule{0pt}{3ex}
$\frac{p_{\textnormal{compromised}}}{p_s}$ & $[0.0053,0.016]$ & $[0.0023,0.0053]$ & $[0.0009,0.0023]$ & $\ldots$  \\[2pt] \hline \rule{0pt}{2ex}
$q^{\#}$ (i.e., optimal $q$)                                & 6               & 7             & 8             & $\ldots$             \\[2pt] \hline
\end{tabular}}
\end{table}

All of our results above will be proved in Section \ref{sec:lem:pcomp2}. We now discuss Theorem \ref{thm:pcomp2} in particular since it has a set of seemingly complex conditions. Results other than Theorem \ref{thm:pcomp2} are less involved and have already been well explained.

\subsection{Further Explanations of Theorem \ref{thm:pcomp2}}

We will interpret conditions and results of Theorem \ref{thm:pcomp2} for a better understanding.

\textbf{Interpreting (\ref{moPnKn-sqr}) (i.e., $m = o\big(\sqrt{\frac{P_n}{{K_n}^2}} \hspace{2pt}\big)$).} To provide a more intuitive understanding, we discuss (\ref{moPnKn-sqr}) below. To compute the probability $p_{\textnormal{compromised}}$ of link compromise under $m$ captured nodes, Bayes' rule is used by considering the number of different keys in the key rings of  the $m$ captured nodes. Specifically, after $A_{\tau}$ denotes the probability that
the $m$ captured nodes have $\tau$ different keys in total, and $B_{\tau}$ denotes the conditional probability of link compromise conditioning on the event that the $m$ captured nodes have $\tau$ different keys in total, it is straightforward from Bayes' rule that $p_{\textnormal{compromised}} = \sum_{\tau} (A_{\tau} B_{\tau})$, where $\tau$ iterates its possible values; i.e., $K_n \leq \tau \leq \min\{mK_n, P_n\}$, which holds from the following three points. First, $A_{\tau}$ is at least $K_n$ because each node already has $K_n$ keys. Second, $A_{\tau}$ may take the maximum $m K_n$ when the key rings of the $m$ captured nodes are completely different. Third, an additional note is that the above maximum cannot exceed $P_n$ since the whole key pool has $P_n$ keys. Now that given $p_{\textnormal{compromised}} = \sum_{\tau} (A_{\tau} B_{\tau})$, because $A_{\tau}$ and $B_{\tau}$ have complex expressions, it makes difficult for us to understand how $p_{\textnormal{compromised}}$ changes with respect to network parameters and how we can choose parameters to minimize $p_{\textnormal{compromised}}$. In order to simplify computing $p_{\textnormal{compromised}}$, we want to provide a simple yet rigorous approximation. To start with, we consider the distribution of $\tau$, the number of different keys in the key rings of  the $m$ captured nodes. Since we are mostly concerned with the practical case where the key pool size $P_n$ is much larger than the key ring size $K_n$ and the captured-node number $m$, it is reasonable to consider that the key rings of the $m$ captured nodes are completely different so that $\tau= m K_n$. We want to ensure this event with high probability; i.e., the probability $A_{m K_n}$ of this event can be written as $1-o(1)$ (arbitrarily close to $1$ for large $n$). If $A_{m K_n}=1-o(1)$, then any $A_{\tau}$ with $\tau < m K_n$ will be $o(1)$ (i.e., negligible) since all $A_{\tau}$ sums to $1$. In addition, clearly $B_{\tau}$ increases as $\tau$ increases because link compromise is more likely conditioning on that  the $m$ captured nodes have more different keys in total. Given the above, if $A_{m K_n}=1-o(1)$, recalling $p_{\textnormal{compromised}} = \sum_{\tau} (A_{\tau} B_{\tau})$, we can use $A_{m K_n} B_{m K_n}$ and furthermore $B_{m K_n}$ to approximate $p_{\textnormal{compromised}}$. Evaluating $B_{m K_n}$ is much simpler than computing all $A_{\tau}$ and $B_{\tau}$, and we defer the discussion of $B_{m K_n}$ next. Note that here our explanation is intuitive (sometimes not rigorous), but we present a rigorous analysis in Section \ref{sec-derive-pcCapture}.

 Now we return to how we can find a condition to ensure $A_{m K_n}=1-o(1)$. The probability $A_{m K_n}$ is the event that after each captured node selects $K_n$ keys uniformly at random from the key pool of $P_n$ keys, the total number of different keys selected equals $m K_n$. We first discuss a simple intuition using the renowned ``birthday paradox'' \cite{suzuki2006birthday}. In ``birthday paradox'', each of $m$ persons has a birthday selected randomly from a year (say $P$ days), then it is well known that if $m$ is on the order of $\sqrt{P}$, at least two people celebrate the same birthday with high probability; in contrast, if $m$ is much smaller than $\sqrt{P}$, all people have different birthdays with high probability. This ``birthday paradox'' example corresponds to the special case of key distribution where each node selects just one key from the key pool (note that the notions of node and key correspond to person and birthday). Hence, if $K_n$ is just $1$, we know from the ``birthday paradox'' that if $m$ is much smaller than $\sqrt{P_n}$, all $m$ captured nodes will have different keys with high probability. Now for the general case where $K_n$ is not $1$, the intuition thus becomes that if $m K_n$ is much smaller than $\sqrt{P_n}$ (formally $m K_n = o(\sqrt{P_n})$), all $m$ captured nodes will have different keys with high probability. In fact, this can also be seen from a simple analysis. Without any restriction, there are $\binom{P_n}{K_n}$ ways to construct a key ring of each node, and thus $\left[\binom{P_n}{K_n}\right]^m$ ways to construct the key rings of $m$ nodes. Now for the case where $m$ captured nodes together have $m K_n$ different keys, there are $\binom{P_n}{K_n}\binom{P_n-K_n}{K_n}\ldots\binom{P_n-(m-1)K_n}{K_n}$ ways to construct the key rings, since the first node has $\binom{P_n}{K_n}$ choices, the second node has $\binom{P_n-K_n}{K_n}$ choices in order to not share any key with the first node, and so on. Then $A_{m K_n}$ can be given by $\frac{\prod_{j=0}^{m-1}\binom{P_n-jK_n}{K_n}}{\left[\binom{P_n}{K_n}\right]^m}$, and the result $A_{m K_n}=1-o(1)$ under $m K_n = o(\sqrt{P_n})$ can be formally proved. The above condition $m K_n = o(\sqrt{P_n})$ is precisely (\ref{moPnKn-sqr}) after simple arrangements. \label{analyzeAmKn}

  To summarize, we introduce (\ref{moPnKn-sqr}) so that the $m$ captured nodes together will have $m K_n$ different keys with high probability and then we only need to conditioning on this event to provide a rigorous approximation for the probability of link compromise.

\textbf{Interpreting  (\ref{ptnc}) (i.e., $p_{\textnormal{compromised}}   \sim
\big(\frac{mK_n}{P_n} \big)^q$).} We have explained above that the $m$ captured nodes together will have $m K_n$ different keys with high probability so that $p_{\textnormal{compromised}} = \sum_{\tau} (A_{\tau} B_{\tau})$ can be approximated by $B_{m K_n}$, where $B_{m K_n}$ is the  probability of event $\mathcal{B}_{m K_n}$, with $\mathcal{B}_{m K_n}$ denoting the event of link compromise conditioning on the event that the $m$ captured nodes have $m K_n$ different keys in total. We now analyze $B_{m K_n}$  by Bayes' rule. Let $C_{mK_n, u}$ denotes the probability of event $\mathcal{B}_{m K_n}$ conditioning on the event that two benign nodes share $u$ keys in their key rings. Recall that $p_s$ is the probability that two benign nodes share at least $q$ keys in their key rings (i.e., $p_s$ is the link-setup probability in the $q$-composite scheme), and $\rho_u$ denotes the
probability that two benign nodes share exactly $u$ keys in their key rings. Given there is already a link between two benign nodes, the probability that they share exactly $u$ keys is $\frac{\rho_u}{p_s}$. Given the above, we obtain from Bayes' rule that $B_{m K_n} = \sum_{u=q}^{K_n} (C_{mK_n, u} \cdot \frac{\rho_u}{p_s})$. In our studied parameter range, $\frac{\rho_q}{p_s}$ is much greater than $\frac{\rho_u}{p_s}$  with $u > q$; i.e., given there is already a link between two benign nodes, then it is most likely that they share exactly $q$ keys rather than more than $q$ keys. Hence, we only need to evaluate link compromise in the case where the two benign nodes share exactly $q$ keys. Without any restriction, there are $\binom{P_n}{q}$ to select these $q$ keys. To compromise this link, these $q$ keys must fall into the $m K_n$ keys that the $m$ captured nodes together have, and there are $\binom{m K_n}{q}$ choices to select the $q$ keys. Then the link compromise probability becomes $\frac{\binom{m K_n}{q}}{\binom{P_n}{q}}$, whose numerator and denominator become roughly $\frac{1}{q!}(mK_n)^q$ and  $\frac{1}{q!}{P_n}^q$ respectively for $mK_n$ and $P_n$ much greater than $q$. Then we finally obtain the expression $\big(\frac{mK_n}{P_n} \big)^q$ for $p_{\textnormal{compromised}}$; i.e., (\ref{ptnc}) follows.

\textbf{Interpreting (\ref{Knomega1}) (i.e., $K_n = \omega(1)$).} We need $K_n = \omega(1)$ so that $\binom{m K_n}{q}$ above becomes $\frac{1}{q!}(mK_n)^q$ asymptotically.

\textbf{Interpreting (\ref{fq}) (i.e., $\frac{p_{\textnormal{compromised}}}{p_s}  \sim q!
  \big(\frac{m}{K_n}\big)^q$).} We have already interpreted (\ref{ptnc}) (i.e., $p_{\textnormal{compromised}}   \sim
\big(\frac{mK_n}{P_n} \big)^q$) above. Then (\ref{fq}) (i.e., $\frac{p_{\textnormal{compromised}}}{p_s}  \sim q!
  \big(\frac{m}{K_n}\big)^q$) follows from  (\ref{ptnc}) and (\ref{pssimq}) on Page \pageref{pssimq} (i.e., $p_s  \sim \frac{1}{q!} \big( \frac{{K_n}^2}{P_n} \big)^{q}$).


\subsection{Practicality of the Conditions in the Results}\vspace{-3pt}

We check the practicality of (\ref{PnKnn}). Recall that $n$ is the
number of sensors, the key ring size $K_n$ controls the number of
keys in each sensor's memory and $P_n$ is the key pool size. In
real-world implementations of sensor networks, $K_n$ is several
orders of magnitude smaller than $P_n$ and $n$ due to limited memory
and computational capability of sensors, and $P_n$ is larger
\cite{virgil,adrian,YaganThesis} than $n$. Therefore, (\ref{PnKnn})
is practical. In fact, Di Pietro \emph{et al.} \cite{DiPietroTissec}
also consider the range of $\frac{P_n}{K_n} = \Omega(n)$ in the
special case of $q=1$. In addition, under (\ref{PnKnn}) and a
reasonable assumption that only a small number (in particular
$o(n)$) of sensors are captured by the adversary, (\ref{moPnKn})
follows since any $o(n)$ can be expressed as
$o\big(\frac{P_n}{K_n}\big)$ by (\ref{PnKnn}).\vspace{-1pt}

For $m \geq 1$, condition (\ref{moPnKn-sqr}) implies
$\frac{P_n}{{K_n}^2} = \omega(1)$; i.e.,
$\frac{{K_n}^2}{P_n}   =  o(1)$. We check the practicality of (\ref{Knomega1}) (i.e., $K_n   = \omega(1)$) and $\frac{{K_n}^2}{P_n}   =  o(1)$.
Since $K_n$ is often larger than $\ln n$
 in real-world implementations \cite{YaganThesis,QcompTech14},
the practicality of $K_n   = \omega(1)$ follows. As mentioned, $P_n$ is
much larger compared to $K_n$, so $\frac{{K_n}^2}{P_n}   =  o(1)$ also holds in
practice. In addition, to have an idea of $m$ in (\ref{moPnKn-sqr}),
we observe $K_n$ is much less than $n$ and $P_n$ is larger (but not
too much) than $n$ \cite{virgil,adrian,YaganThesis}, so
$\sqrt{\frac{P_n}{{K_n}^2}}$ in (\ref{moPnKn-sqr}) is often a
fractional order of $n$. Under small-scale node-capture attacks, $m$
satisfies (\ref{moPnKn-sqr}).\vspace{-7pt}

\setlist{leftmargin=4pt}

\subsection{Experiments}\vspace{-3pt}

We provide experiments to confirm our theoretical results.

\subsubsection{\textbf{Impact of scheme parameters to link compromise by an adversary with a given number of captured nodes}}~\vspace{-1pt}

\begin{figure*}
\vspace{0pt}
\addtolength{\subfigcapskip}{-4pt}\centering     
\hspace{0pt}\subfigure[]{\label{fig:capture_vary_q_with_m_K20_ps005}\includegraphics[height=0.165\textwidth]{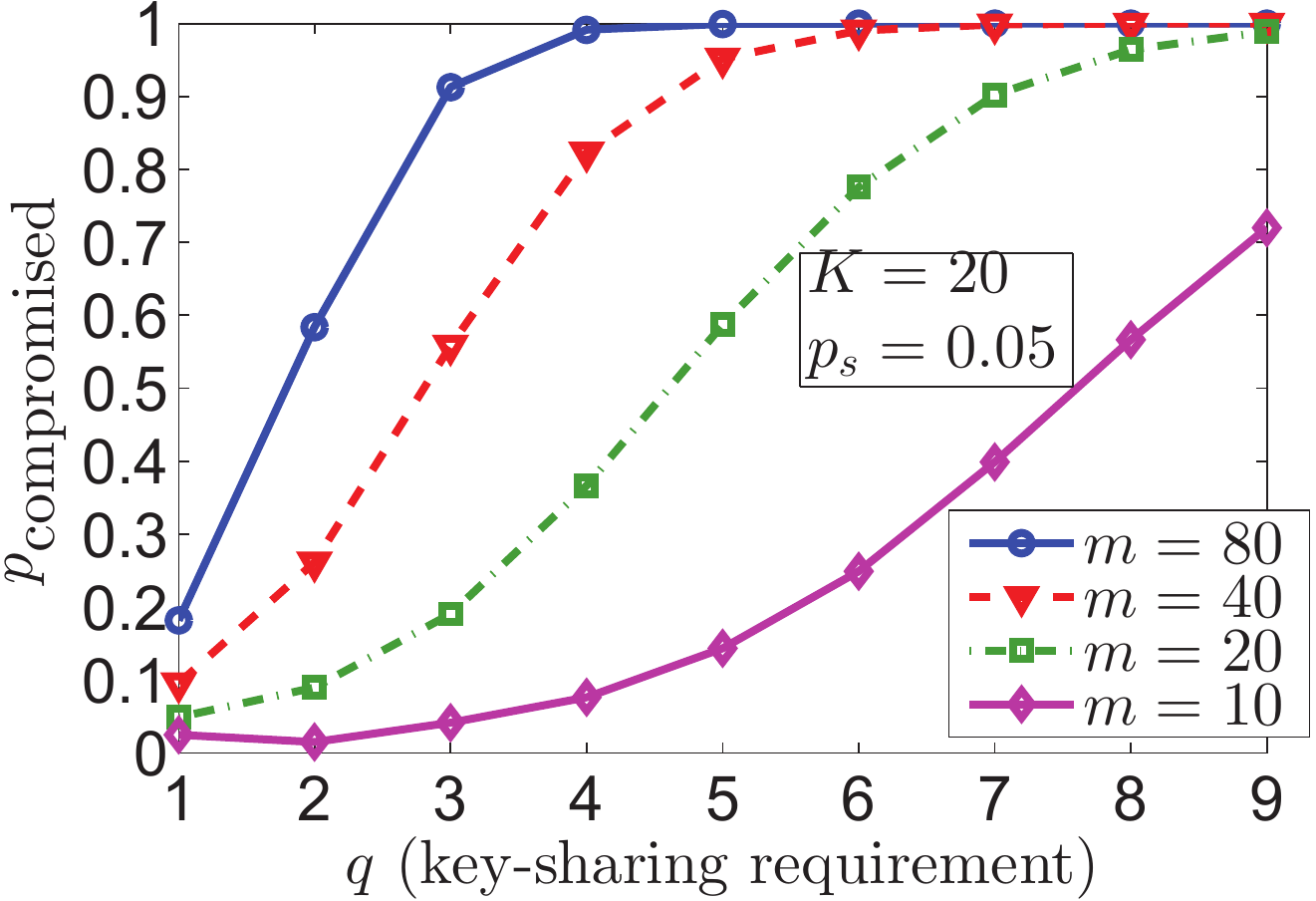}}
\hspace{0pt}\subfigure[]{\label{fig:capture_vary_q_with_m_K40_ps005}\includegraphics[height=0.165\textwidth]{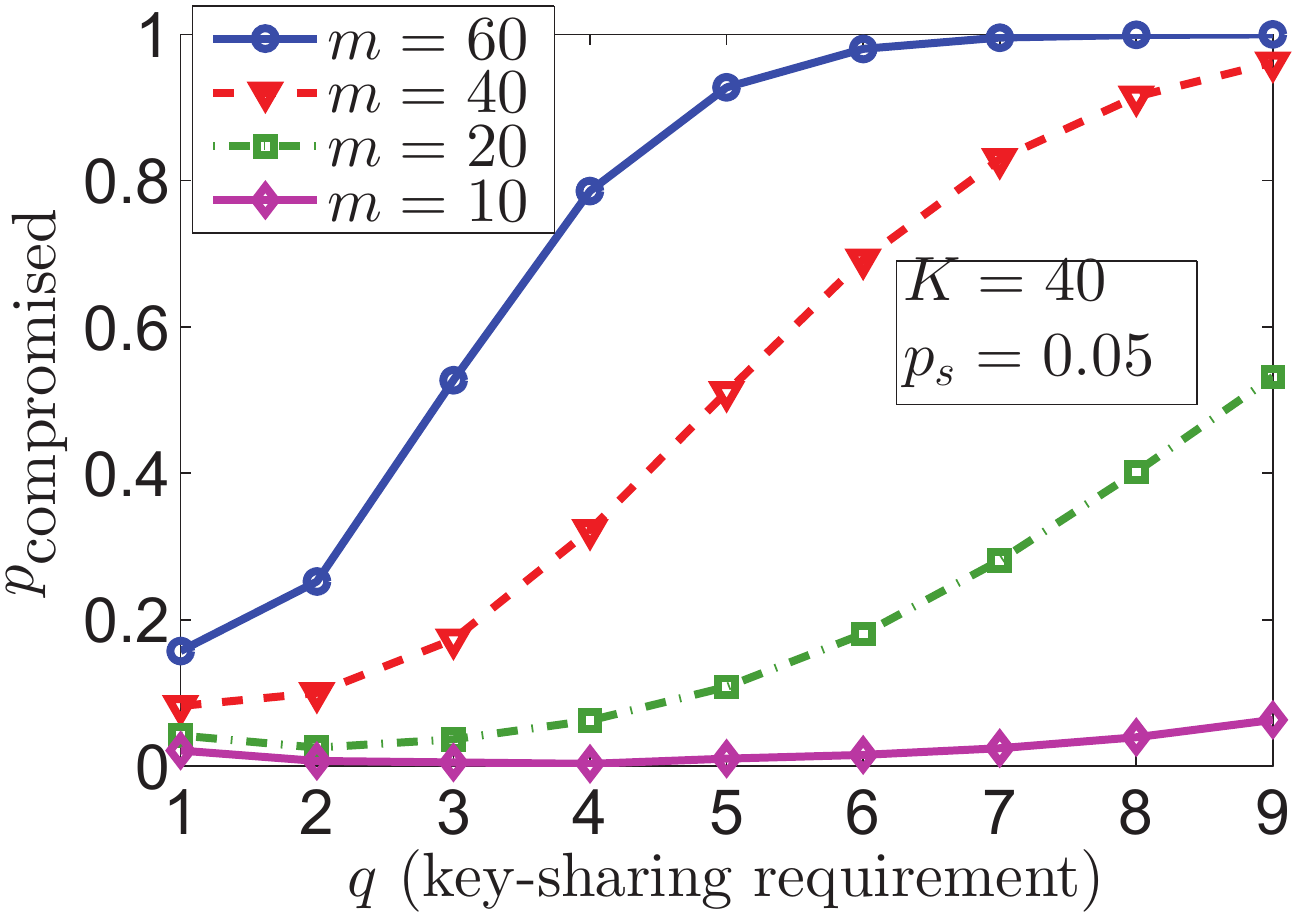}}
 \hspace{0pt}\subfigure[]{\label{fig:capture_vary_q_with_m_K80_ps005}\includegraphics[height=0.165\textwidth]{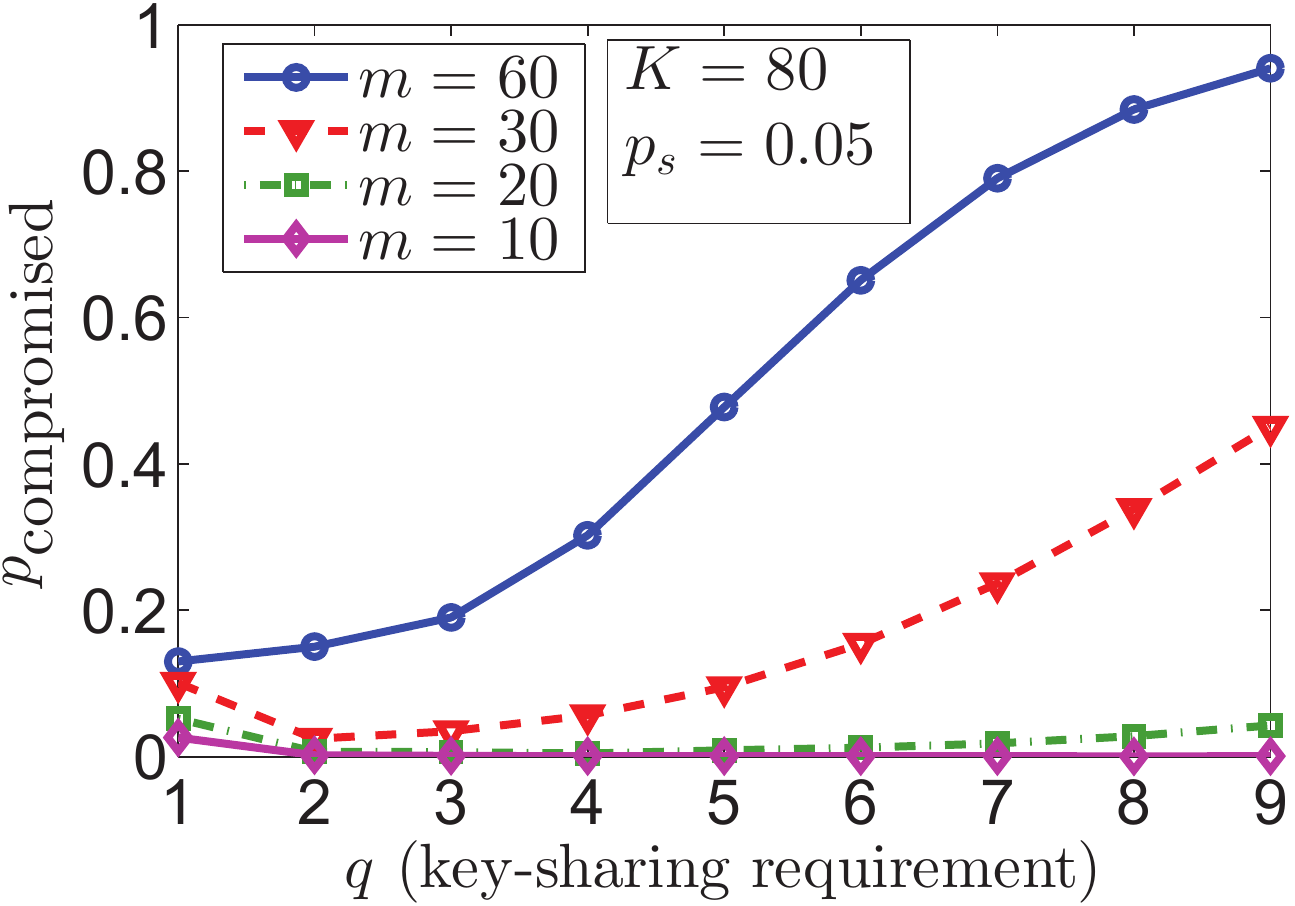}}
 \hspace{0pt}\subfigure[]{\label{fig:capture_vary_q_with_m_K120_ps005}\includegraphics[height=0.165\textwidth]{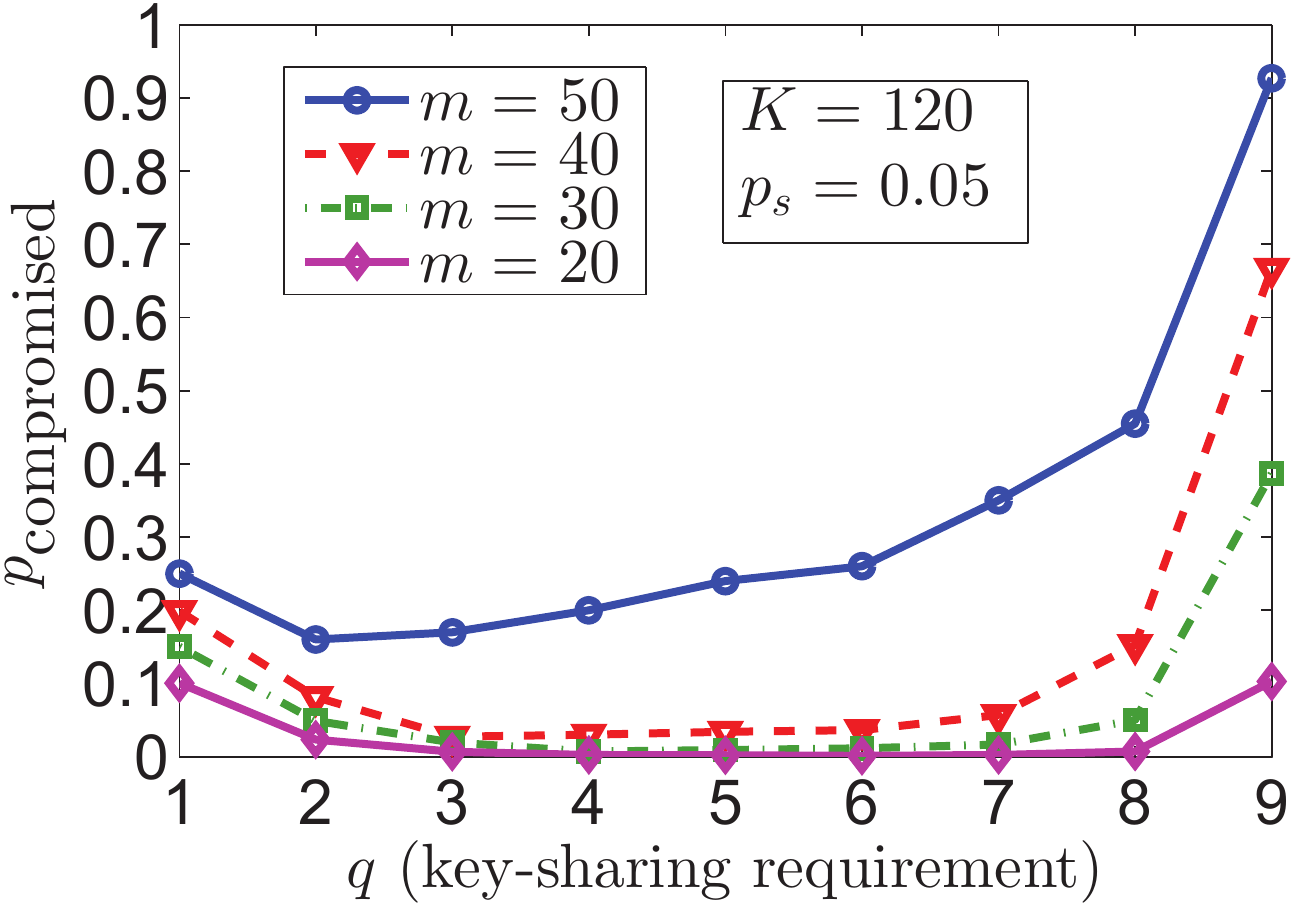}}
\vspace{-9pt}\caption{Each subfigure plots the fraction of
compromised communications between non-captured nodes, $p_{\textnormal{compromised}}$, with respect to $q$ when $m$ nodes are captured, and we fix $K$ and $p_s$ in each subfigure. In different subfigures, we have different $K$ and fix $p_s$ as $0.05$. We observe that the $q$ that minimizes $p_{\textnormal{compromised}}$ is in accordance with Corollary \ref{cor:capture_vary_q_with_m}; more specifically, the optimal $q$ is $\max\{\big\lfloor\frac{K_n}{m}\big\rfloor, 1\}$, where the floor function $\lfloor x \rfloor$ maps a real number $x$ to the largest integer that is no greater than $x$. \vspace{-10pt}} \label{fig:capture_vary_q_with_m_ps005}
\end{figure*}

\begin{figure*}
\vspace{0pt}
\addtolength{\subfigcapskip}{-4pt}\centering     
\hspace{0pt}\subfigure[]{\label{fig:capture_vary_q_with_m_K20_ps01}\includegraphics[height=0.165\textwidth]{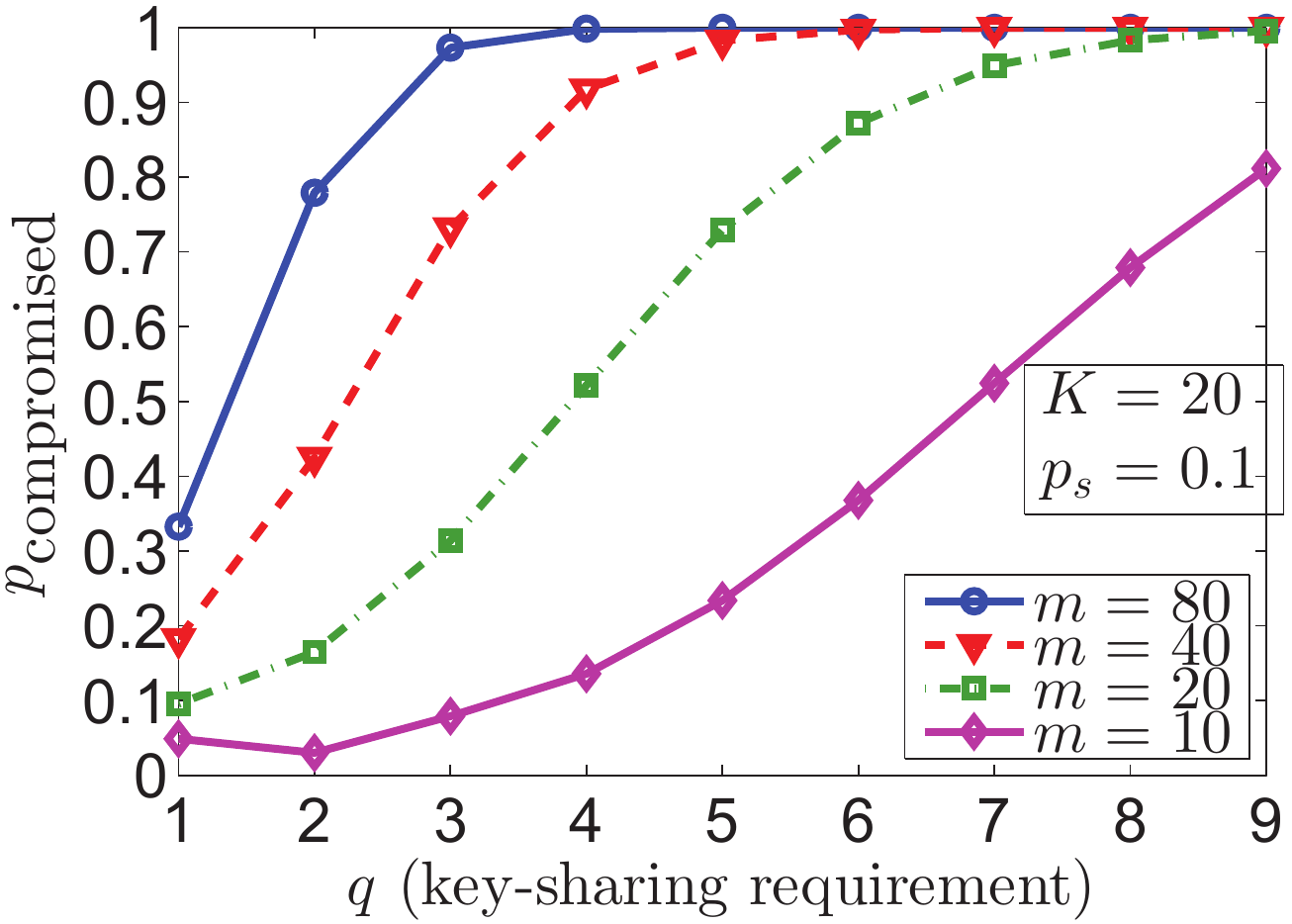}}
\hspace{0pt}\subfigure[]{\label{fig:capture_vary_q_with_m_K40_ps01}\includegraphics[height=0.165\textwidth]{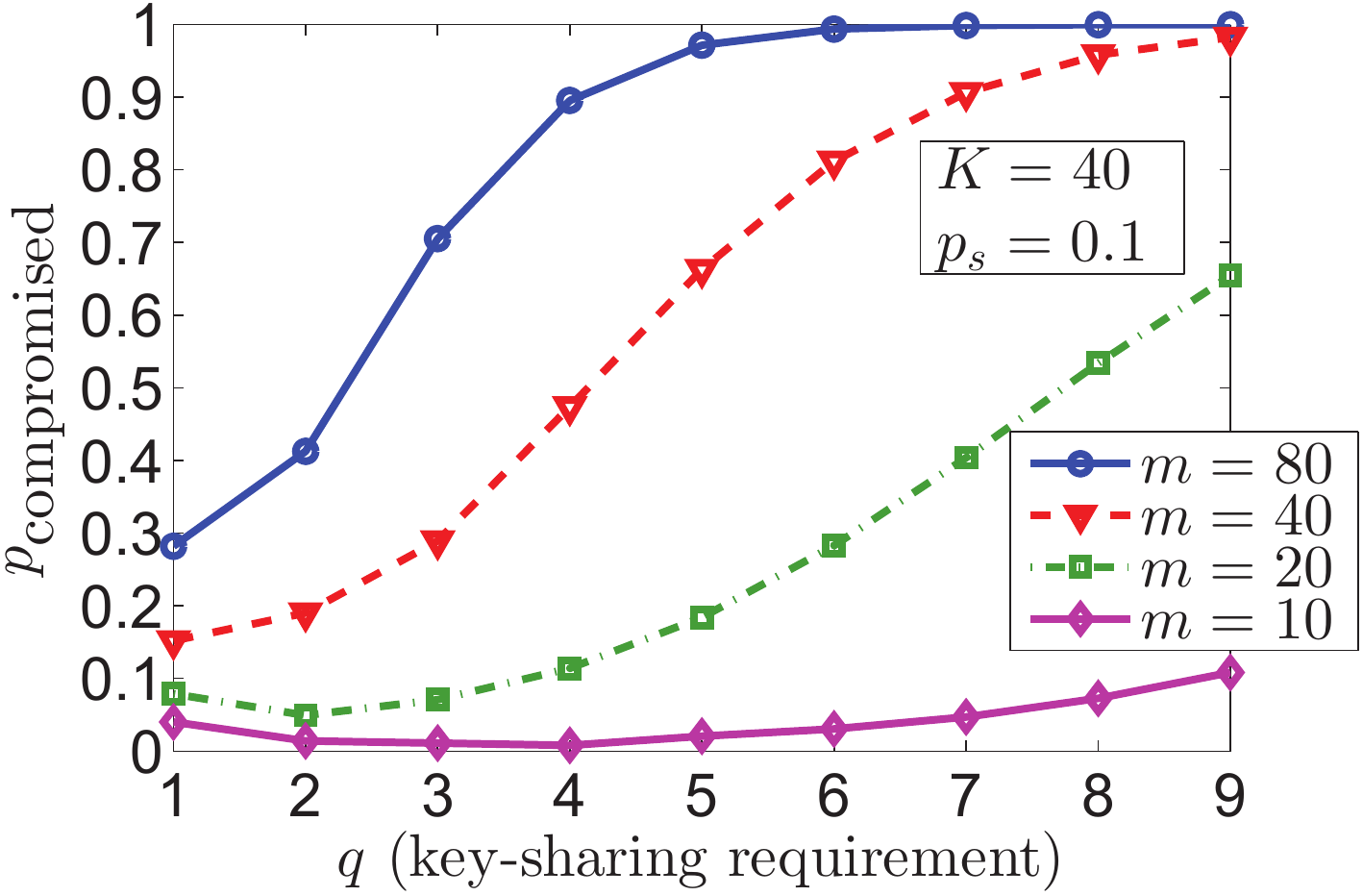}}
 \hspace{0pt}\subfigure[]{\label{fig:capture_vary_q_with_m_K80_ps01}\includegraphics[height=0.165\textwidth]{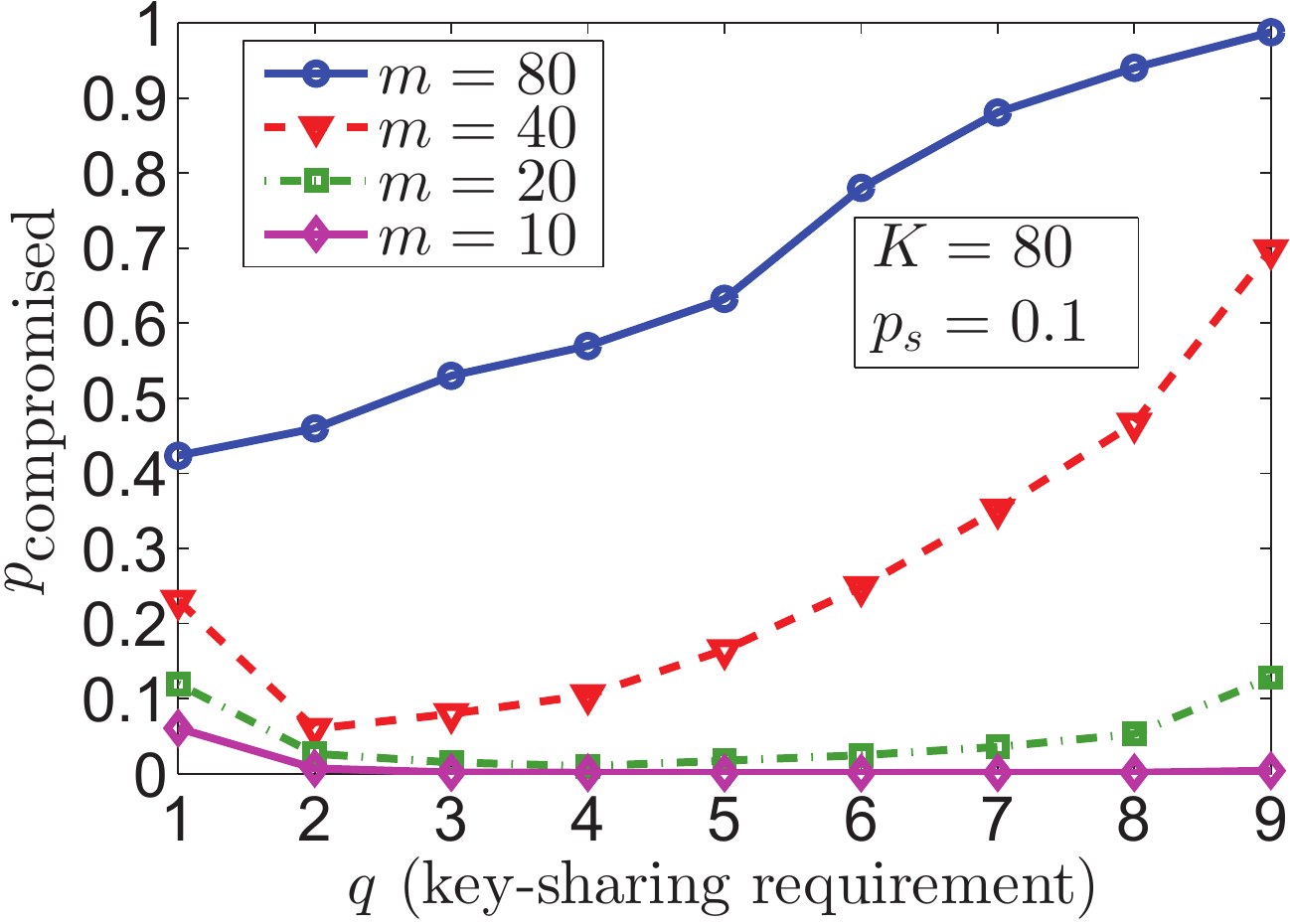}}
 \hspace{0pt}\subfigure[]{\label{fig:capture_vary_q_with_m_K120_ps01}\includegraphics[height=0.165\textwidth]{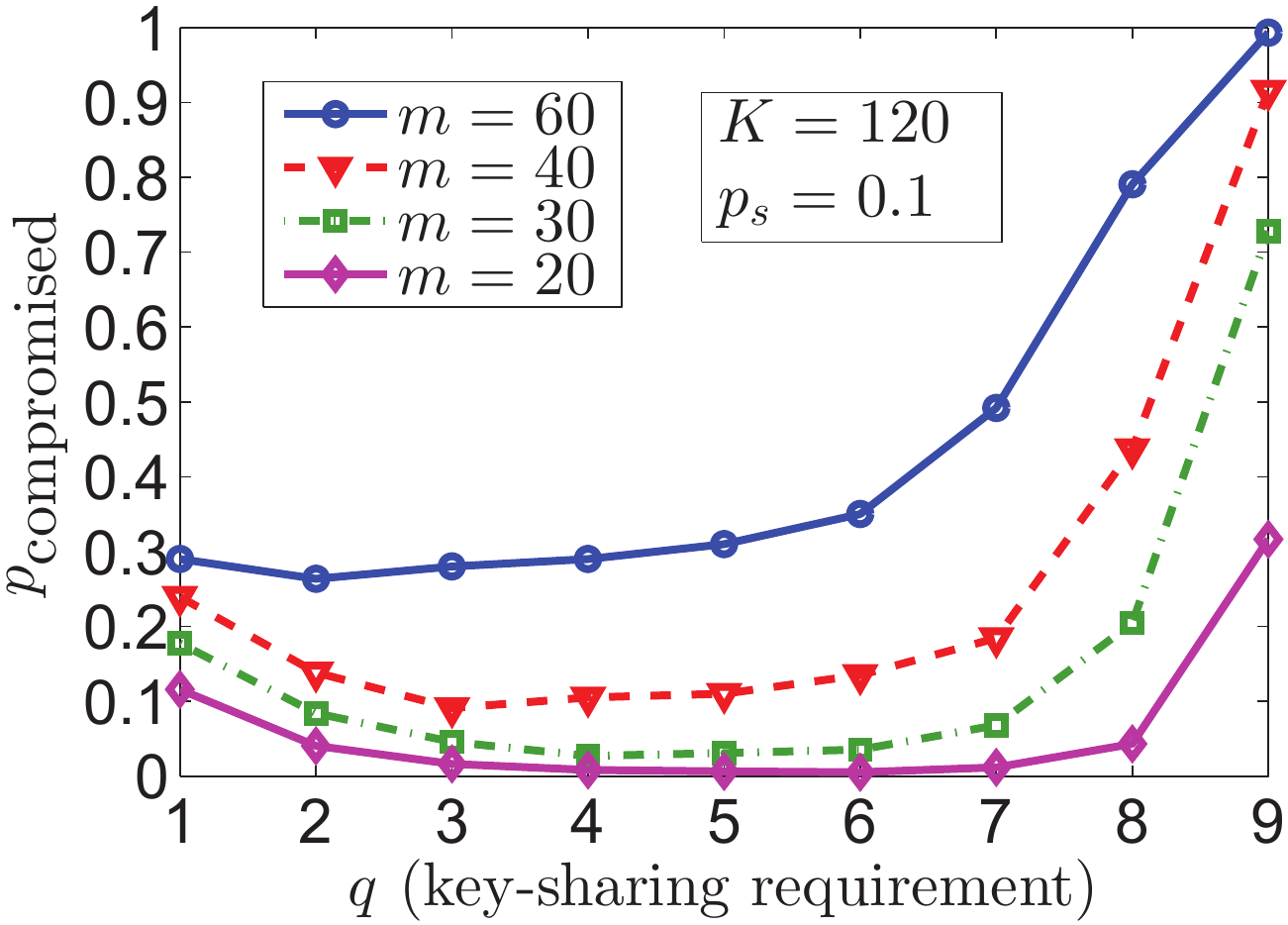}}
\vspace{-9pt}\caption{Each subfigure plots the fraction of
compromised communications between non-captured nodes, $p_{\textnormal{compromised}}$, with respect to $q$ when $m$ nodes are captured, and we fix $K$ and $p_s$ in each subfigure. In different subfigures, we have different $K$ and fix $p_s$ as $0.1$. Similar to Figure \ref{fig:capture_vary_q_with_m_ps005} above, we see that the $q$ that minimizes $p_{\textnormal{compromised}}$ is $\max\{\big\lfloor\frac{K_n}{m}\big\rfloor, 1\}$ and thus in agreement with Corollary \ref{cor:capture_vary_q_with_m}.\vspace{-2pt}} \label{fig:capture_vary_q_with_m_ps01}
\end{figure*}

We present Figures \ref{fig:capture_vary_q_with_m_ps005} and \ref{fig:capture_vary_q_with_m_ps01} on Page \pageref{fig:capture_vary_q_with_m_ps01} to illustrate the impact of scheme parameters to link compromise by an adversary with a given number of captured nodes. Each figure has four subfigures. In each subfigure, we fix $K$ and $p_s$, and study how $p_{\textnormal{compromised}}$ changes with respect to $q$ for different $m$ (we often write $K_n$ and $P_n$ as $K$ and $P$ when not considering their scalings with $n$). Across subfigures of either Figure \ref{fig:capture_vary_q_with_m_ps005} or \ref{fig:capture_vary_q_with_m_ps01}, we consider different $K$, and across some subfigures, we also have different sets of $m$ for better discussion (explained soon). Between Figures \ref{fig:capture_vary_q_with_m_ps005} and \ref{fig:capture_vary_q_with_m_ps01}, we consider different $p_s$. Hence, our experiments and the resulting figure plots provide a comprehensive study to validate our theoretical results.

In all 8 subfigures, we observe that:\vspace{-1pt}
 \begin{itemize}
   \item[\ding{172}] for each curve, $p_{\textnormal{compromised}}$ is indeed minimized at $q=\max\{\big\lfloor\frac{K}{m}\big\rfloor, 1\}$, which confirms Corollary \ref{cor:capture_vary_q_with_m}.
 \end{itemize}
For the above point, our plots address the case of $\big\lfloor\frac{K}{m}\big\rfloor$ being an integer (e.g., $K=40$ with $m=40,20,10$ in Figure \ref{fig:capture_vary_q_with_m_K40_ps005}) as well as the case where $\big\lfloor\frac{K_n}{m}\big\rfloor$ is not an integer (e.g., $K=80$ with $m=30$ in Figure \ref{fig:capture_vary_q_with_m_K80_ps005}, and $K=120$ with $m=50$ in Figure \ref{fig:capture_vary_q_with_m_K120_ps005}). For this reason, some subfigures have different sets of $m$. Furthermore, as explained below, we find that the plots are in agreement with our asymptotic result (\ref{fq}) (i.e., $\frac{p_{\textnormal{compromised}}}{p_s}  \sim q!
  \big(\frac{m}{K_n}\big)^q$):
\begin{itemize}
  \item[\ding{173}] Given $q$, $K$ and $p_s$, $p_{\textnormal{compromised}}$ increases with $m$. This can be seen from that in each subfigure, curves with larger $m$ are always above curves with lower $m$.
  \item[\ding{174}] Given $q$, $m$ and $p_s$, $p_{\textnormal{compromised}}$ decreases with $K$. This can be seen from that for either Figure \ref{fig:capture_vary_q_with_m_ps005} or \ref{fig:capture_vary_q_with_m_ps01}, from subfigure (a) to (d), the curves become lower and lower as $K$ increases (we compare curves with the same $m$).
  \item[\ding{175}] Given $q$, $K$ and $m$, $p_{\textnormal{compromised}}$ increases with $p_s$, This can be seen from that the curves in Figure \ref{fig:capture_vary_q_with_m_K20_ps005} (resp., (b) (c) (d)) are lower than those of Figure \ref{fig:capture_vary_q_with_m_K20_ps01} (resp., (b) (c) (d)) (we compare curves with the same $m$).
\end{itemize}

\subsubsection{\textbf{Impact of scheme parameters to the required number of captured nodes by an adversary with a given fraction of link compromise as its goal}}~

\begin{figure*}
\vspace{0pt}
\addtolength{\subfigcapskip}{-4pt}\centering     
\subfigure[]{\label{fig:capture_vary_q_with_pc1}\includegraphics[height=0.165\textwidth]{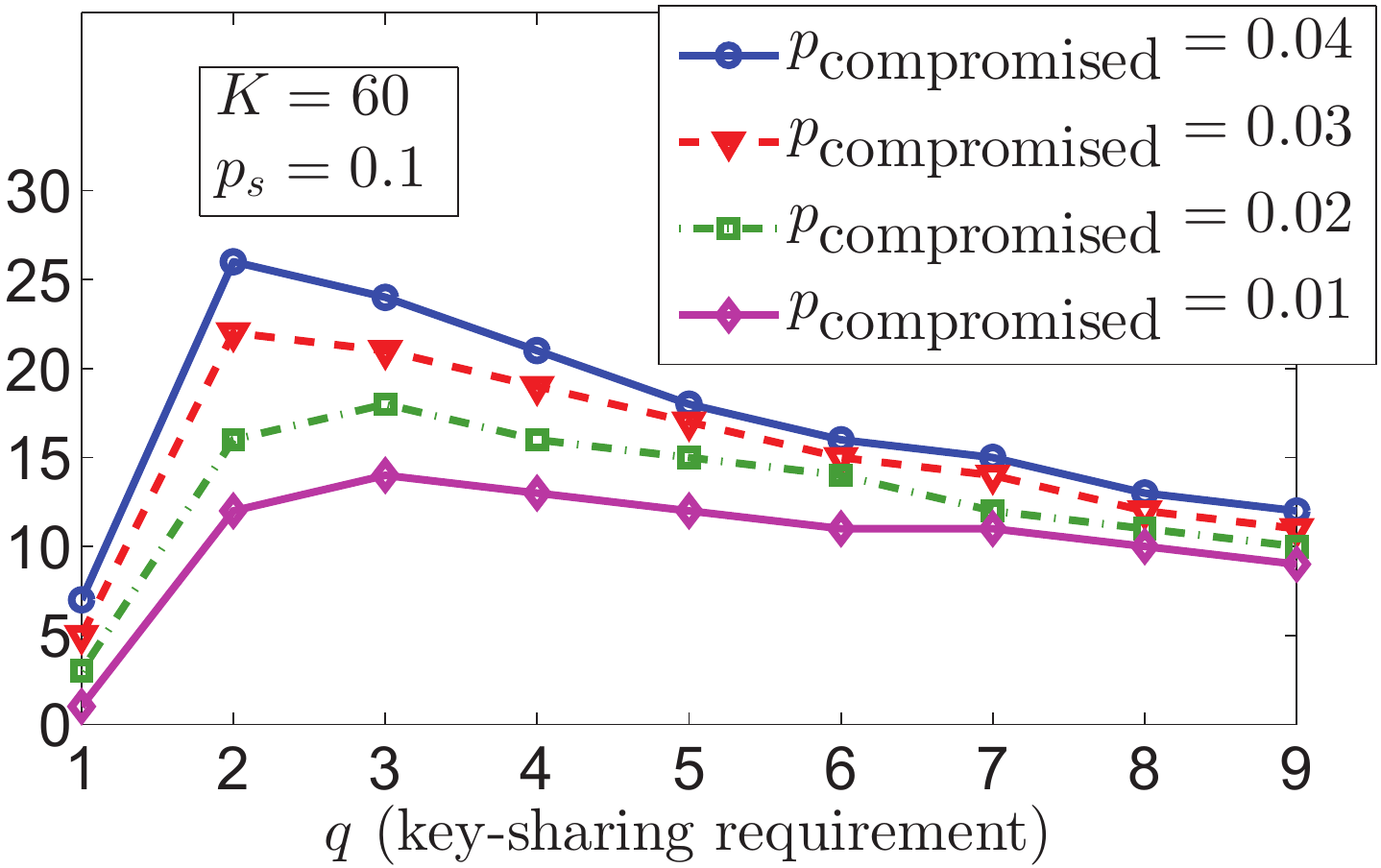}}
\hspace{-2pt}\subfigure[]{\label{fig:capture_vary_q_with_pc2}\includegraphics[height=0.165\textwidth]{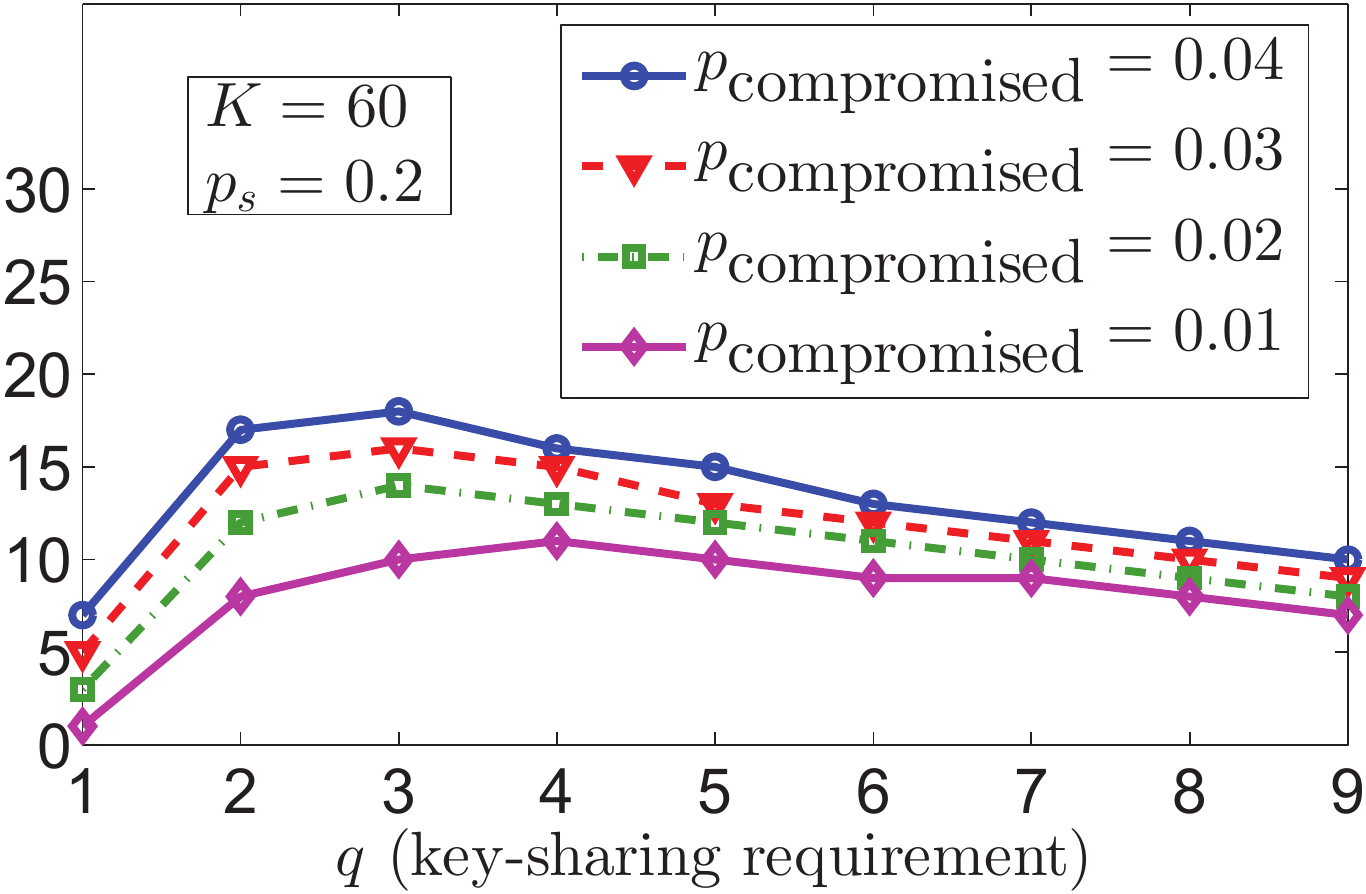}}
\hspace{-2pt}\subfigure[]{\label{fig:capture_vary_q_with_pc3}\includegraphics[height=0.165\textwidth]{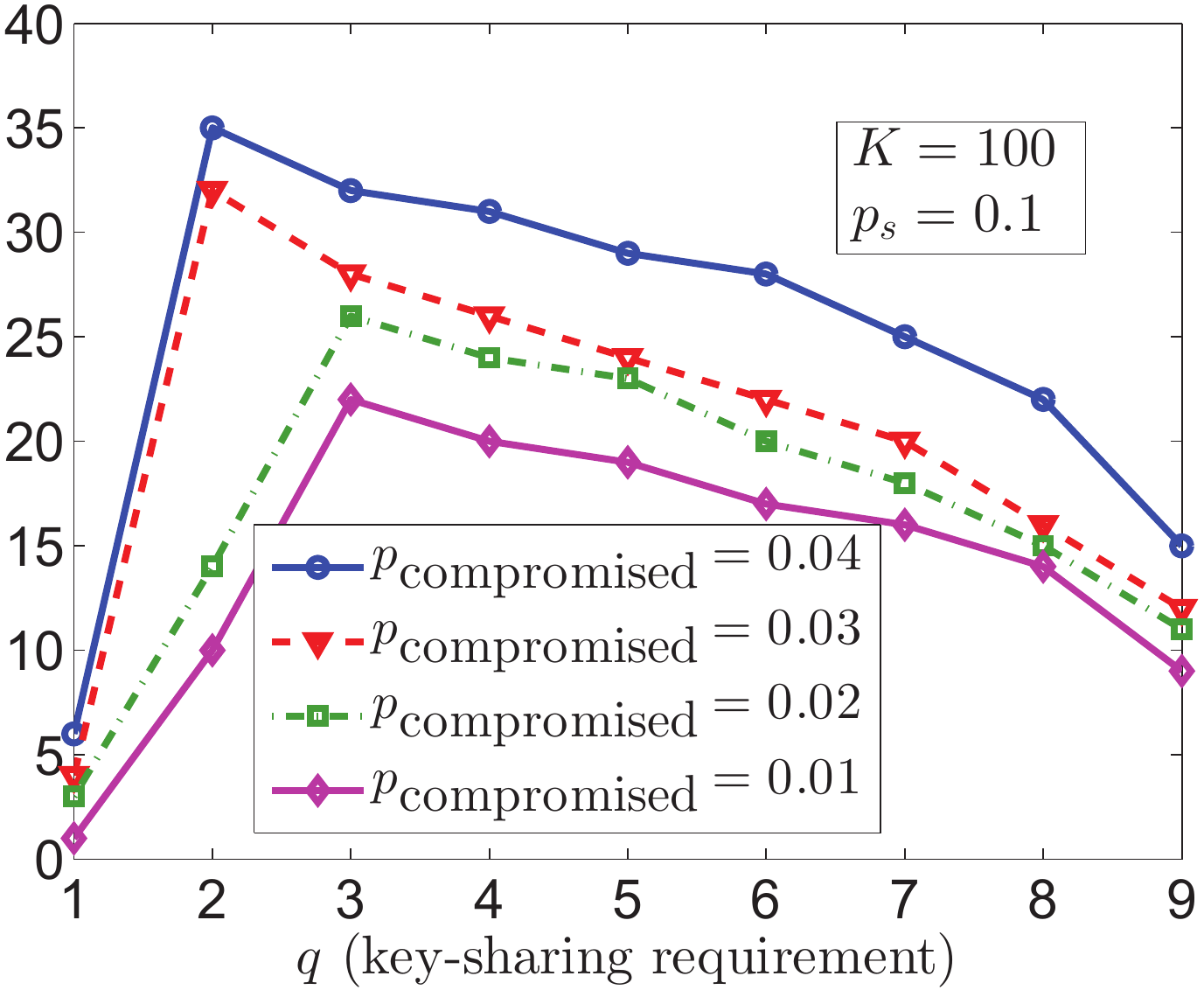}}
 \hspace{-2pt}\subfigure[]{\label{fig:capture_vary_q_with_pc4}\includegraphics[height=0.165\textwidth]{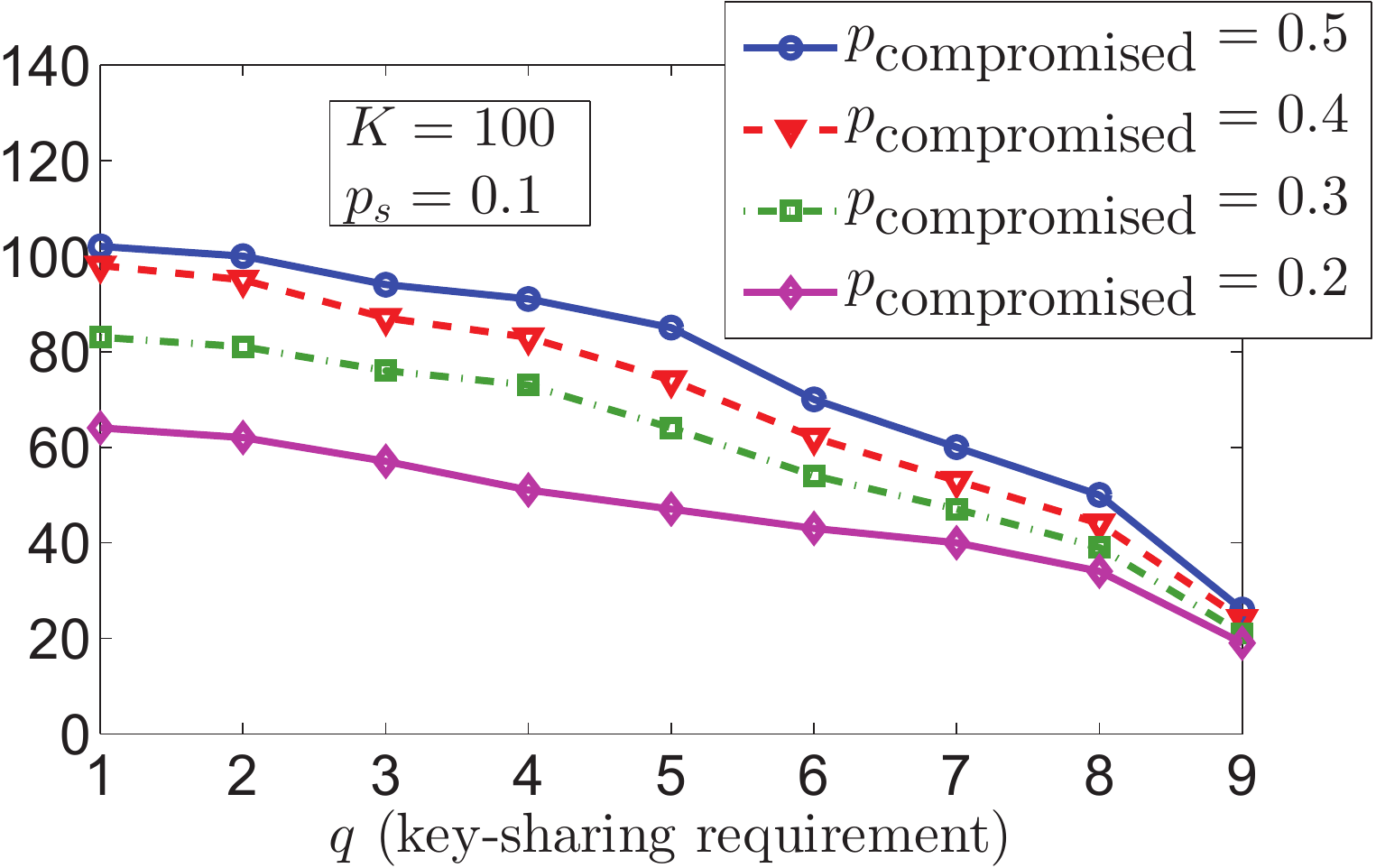}}
\vspace{-10pt}\caption{The $y$ axis stands for the  number of nodes to be captured to ensure $p_{\textnormal{compromised}}$ fraction of
compromised communications between non-captured nodes. In each subfigure, different lines have different $p_{\textnormal{compromised}}$, while different points in each line vary $q$ while fixing $K$ and $p_{\textnormal{compromised}}$. We observe that the $q$ that maximizes $m$ is in accordance with Corollary \ref{cor:capture_vary_q_with_pc}; more specifically, given $\frac{p_{\textnormal{compromised}}}{p_s}$, the optimal $q$ is determined by Table \ref{table-optimal-q-pound}. \vspace{-11.0pt}} \label{fig:capture_vary_q_with_pc}
\end{figure*}

We further present Figure \ref{fig:capture_vary_q_with_pc} on Page \pageref{fig:capture_vary_q_with_m_ps01} to illustrate the impact of scheme parameters to the required number of captured nodes by an adversary whose goal is to compromise a given fraction of links. Figure \ref{fig:capture_vary_q_with_pc} has four subfigures. In each subfigure, we fix $K$ and $p_s$, and study how $m$ changes with respect to $q$ for different $p_{\textnormal{compromised}}$. Between subfigures (a) and (b) of Figure \ref{fig:capture_vary_q_with_pc}, we consider different $p_s$ with the same $K$ and the same set of $p_{\textnormal{compromised}}$. Compared with subfigures (a) and (b), we then consider a partially different set of $p_{\textnormal{compromised}}$ as well as different $p_s$ in Figure \ref{fig:capture_vary_q_with_pc3}. We further look at a different $K$ in Figure \ref{fig:capture_vary_q_with_pc4}. Hence, our experiments here also provide a comprehensive study.

In all 4 subfigures, we observe that:
 \begin{itemize}
   \item[\ding{176}] for each curve, $m$ is  maximized at Corollary \ref{cor:capture_vary_q_with_pc}'s $q^{\#}$, which is determined by Table \ref{table-optimal-q-pound} or Figure \ref{figure-optimal-q-pound} on Page \pageref{table-optimal-q-pound} given $\frac{p_{\textnormal{compromised}}}{p_s}$.
 \end{itemize}

From Corollary \ref{cor:capture_vary_q_with_pc},   $m$ denoting
the number of captured nodes needed to compromise $p_{\textnormal{compromised}}$ fraction of communications  equals
$K \times (\frac{p_{\textnormal{compromised}}/p_s}{q!})^{1/q}$ asymptotically. Then
\begin{itemize}
  \item[\ding{177}] Given $K$ and $p_s$, $m$ increases with $ p_{\textnormal{compromised}}$. This can be seen from that in each subfigure, curves with larger $ p_{\textnormal{compromised}}$ are always above curves with lower $ p_{\textnormal{compromised}}$.
  \item[\ding{178}] Given $K$ and $ p_{\textnormal{compromised}}$, $m$ decreases with $p_s$. This can be seen from that between subfigures (a) and (b) of Figure \ref{fig:capture_vary_q_with_pc}, the curves become lower as $p_s$ increases (we compare curves with the same $ p_{\textnormal{compromised}}$).
  \item[\ding{179}] Given $p_s$ and $ p_{\textnormal{compromised}}$, $m$ decreases with $K$. This can be seen from that between subfigures (a) and (c) of Figure \ref{fig:capture_vary_q_with_pc}, the curves become higher as $K$ increases (we compare curves with the same $ p_{\textnormal{compromised}}$).
\end{itemize}

Summarizing \ding{172}--\ding{179} above, we conclude that the experiments are in accordance with our theoretical findings.

\section{Connectivity under Transmission Constraints} \label{sec:main:res}

In this section, we investigate connectivity of secure sensor network under transmission constraints.

\subsection{Modeling a Secure Sensor Network with the $q$-Composite
Scheme under Transmission Constraints} \label{mds}

We consider an $n$-size secure WSN implementing the $q$-composite
key predistribution scheme. Let $\mathcal {V}
\hspace{-1pt}=\hspace{-1pt} \{\hspace{-1pt}v_1,
\hspace{-1pt}v_2,\hspace{-1pt} \ldots,\hspace{-1.5pt} v_n \}$ denote
the set of sensors. Prior to deployment, each sensor $v_i$
($1\leq i \leq n$) is assigned a key ring independently and
uniformly selected from all $K_n$-size subsets of a key pool with
size $P_n$. The $q$-composite scheme induces a key graph $G_q(n, K_n, P_n)$
 (also known to a random intersection graph \cite{QcompTech14,bloznelis2013,Fill:2000:RIG:340808.340814}), defined on node
set $\mathcal{V}$ such that any two distinct nodes $v_i$ and $v_j$
($1\leq i < j \leq n$) establish an edge (an event denoted by
$K_{ij}$) if and only if they possess at least $q$ key(s) in common.

After the key predistribution, sensors are deployed uniformly and
independently in some network field $\mathcal {A}$. To model transmission constraints, we use the
classic disk model~\cite{ISIT_RKGRGG,Krzywdzi,YaganThesis,pietro2004connectivity}
in which nodes $v_i$ and $v_j$ can communicate (not necessarily
securely) if and only if they are within distance $r_n$ (an event
denoted by $R_{ij}$). The topology resulted from the disk model is
represented by a \emph{random geometric graph}
\cite{ISIT_RKGRGG,Krzywdzi,penrose2016connectivity}
$G_{RGG}(n, r_n, \mathcal{A})$. 

In secure WSNs above with the $q$-composite scheme under the disk
model, nodes $v_i$ and $v_j$ establish a secure link if and only if
events $K_{ij}$ and $R_{ij}$ occur at the same time. Thus, 
 the intersection of
graphs $G_q(n, K_n, P_n)$ and $G_{RGG}(n, r_n, \mathcal{A})$,
denoted by $\mathbb{G}_q(n,K_n, P_n, r_n, \mathcal {A})$, represents
the topology of secure links. We consider $\mathcal {A}$ as a square
of unit area and ignore the boundary effect, so the square is essentially
a torus $\mathcal{T}$. Then the network is modeled by graph
$\mathbb{G}_q(n,K_n, P_n, r_n, \mathcal {T})$. In the rest of the paper, $\mathbb{P}[\mathcal
{E}]$ denotes the probability that event $\mathcal {E}$ happens.

\subsection{Connectivity in the Absence of Node Capture}

In the following Theorem \ref{thm:gqt}, we present connectivity
results in the absence of node capture.

%
%
%

\begin{thm} \label{thm:gqt}


\hspace{1pt}In \hspace{1pt}graph\hspace{1pt} $\mathbb{G}_q(n,
\hspace{1pt} K_n, \hspace{1pt} P_n, \hspace{1pt} r_n, \hspace{1pt}
\mathcal {T})$ with $r_n \hspace{1pt}\leq \hspace{1pt} \frac{1}{2}$,
\\$K_n = \omega(\ln n)$,
$\frac{{K_n}^2}{P_n}  =  o(1)$ and $\frac{K_n}{P_n} =
o\big(\frac{1}{n}\big)$, assume that
\begin{align}
p_s \cdot \pi {r_n}^{2} & \sim a\cdot \frac{\ln n}{n} \label{psc}
\end{align}
holds for some constant $a>0$. Then as $n \to \infty$, we have
\begin{subnumcases}  {\hspace{-17pt}\mathbb{P}\bigg[
   \begin{array}{l} \mathbb{G}_q(n,K_n, P_n, r_n, \mathcal {T})\\ \textrm{ is
   connected.} \end{array}\hspace{-2pt}
\bigg] \to} \hspace{-2pt} \normalsize \selectfont 0, \textrm{~~if
$a
<1$}, \label{thm:gqt:010} \\
\hspace{-2pt} 1, \textrm{~~if $a
>1$}.  \label{thm:gqt:011}
 \end{subnumcases}

 \end{thm}

\begin{rem}
The term (\ref{psc}) is the edge probability of the graph $\mathbb{G}_q(n,
\hspace{1pt} K_n, \hspace{1pt} P_n, \hspace{1pt} r_n, \hspace{1pt}
\mathcal {T})$ (i.e., the probability of a secure link between two sensors).
(\ref{thm:gqt:010}) and (\ref{thm:gqt:011}) constitute a sharp zero--one
law \cite{YaganThesis,citeulike:505396} for connectivity, in the sense that if the edge probability is slightly greater (resp., smaller) than $\frac{\ln n}{n}$, the graph becomes connected (resp., disconnected) quickly. This also implies that in the graph intersection $\mathbb{G}_q(n,
\hspace{1pt} K_n, \hspace{1pt} P_n, \hspace{1pt} r_n, \hspace{1pt}
\mathcal {T})$, a critical threshold for connectivity is given by the edge probability being $\frac{\ln n}{n}$. This result is similar to that for the intersection  of an Erd\H{o}s--R\'{e}nyi
graph and a random geometric graph discussed on Page \pageref{Penroseground}. In fact, we will prove the one-law (\ref{thm:gqt:011}) by bridging these two graph intersections (the details are given in Section VI of the full version \cite{full} due to space limitation).

 .

 \end{rem}

\begin{rem} \label{cor1}

With $p_s \cdot \pi {r_n}^{2}$ in (\ref{psc}) replaced by
$\frac{1}{q!} \Big( \frac{{K_n}^2}{P_n} \Big)^{q}$; i.e., with
(\ref{psc}) replaced by
\begin{align}
\frac{1}{q!} \bigg( \frac{{K_n}^2}{P_n} \bigg)^{q} \cdot \pi
{r_n}^{2} & \sim a \cdot \frac{\ln n}{n},  \label{KnPnqlnn}
\end{align}
and with all other conditions in Theorem \ref{thm:gqt} unchanged,
then as $n \to \infty$, (\ref{thm:gqt:010}) and (\ref{thm:gqt:011})
still follow. This result is true since
(\ref{KnPnqlnn}) and (\ref{pssimq}) on Page \pageref{pssimq} clearly induce (\ref{psc}).

 \end{rem}

The proof of Theorem \ref{thm:gqt} is given in Section VI of the full version \cite{full} due to space limitation.

\subsection{Connectivity in the Presence of Node Capture}

We now consider that the adversary has captured some \emph{random}
set of $m$ nodes, where $m < n$. Let the topology formed by the
$(n-m)$ non-captured nodes be graph $\mathbb{H}_q(n, m, K_n, P_n,
r_n, \mathcal {T})$, which is statistically equivalent to the random
graph $\mathbb{G}_q(n-m,K_n, P_n, r_n, \mathcal {T})$ since the
non-captured nodes are uniformly distributed on $\mathcal {T}$.
Hence, by replacing $\frac{\ln n}{n}$ with $\frac{\ln (n-m)}{n-m}$
in (\ref{psc}) of Theorem \ref{thm:gqt}, we obtain Theorem
\ref{thm:gqt:comp} below.

\begin{thm} \label{thm:gqt:comp}

\hspace{1pt}In graph $\mathbb{H}_q(n, m, K_n, P_n, r_n, \mathcal
{T})$ with $r_n \hspace{1pt}\leq \hspace{1pt} \frac{1}{2}$,
\\$K_n = \omega(\ln n)$,
$\frac{{K_n}^2}{P_n}  =  o(1)$ and $\frac{K_n}{P_n} =
o\big(\frac{1}{n}\big)$, assume
that
\begin{align}
p_s \cdot \pi {r_n}^{2} & \sim a\cdot \frac{\ln
(n-m)}{n-m}\label{psc:comp}
\end{align}
holds for some constant $a>0$. Then as $n \to \infty$, we have\vspace{2pt}
\begin{subnumcases}  {\hspace{-19pt}\mathbb{P}\bigg[
   \begin{array}{l} \mathbb{H}_q(n,m,K_n, P_n, r_n, \mathcal {T})\\ \textrm{ is
   connected.} \end{array}\hspace{-2pt}
\bigg] \to  } \hspace{-2pt} \normalsize \selectfont 0, \textrm{~~if
$a
<1$}, \label{thm:gqt:010:comp} \\[-2pt]
\hspace{-2pt} 1, \textrm{~~if $a
>1$}.  \label{thm:gqt:011:comp}
 \end{subnumcases}


 \end{thm}

\begin{rem} \label{cor1:compx}

Similar to Remark \ref{cor1}, with $p_s \cdot \pi {r_n}^{2}$ in
(\ref{psc:comp}) replaced by $\frac{1}{q!} \Big( \frac{{K_n}^2}{P_n}
\Big)^{q}$; i.e., with (\ref{psc:comp}) replaced by \vspace{-3pt}
\begin{align}
\frac{1}{q!} \bigg( \frac{{K_n}^2}{P_n} \bigg)^{q} \cdot \pi
{r_n}^{2} & \sim a \cdot\frac{\ln (n-m)}{n-m} ,
\label{KnPnqlnn:compx}
\end{align}
and with all other conditions in Theorem \ref{thm:gqt:comp}
unchanged, then as $n \to \infty$, (\ref{thm:gqt:010:comp}) and
(\ref{thm:gqt:011:comp}) still hold. This result follows because
 (\ref{KnPnqlnn:compx}) and (\ref{pssimq}) on Page \pageref{pssimq} yield (\ref{psc:comp}).

 \end{rem}

 Based on our   results, now we provide design guidelines
of secure sensor networks for connectivity.

\textbf{Design guideline for connectivity in the presence/absence of node capture:} In an $n$-size secure wireless
sensor network employing the $q$-composite scheme with key ring size
$K_n$ and key pool size $P_n$ and working under the disk model in
which two sensors have to be within distance $r_n $
for communication,
\begin{itemize}
  \item in the absence of node capture,
 from $\frac{1}{q!} \big( \frac{{K_n}^2}{P_n} \big)^{q} \cdot \pi
{r_n}^{2}   = \frac{\ln n}{n}$ (i.e., replacing ``$\sim$'' with ``$=$'' and letting $a=1$ in (\ref{KnPnqlnn})),
\begin{itemize}
  \item the
\emph{critical} key ring size for connectivity is given by
\begin{align}
(q!/\pi)^{\frac{1}{2q}}
 \big(n^{-1}\ln n\big)^{\frac{1}{2q}} {P_n}^{\frac{1}{2}}
{r_n}^{-\frac{1}{q}},\label{Knstarqnewx}
\end{align}
  \item the
\emph{critical} key pool size for connectivity equals
\begin{align}
 (\pi / q!)^{\frac{1}{q}}(n/\ln n)^{\frac{1}{q}} {K_n}^2
{r_n}^\frac{2}{q}, \label{Pnstarqnewx}
\end{align}
  \item the
\emph{critical} transmission range for connectivity equals
\begin{align}
 \sqrt{{q!\ln n}/{( \pi n)}} \cdot\big( {P_n}/{{K_n}^2} \big)^{q/2}. \label{rnstarqnewx}
\end{align}
\end{itemize}
  \item and if $m$ nodes have already been captured, from $\frac{1}{q!} \big( \frac{{K_n}^2}{P_n} \big)^{q} \cdot \pi
{r_n}^{2}   =  \frac{\ln (n-m)}{n-m}$ (i.e., replacing ``$\sim$'' with ``$=$'' and letting $a=1$ in (\ref{KnPnqlnn:compx})),
\begin{itemize}
  \item the
\emph{critical} key ring size for connectivity is given by
\begin{align}
(q!/\pi)^{\frac{1}{2q}}
 \big[(n-m)^{-1}\ln (n-m)\big]^{\frac{1}{2q}} {P_n}^{\frac{1}{2}}
{r_n}^{-\frac{1}{q}},\label{Knstarqnewx_nodecapture}
\end{align}
  \item the
\emph{critical} key pool size for connectivity equals
\begin{align}
 (\pi / q!)^{\frac{1}{q}}[(n-m)/\ln (n-m)]^{\frac{1}{q}} {K_n}^2
{r_n}^\frac{2}{q}, \label{Pnstarqnewx_nodecapture}
\end{align}
  \item the
\emph{critical} transmission range for connectivity equals
\begin{align}
 \sqrt{{q!\ln (n-m)}/{[ \pi (n-m)]}} \cdot\big( {P_n}/{{K_n}^2} \big)^{q/2}. \label{rnstarqnewx_nodecapture}
\end{align}
\end{itemize}
\end{itemize}

\subsection{Practicality of Theorem Conditions}\label{sec:pra:con} \vspace{-3pt}

We check the practicality of conditions in Theorems \ref{thm:gqt}
and \ref{thm:gqt:comp}. Since the whole region has a unit area,
clearly the condition $r_n \leq \frac{1}{2}$ holds trivially in practice. $K_n$
controls the number of keys in each sensor's memory. In real-world
implementations, $K_n$ is often larger
\cite{YaganThesis,QcompTech14} than $\ln n$, so $K_n = \omega(\ln
n)$ follows. As concrete examples, we have $\ln 1000 \approx 6.9$,
$\ln 5000 \approx 8.5$ and $\ln 10000 \approx 9.2$. $K_n$ is much
smaller compared to both $n$ and $P_n$ due to constrained memory and
computational resources of sensors \cite{virgil,adrian,YaganThesis}.
Thus, $\frac{{K_n}^2}{P_n} = o(1)$ and $\frac{K_n}{P_n} =
o\big(\frac{1}{n}\big)$ hold in practice. Also, $P_n$ is larger
\cite{virgil,adrian,DiPietroTissec} than $n$, so $P_n = \omega (n
\ln n)$ is also practical. As examples, we have $1000 \ln 1000
\approx 6907$, $ 5000\ln 5000 \approx 42585$ and $10000\ln 10000
\approx 92103$.\vspace{-6pt}

\subsection{Experiments}\vspace{-3pt}

\begin{figure*}
\vspace{0pt}
\addtolength{\subfigcapskip}{-4pt}\centering     
\hspace{-2pt}\subfigure[]{\label{fig:vary_K_under_P}\includegraphics[height=0.175\textwidth]{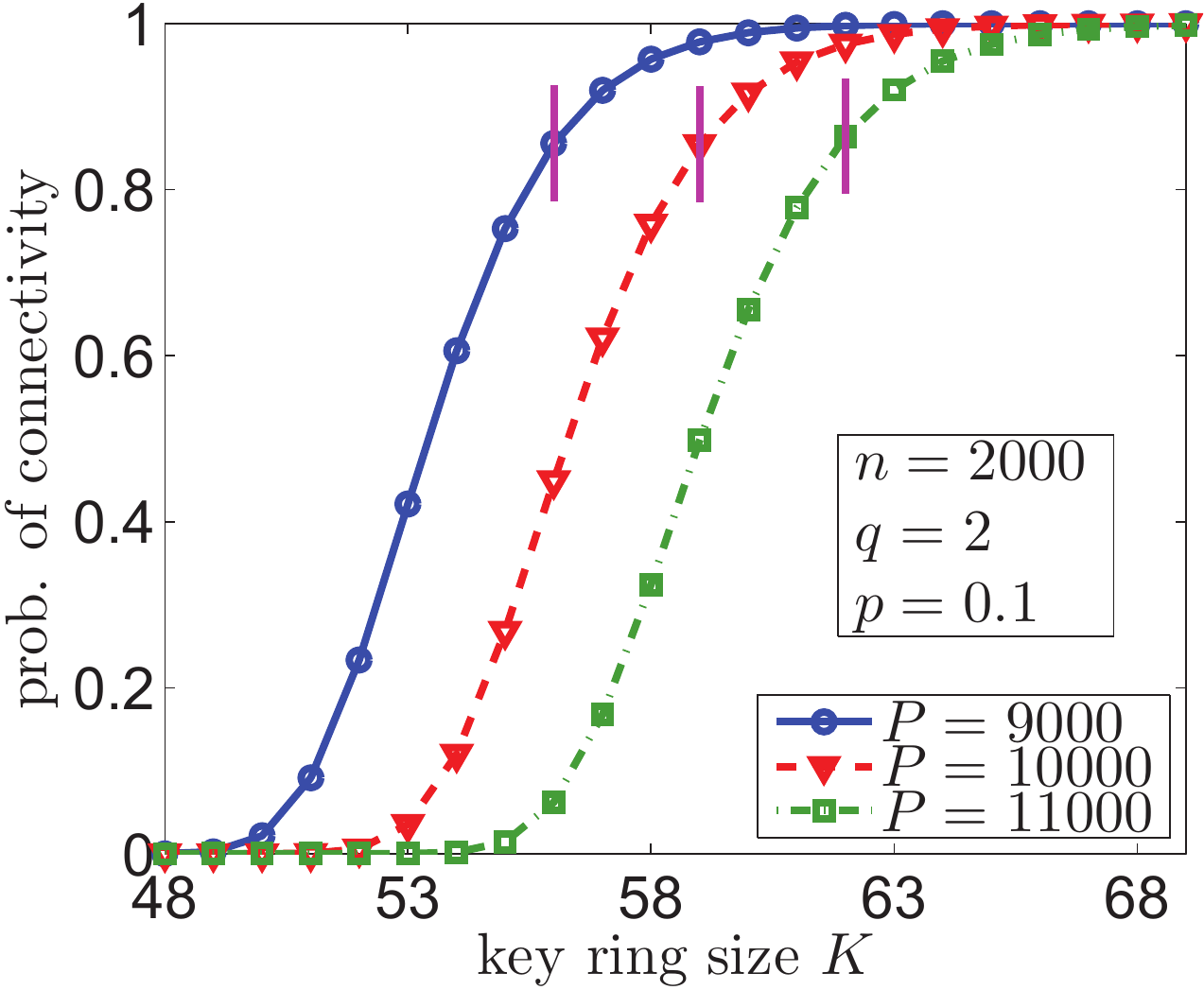}}
\hspace{-2pt}\subfigure[]{\label{fig:vary_K_under_n}\includegraphics[height=0.175\textwidth]{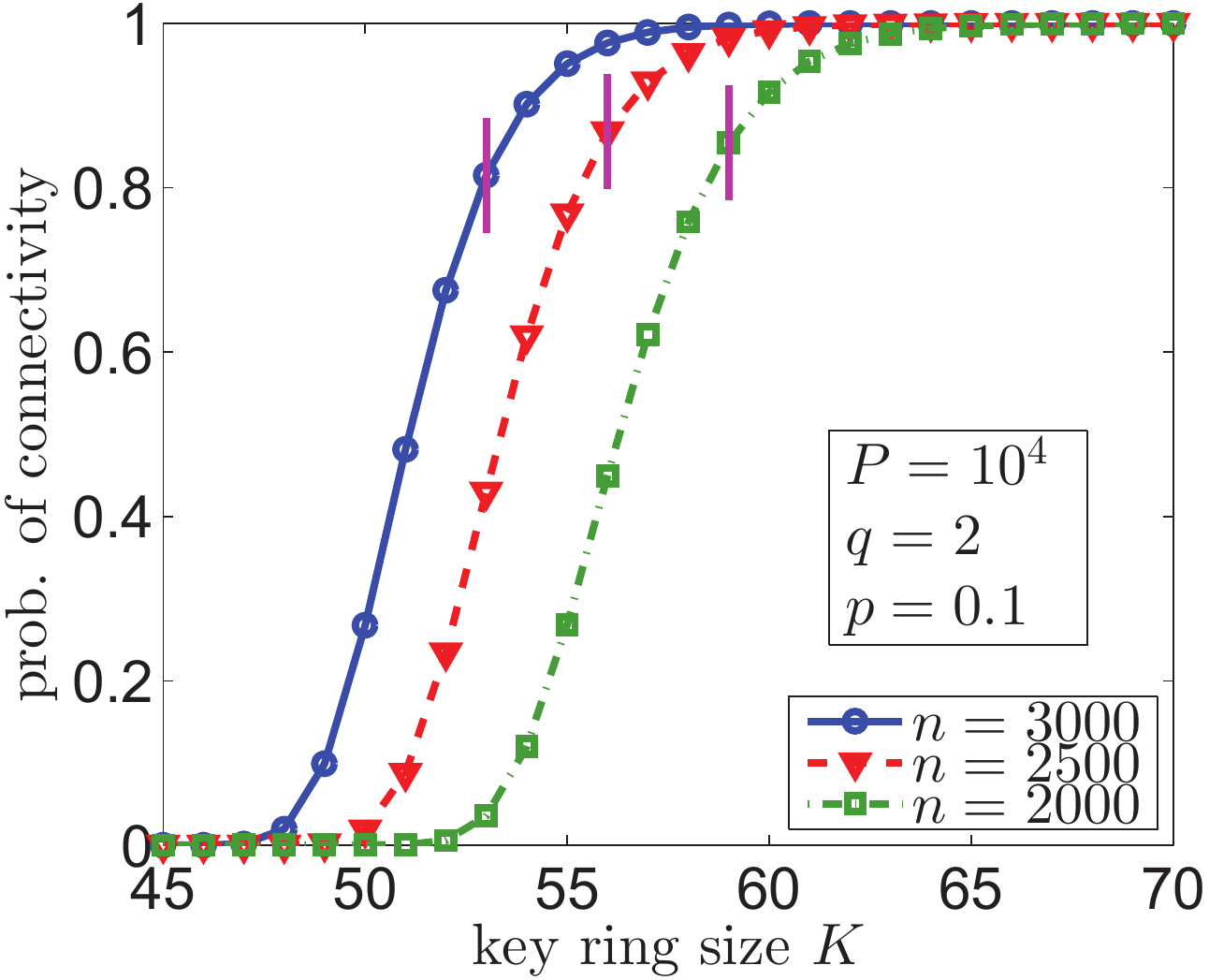}}
\hspace{-2pt}\subfigure[]{\label{fig:vary_K_under_r}\includegraphics[height=0.175\textwidth]{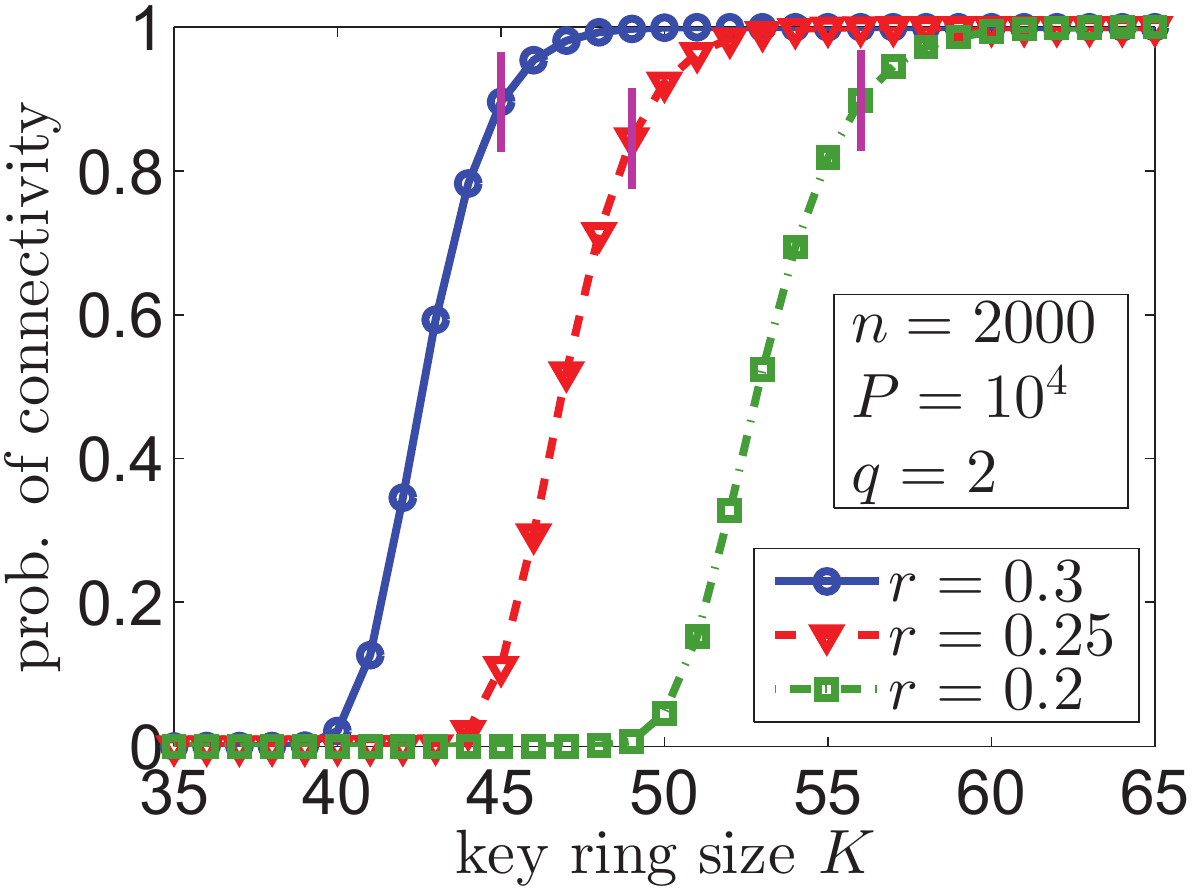}}
\hspace{-2pt}\subfigure[]{\label{fig:vary_K_under_q}\includegraphics[height=0.175\textwidth]{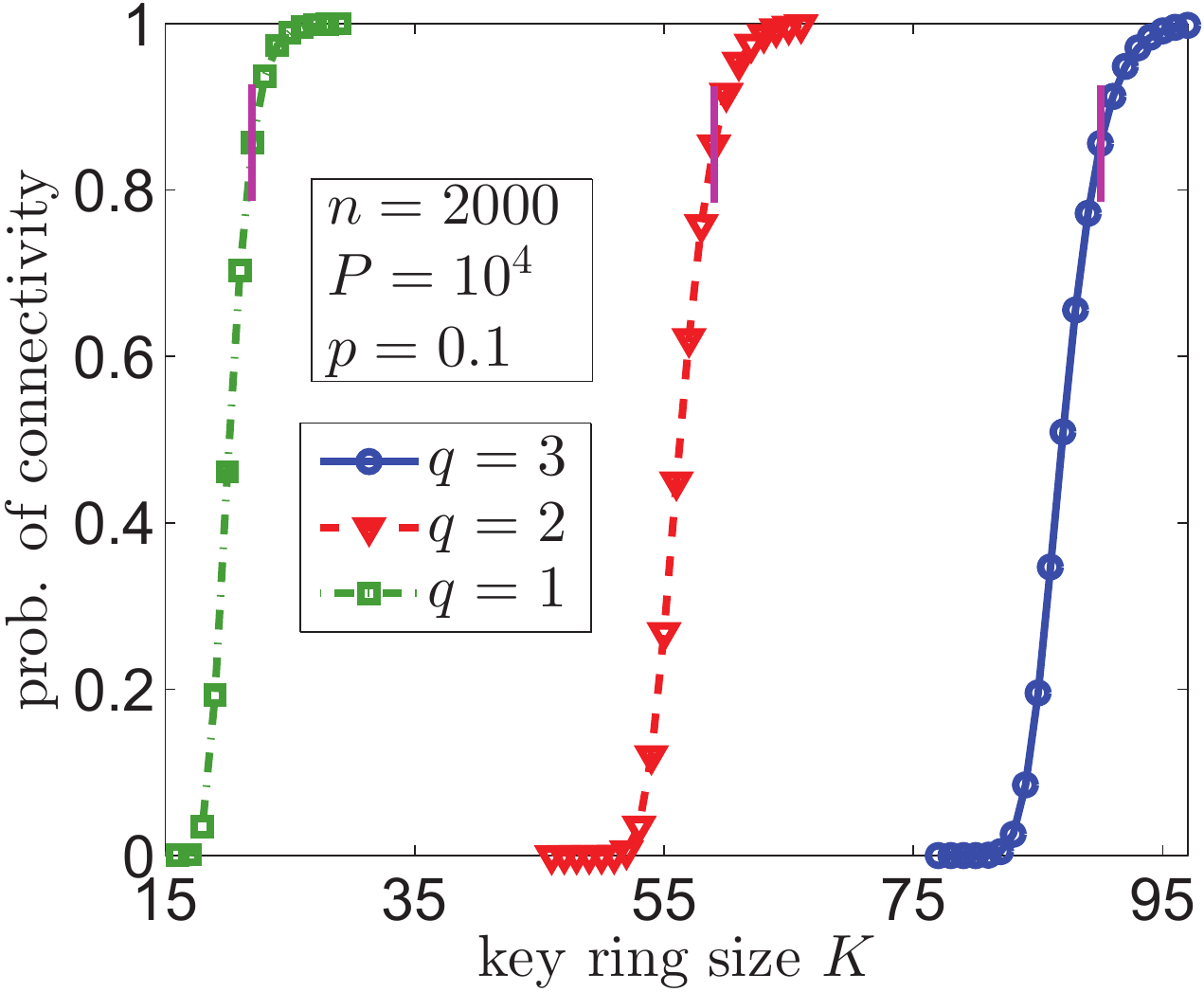}}
\vspace{-9pt}\caption{We plot the connectivity probability of the network $\mathbb {G}_q(n, \hspace{1pt} K, \hspace{1pt} P, \hspace{1pt} r, \hspace{1pt} \mathcal {T})$ with the $q$-composite scheme under transmission constraints, with respect to different key ring sizes $K$. Each vertical line presents the \emph{critical} key ring size given by Eq. (\ref{Knstarqnewx}). \vspace{-10pt}} \label{fig:vary_K}
\end{figure*}

\begin{figure*}
\vspace{0pt}
\addtolength{\subfigcapskip}{-4pt}\centering     
\hspace{-2pt}\subfigure[]{\label{fig:vary_P_under_K}\includegraphics[height=0.165\textwidth]{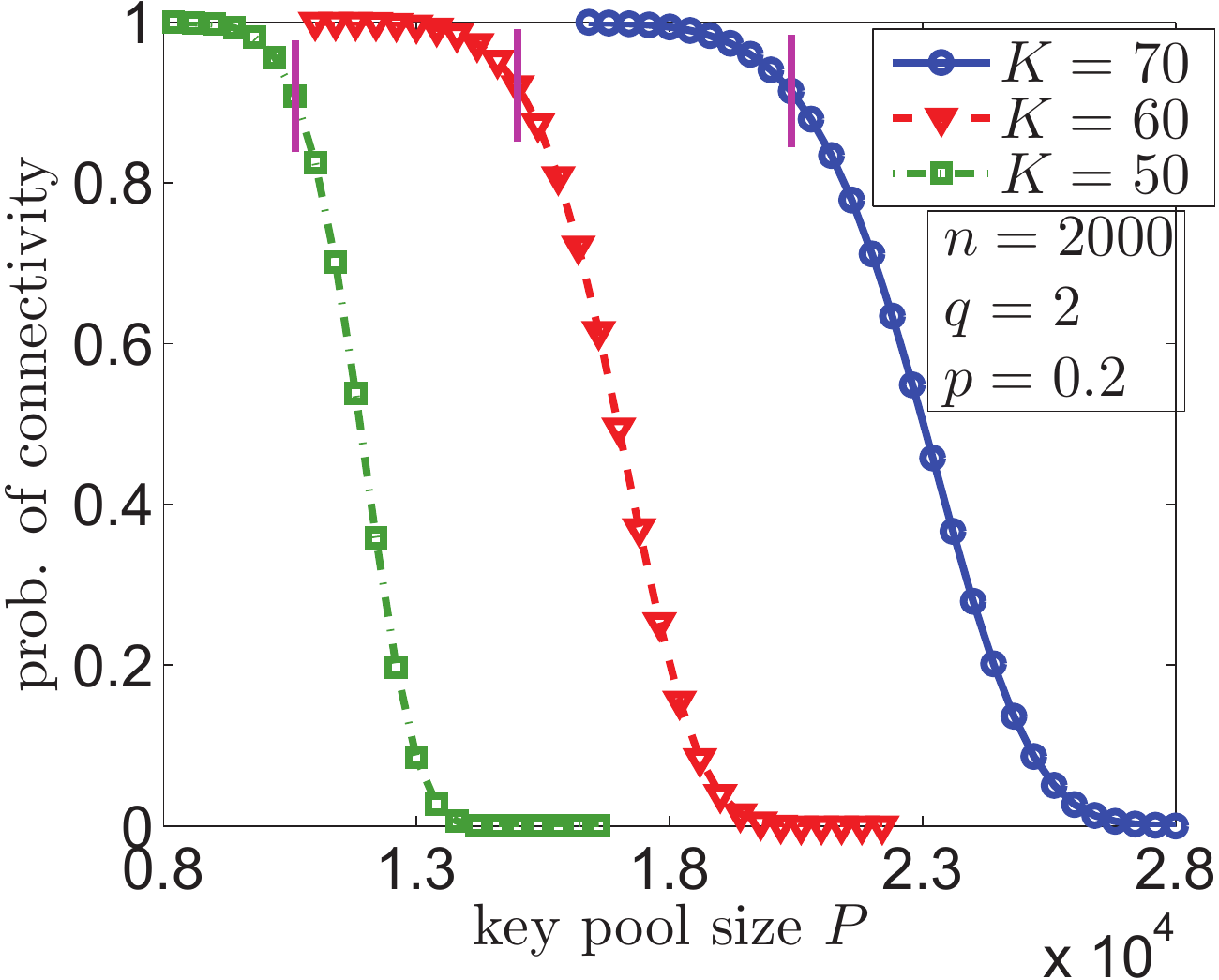}}
 \hspace{-2pt}\subfigure[]{\label{fig:vary_P_under_n}\includegraphics[height=0.165\textwidth]{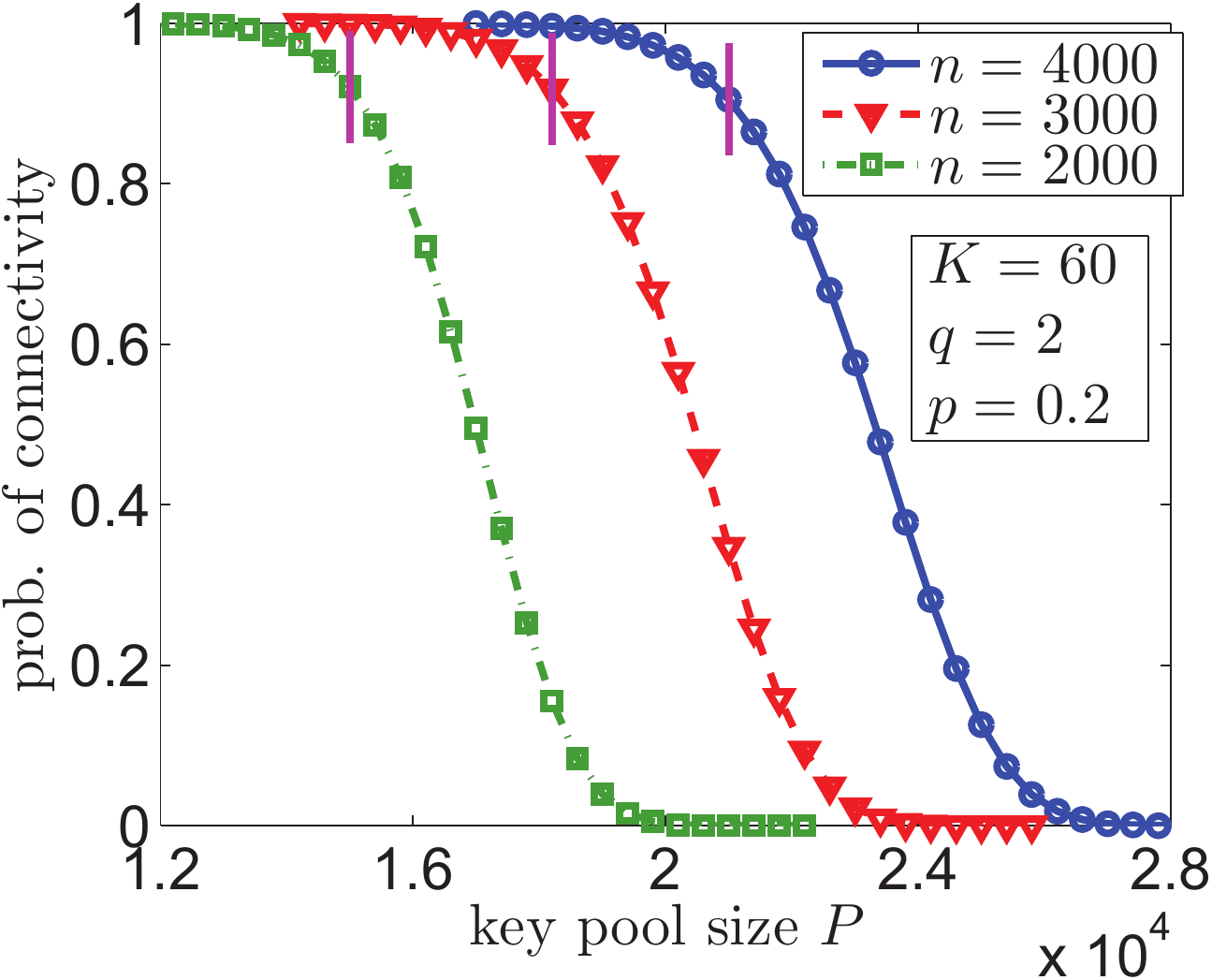}}
\hspace{-2pt}\subfigure[]{\label{fig:vary_P_under_r}\includegraphics[height=0.165\textwidth]{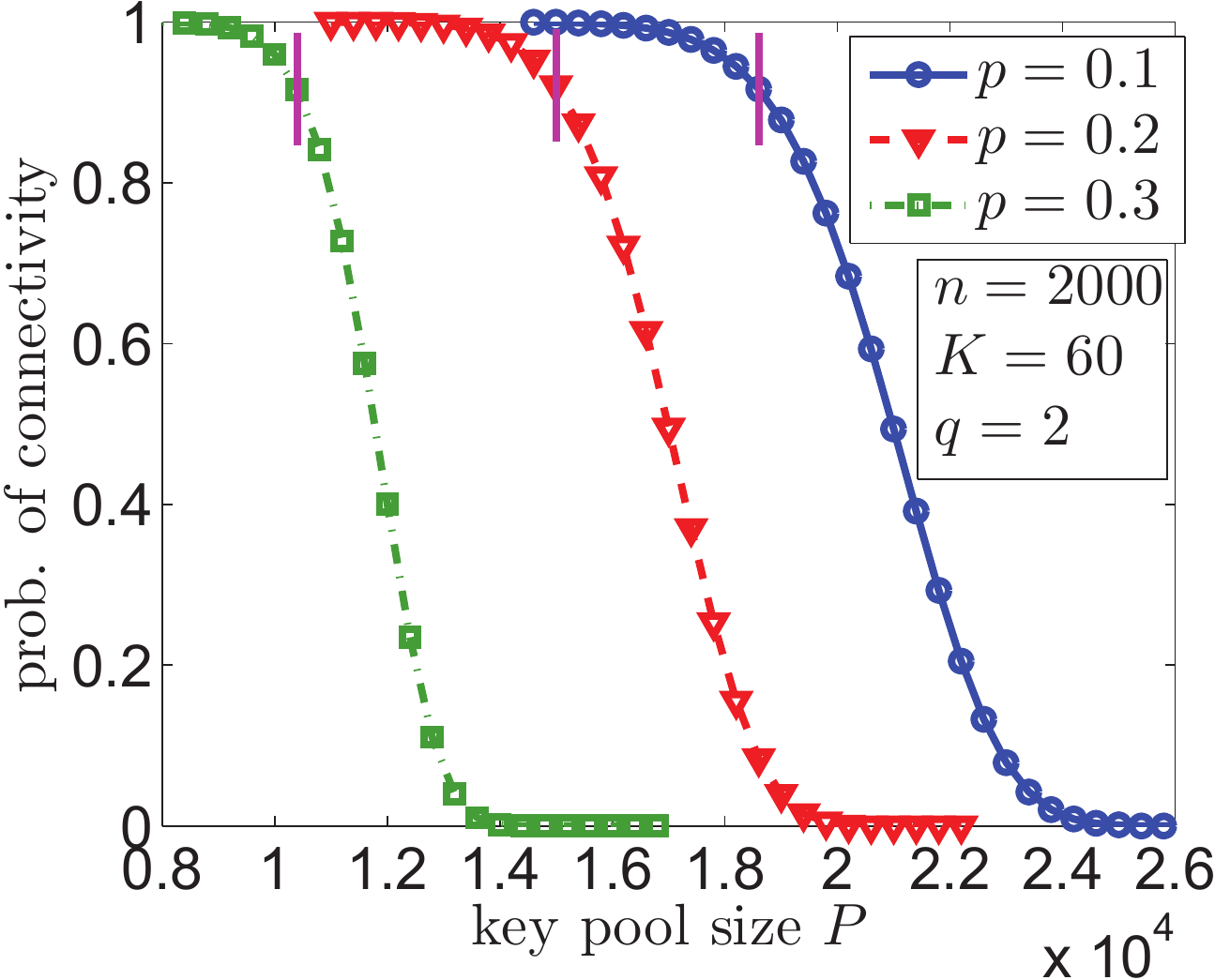}}
 \hspace{-2pt}\subfigure[]{\label{fig:vary_P_under_q}\includegraphics[height=0.165\textwidth]{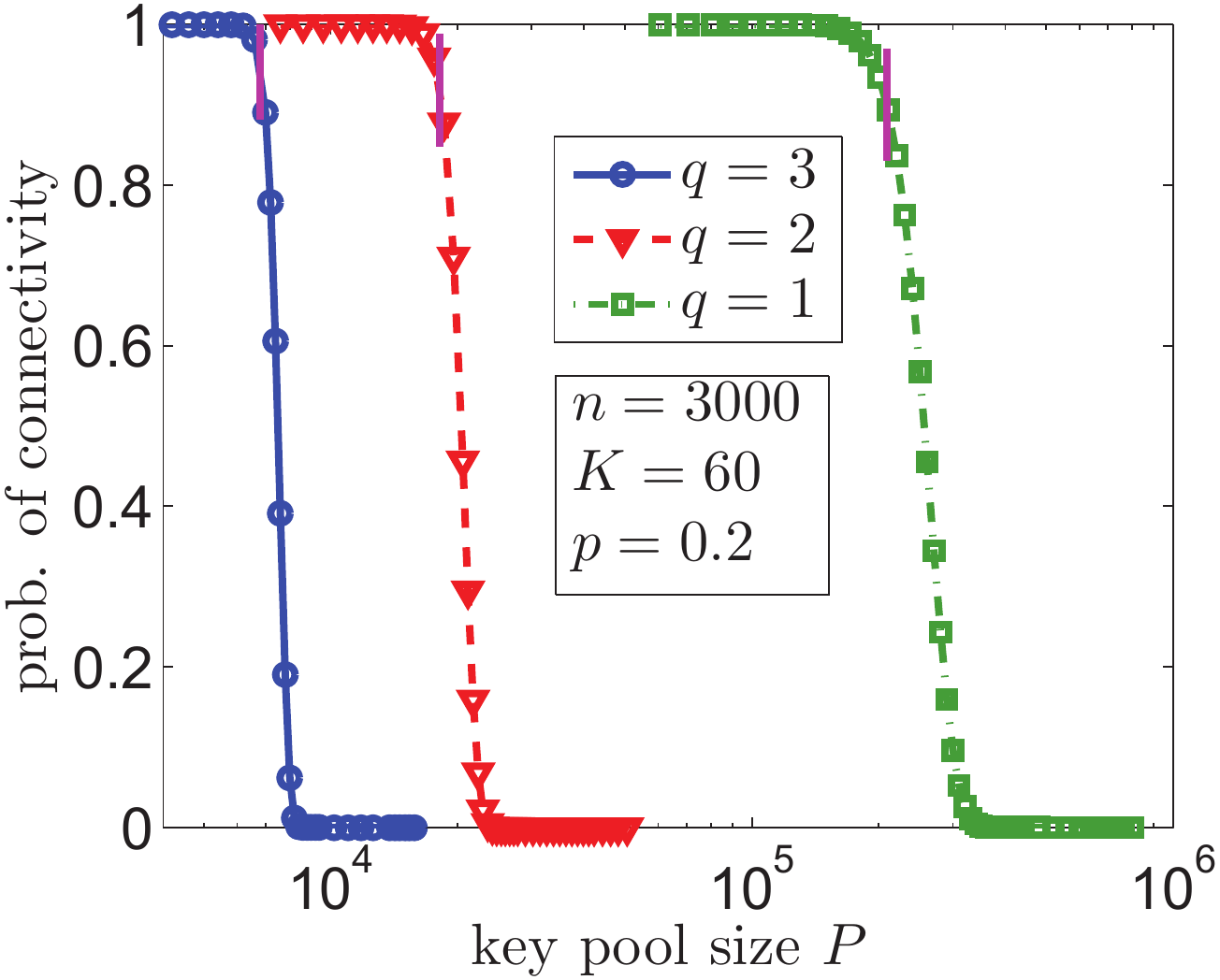}}
\vspace{-9pt}\caption{We plot the connectivity probability of the network $\mathbb{G}_q(n, \hspace{1pt} K, \hspace{1pt} P, \hspace{1pt} r, \hspace{1pt} \mathcal {T})$ with the $q$-composite scheme under transmission constraints, with respect to different key pool sizes $P$. Each vertical line presents the \emph{critical} key pool size given by Eq. (\ref{Pnstarqnewx}).\vspace{-10pt}} \label{fig:vary_P}
\end{figure*}

\begin{figure*}
\vspace{0pt}
\addtolength{\subfigcapskip}{-4pt}\centering     
\hspace{-2pt}\subfigure[]{\label{fig:vary_r_under_K}\includegraphics[height=0.165\textwidth]{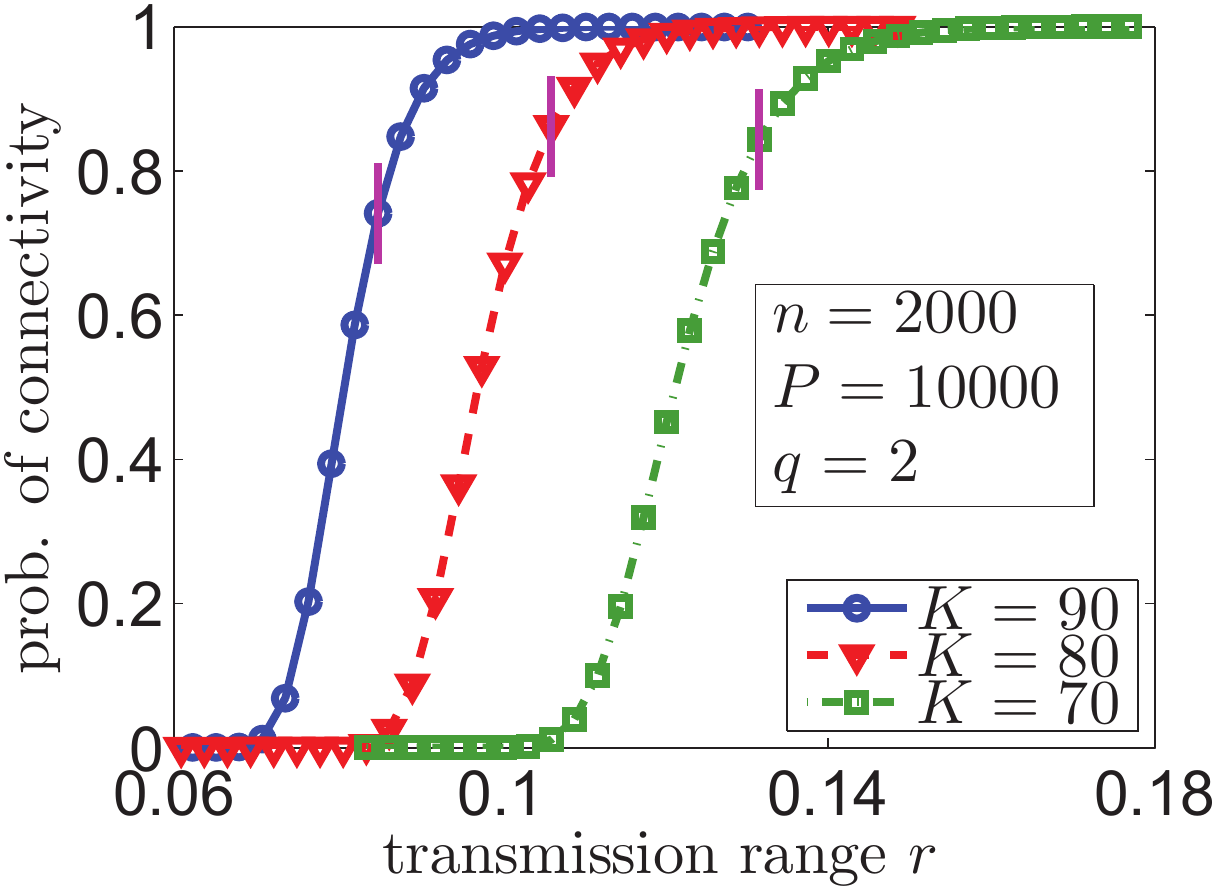}}
\hspace{-2pt}\subfigure[]{\label{fig:vary_r_under_P}\includegraphics[height=0.165\textwidth]{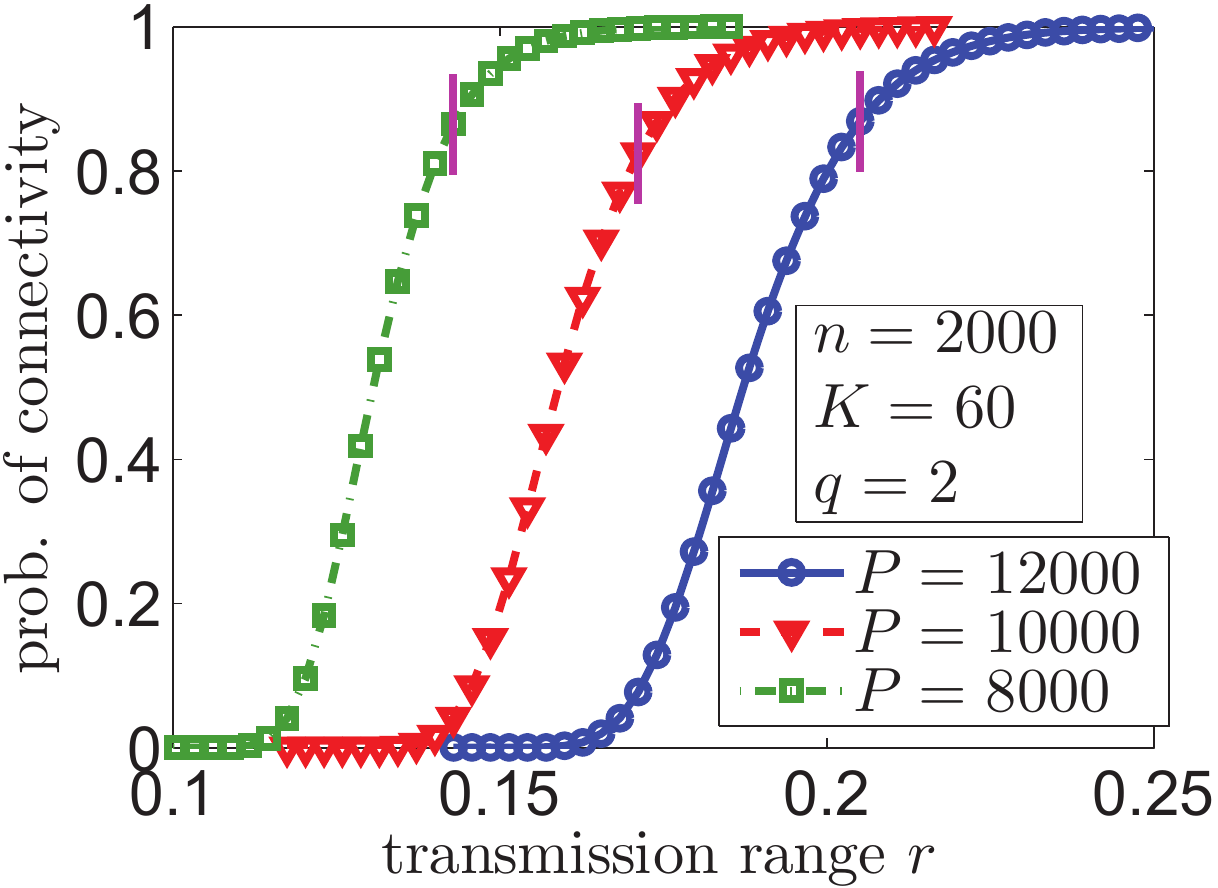}}
 \hspace{-2pt}\subfigure[]{\label{fig:vary_r_under_n}\includegraphics[height=0.165\textwidth]{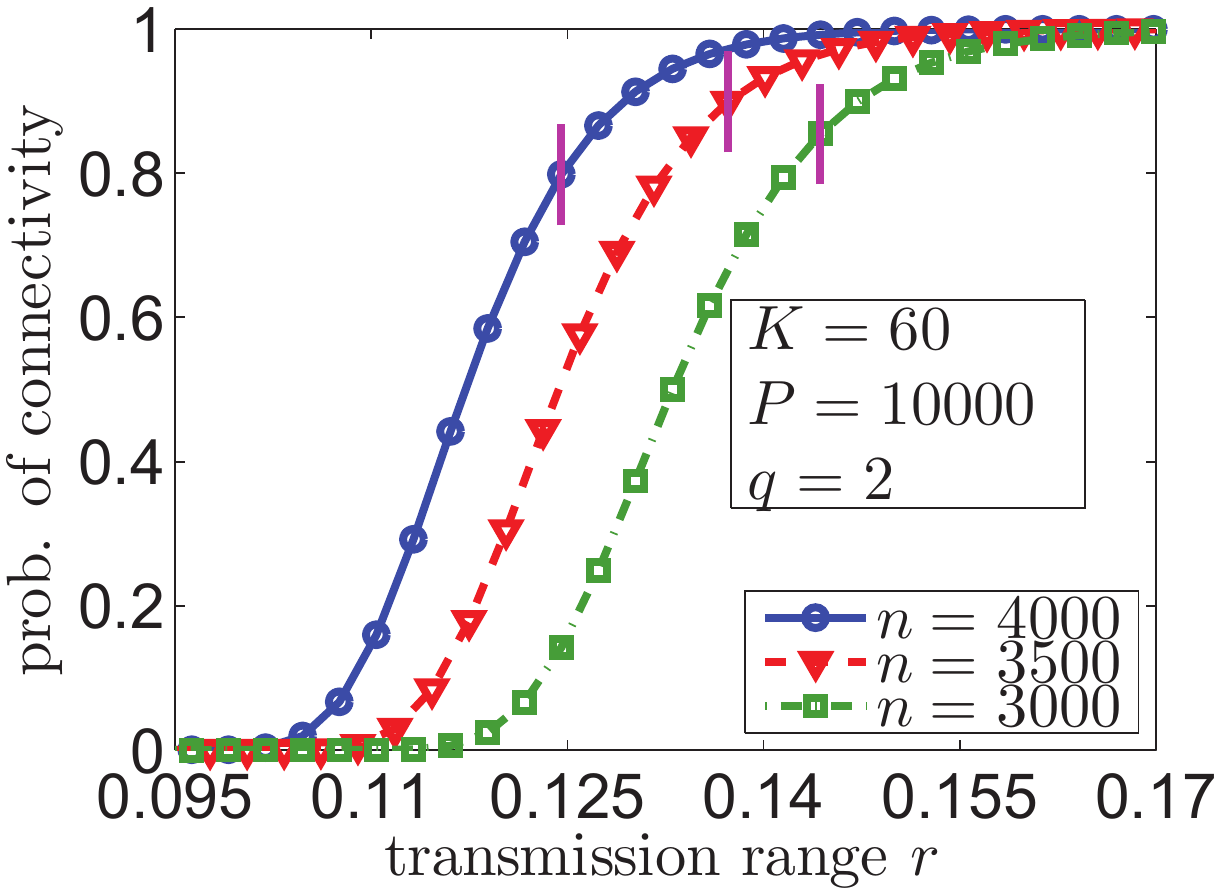}}
 \hspace{-2pt}\subfigure[]{\label{fig:vary_r_under_q}\includegraphics[height=0.165\textwidth]{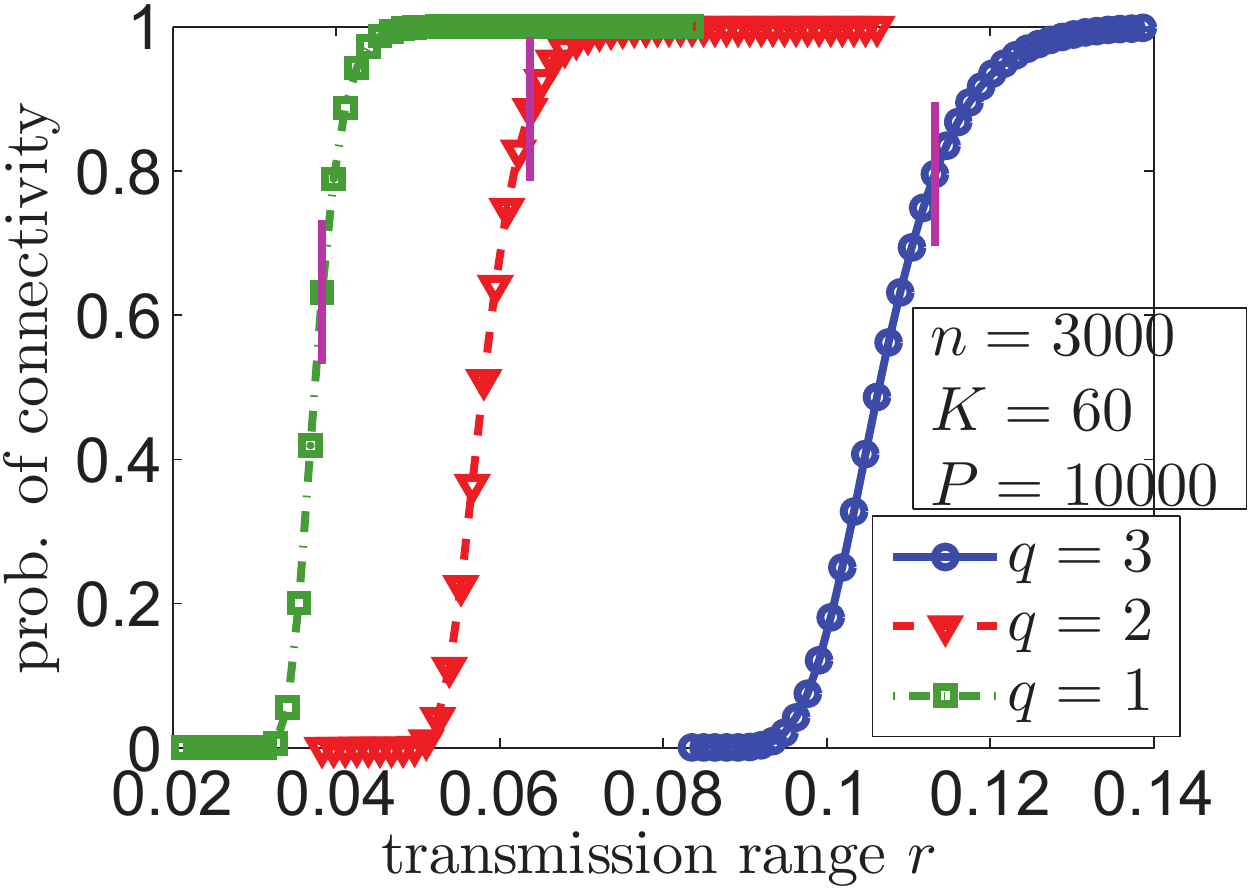}}
\vspace{-5pt}\caption{We plot the connectivity probability of the network $\mathbb{G}_q(n, \hspace{1pt} K, \hspace{1pt} P, \hspace{1pt} r, \hspace{1pt} \mathcal {T})$ with the $q$-composite scheme under transmission constraints, with respect to different transmission ranges $r$. Each vertical line presents the \emph{critical} transmission range given by Eq. (\ref{rnstarqnewx}). \vspace{-10pt}} \label{fig:vary_r}
\end{figure*}

We present experimental results below to confirm our theoretical results of connectivity. First, in the absence of node capture, we study the connectivity behavior
\begin{itemize}
  \item when $K$ varies given different $P$ in Figure \ref{fig:vary_K_under_P} (on Page \pageref{fig:vary_K_under_P}), given different $n$ in Figure \ref{fig:vary_K_under_n}, given different $r$ in Figure \ref{fig:vary_K_under_r}, and given different $q$ in Figure \ref{fig:vary_K_under_q},
  \item when $P$ varies given different $K$ in Figure \ref{fig:vary_P_under_K}, given different $n$ in Figure \ref{fig:vary_P_under_n},  given different $r$ in Figure \ref{fig:vary_P_under_r}, and given different $q$ in Figure \ref{fig:vary_P_under_q},
  \item when $r$ varies given different $K$ in Figure \ref{fig:vary_r_under_K}, given different $P$ in Figure \ref{fig:vary_r_under_P},   given different $n$ in Figure \ref{fig:vary_r_under_n}, and given different $q$ in Figure \ref{fig:vary_r_under_q}.
\end{itemize}
For each data point, we
generate $500$ independent samples of $\mathbb{G}_q(n, K, P,  r,
\mathcal {T})$, record the count that the obtained graph is
connected, and then divide the count by $500$ to obtain the corresponding empirical probability of network connectivity. In each subfigure, we clearly see the transitional behavior of connectivity. Also, we observe that the probability of connectivity
 \begin{itemize}
   \item  increases with $n$ (resp., $K$, $r$) while fixing other parameters,
   \item  decreases with $P$ (resp., $q$) while fixing other parameters,
 \end{itemize}
Moreover, in each subfigure, the vertical line presents the
\emph{critical} parameter for connectivity computed from Equations (\ref{Knstarqnewx}) (\ref{Pnstarqnewx}) and (\ref{rnstarqnewx}): the critical key ring size in Figures \ref{fig:vary_K} (a)--(d), the critical key pool size in Figures \ref{fig:vary_P} (a)--(d), and the critical transmission range in Figures \ref{fig:vary_r} (a)--(d). Summarizing the above, the experiments have confirmed our Theorem  \ref{thm:gqt}.

\begin{figure*}
\vspace{0pt}
\addtolength{\subfigcapskip}{-4pt}\centering     
\hspace{-2pt}\subfigure[]{\label{fig:vary_nodecapture_K_under_m}\includegraphics[height=0.165\textwidth]{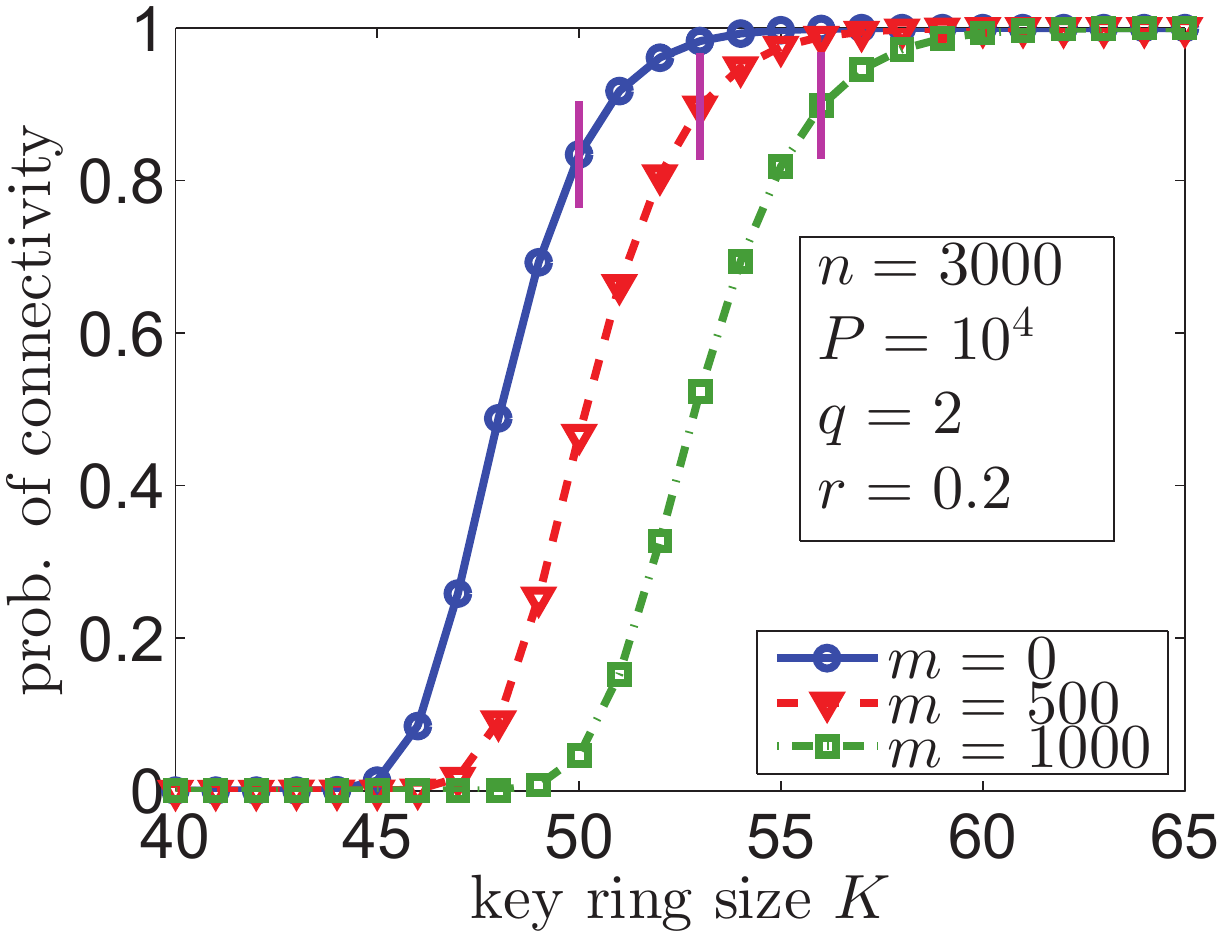}}
\hspace{-2pt}\subfigure[]{\label{fig:vary_nodecapture_P_under_m}\includegraphics[height=0.165\textwidth]{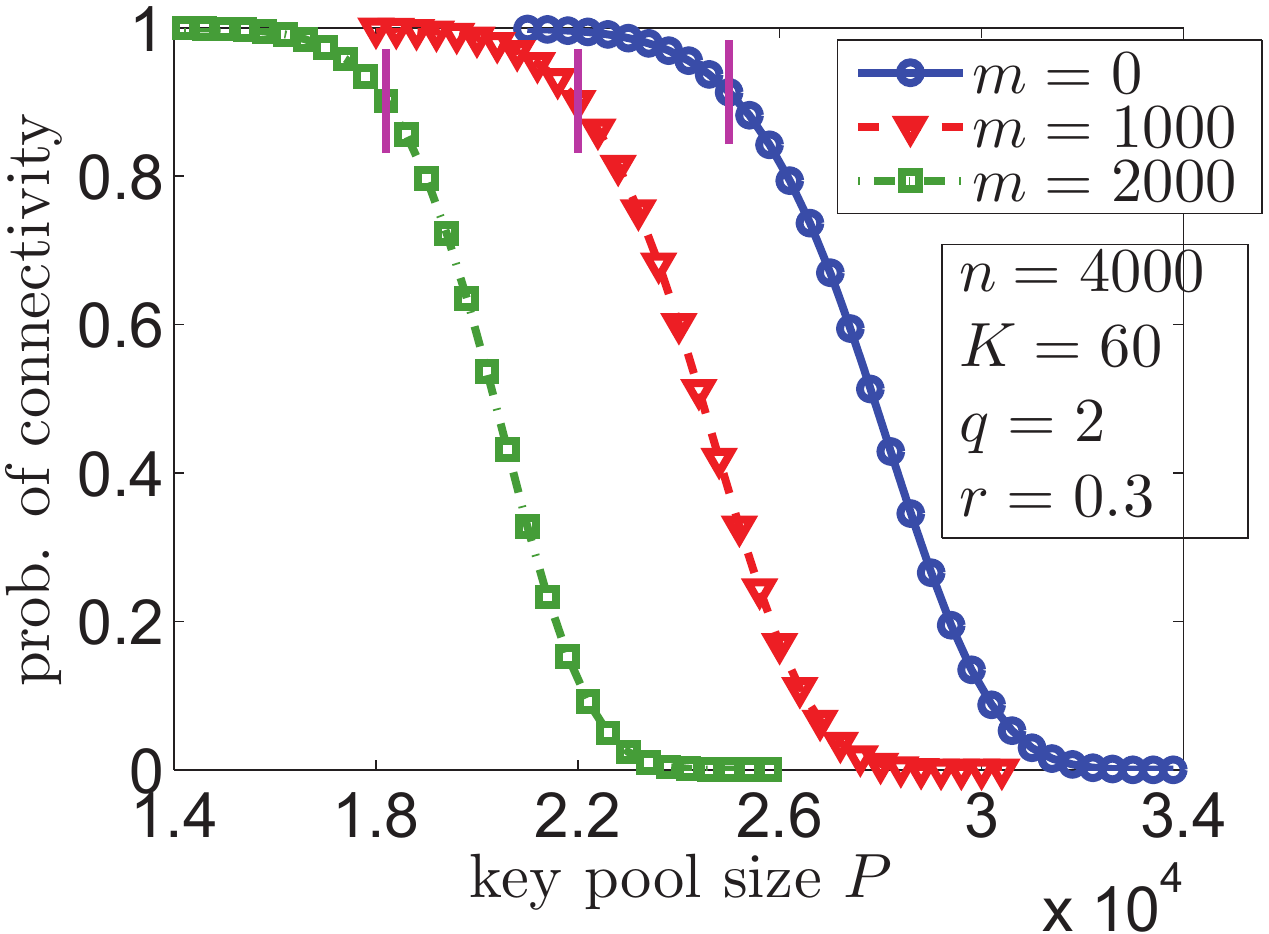}}
 \hspace{-2pt}\subfigure[]{\label{fig:vary_nodecapture_r_under_m}\includegraphics[height=0.165\textwidth]{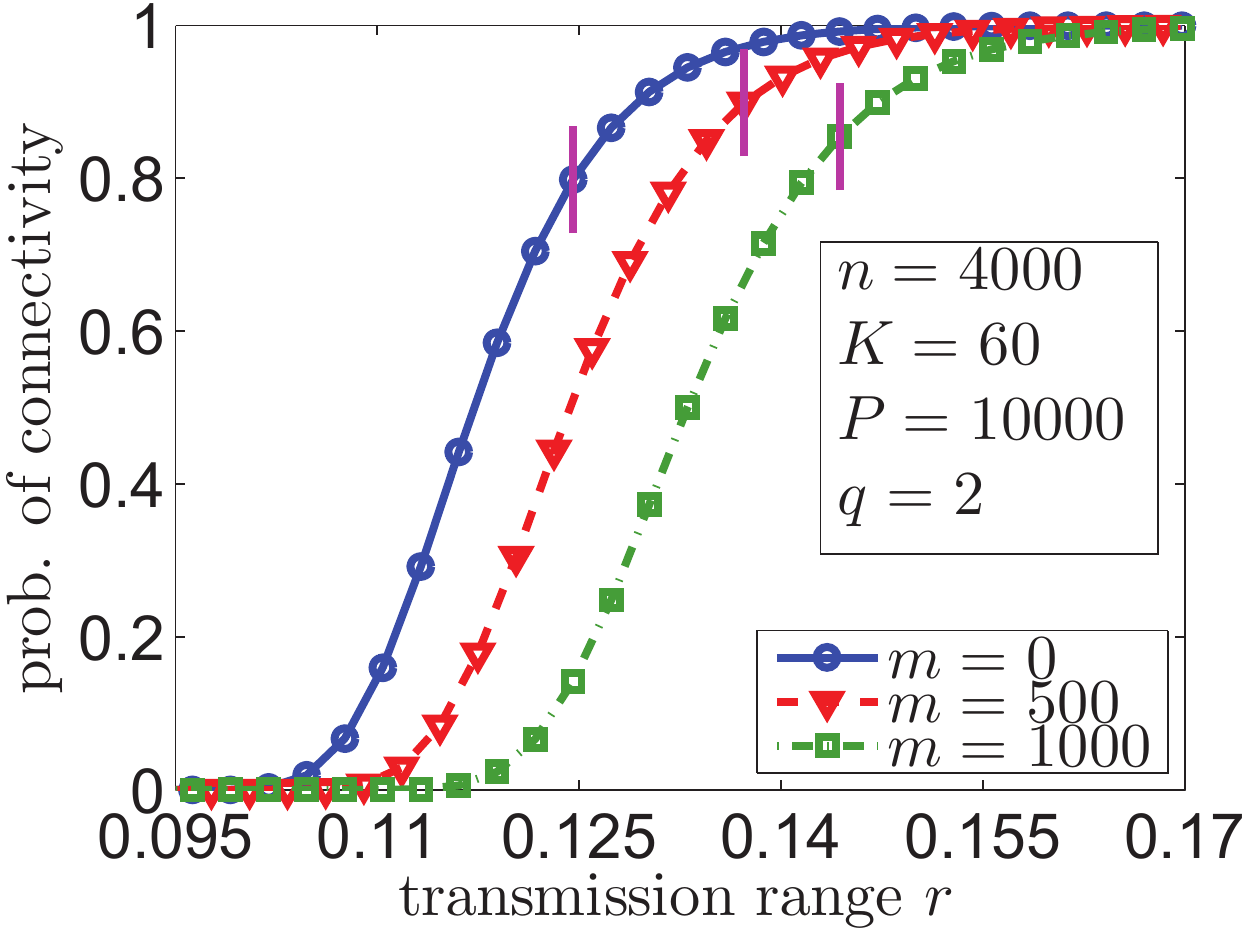}}
 \hspace{-2pt}\subfigure[]{\label{fig:vary_nodecapture_m_under_K}\includegraphics[height=0.165\textwidth]{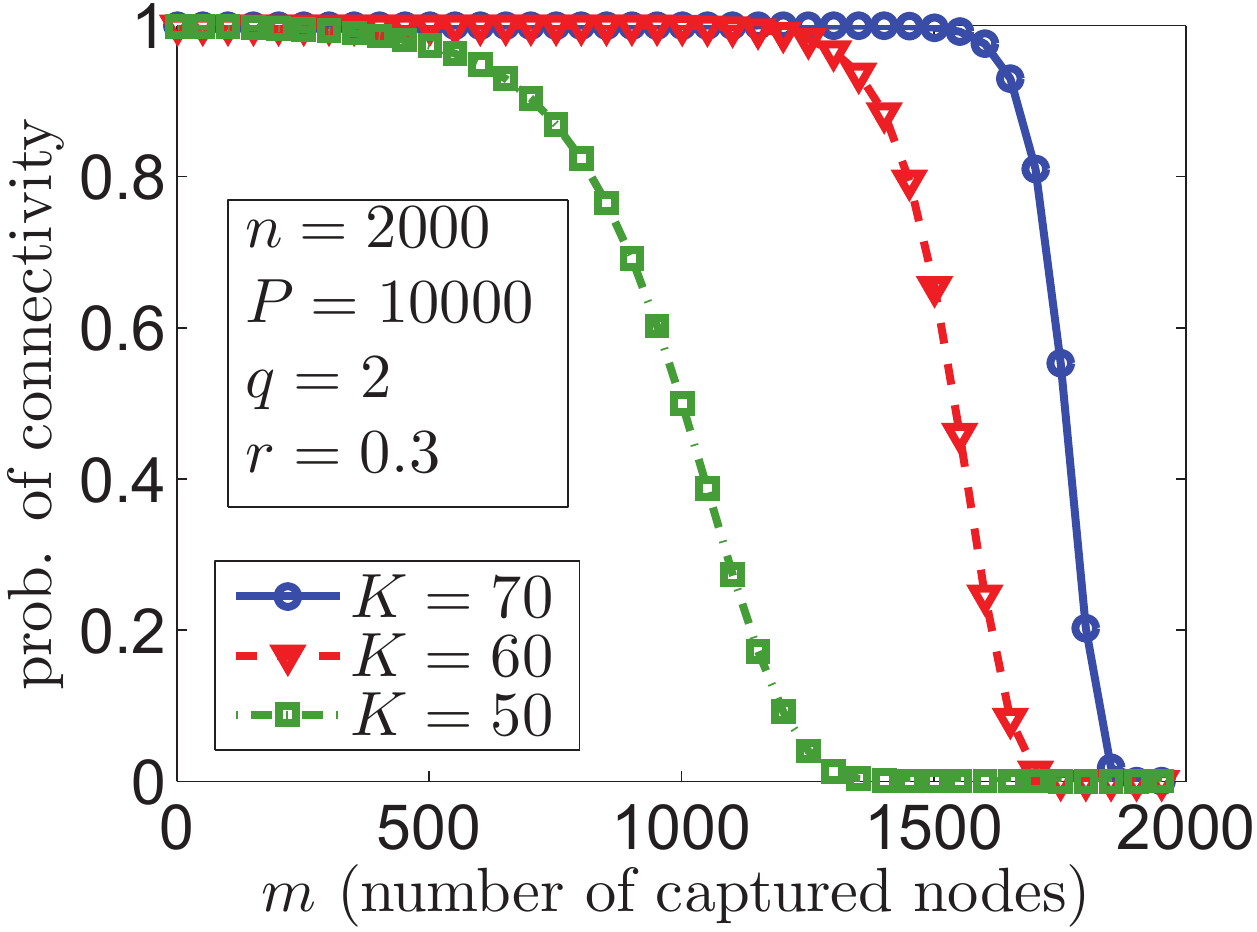}}
\vspace{-9pt}\caption{In the case where $m$ nodes have already been captured, we plot the connectivity probability of the network $\mathbb{G}_q(n, \hspace{1pt} K, \hspace{1pt} P, \hspace{1pt} r, \hspace{1pt} \mathcal {T})$ with the $q$-composite scheme under transmission constraints. Each vertical line in subfigures (a) (b) and (c) presents the \emph{critical} key ring size given by Eq. (\ref{Knstarqnewx}), the \emph{critical} key pool size given by Eq. (\ref{Pnstarqnewx}), and the \emph{critical} transmission range given by Eq. (\ref{rnstarqnewx}), respectively.\vspace{-10pt}} \label{fig:vary_nodecapture}
\end{figure*}

We then consider the presence of node capture to plot Figures \ref{fig:vary_nodecapture} (a)--(d) on Page \pageref{fig:vary_K_under_P}. Given different $m$ (the number of captured nodes), we study the connectivity behavior when $K$ varies  in Figure \ref{fig:vary_nodecapture_K_under_m}, $P$ varies  in Figure \ref{fig:vary_nodecapture_P_under_m}, and $r$ varies  in Figure \ref{fig:vary_nodecapture_r_under_m}, where the data points are generated in the same way descried above (i.e., deriving empirical probabilities from 500 samples). In each subfigure, the vertical line shows the
\emph{critical} parameter for connectivity computed from Equations (\ref{Knstarqnewx_nodecapture}) (\ref{Pnstarqnewx_nodecapture}) and (\ref{rnstarqnewx_nodecapture}): the critical key ring size in Figure \ref{fig:vary_nodecapture_K_under_m}, the critical key pool size in Figure \ref{fig:vary_nodecapture_P_under_m}, and the critical transmission range in Figure \ref{fig:vary_nodecapture_r_under_m}. Furthermore, in Figure \ref{fig:vary_nodecapture_m_under_K}, we vary $m$ given different $K$. In all subfigures, we also observe the  transitional behavior of connectivity. Summarizing the above, the experimental results have confirmed our Theorem  \ref{thm:gqt}. Figures \ref{fig:vary_nodecapture} (a)--(d) have confirmed our Theorem  \ref{thm:gqt:comp}.

\begin{figure*}
\vspace{0pt}
\addtolength{\subfigcapskip}{-4pt}\centering     
 \hspace{-2pt}\subfigure[]{\label{fig:replicationq1}\includegraphics[height=0.17\textwidth]{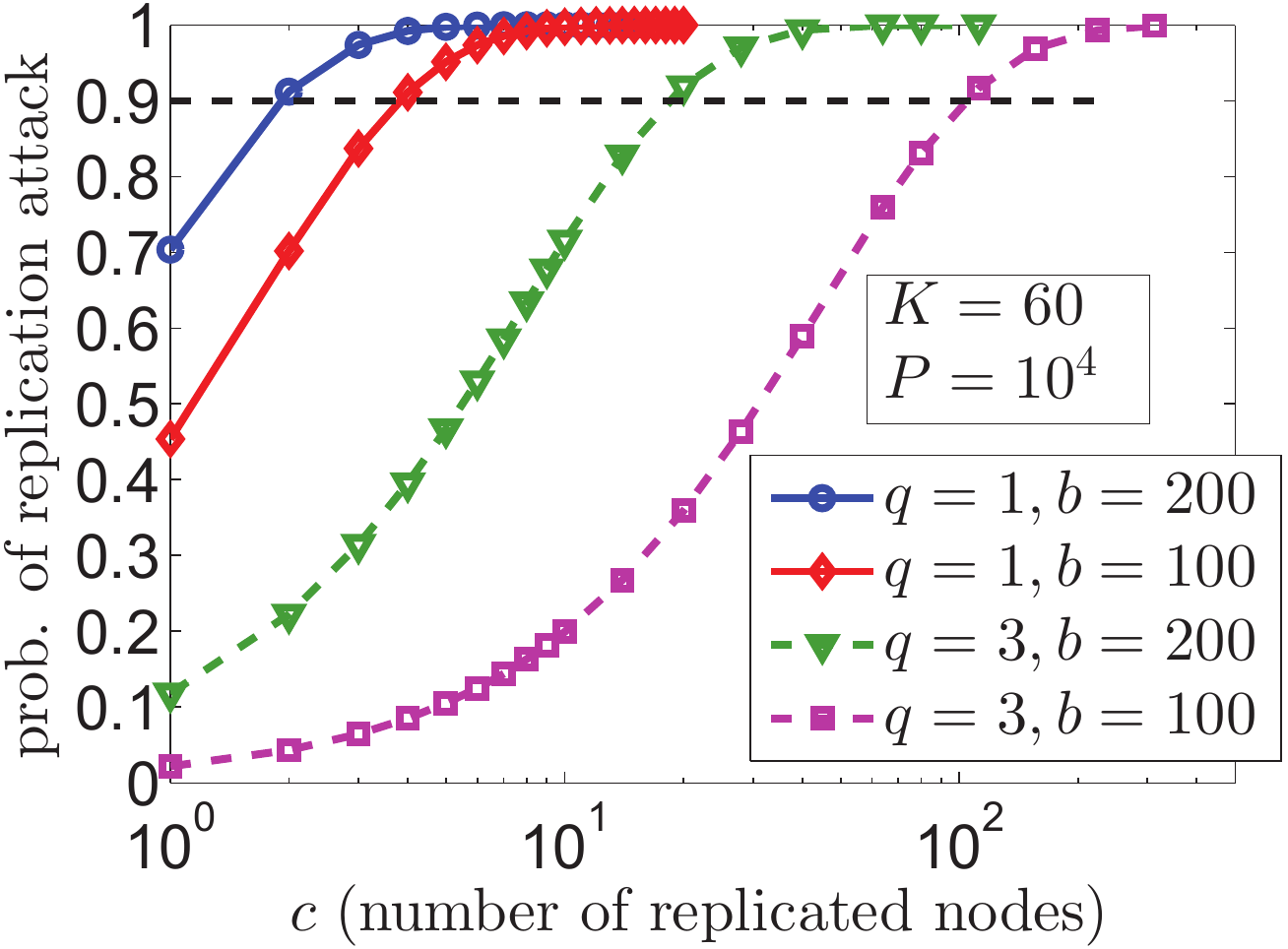}}
\hspace{-2pt}\subfigure[]{\label{fig:replicationq2}\includegraphics[height=0.17\textwidth]{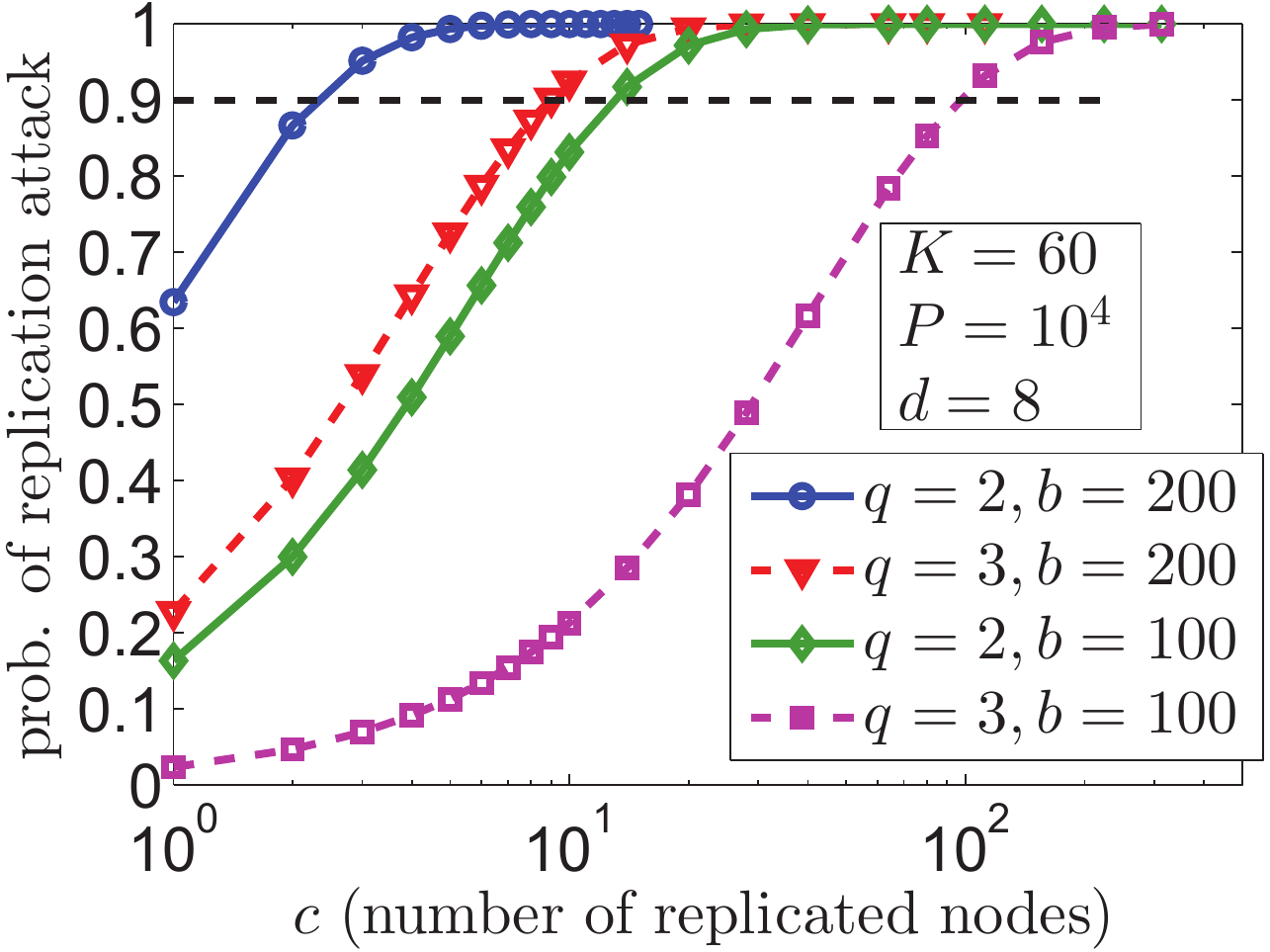}}
\hspace{-2pt}\subfigure[]{\label{fig:replicationq3}\includegraphics[height=0.17\textwidth]{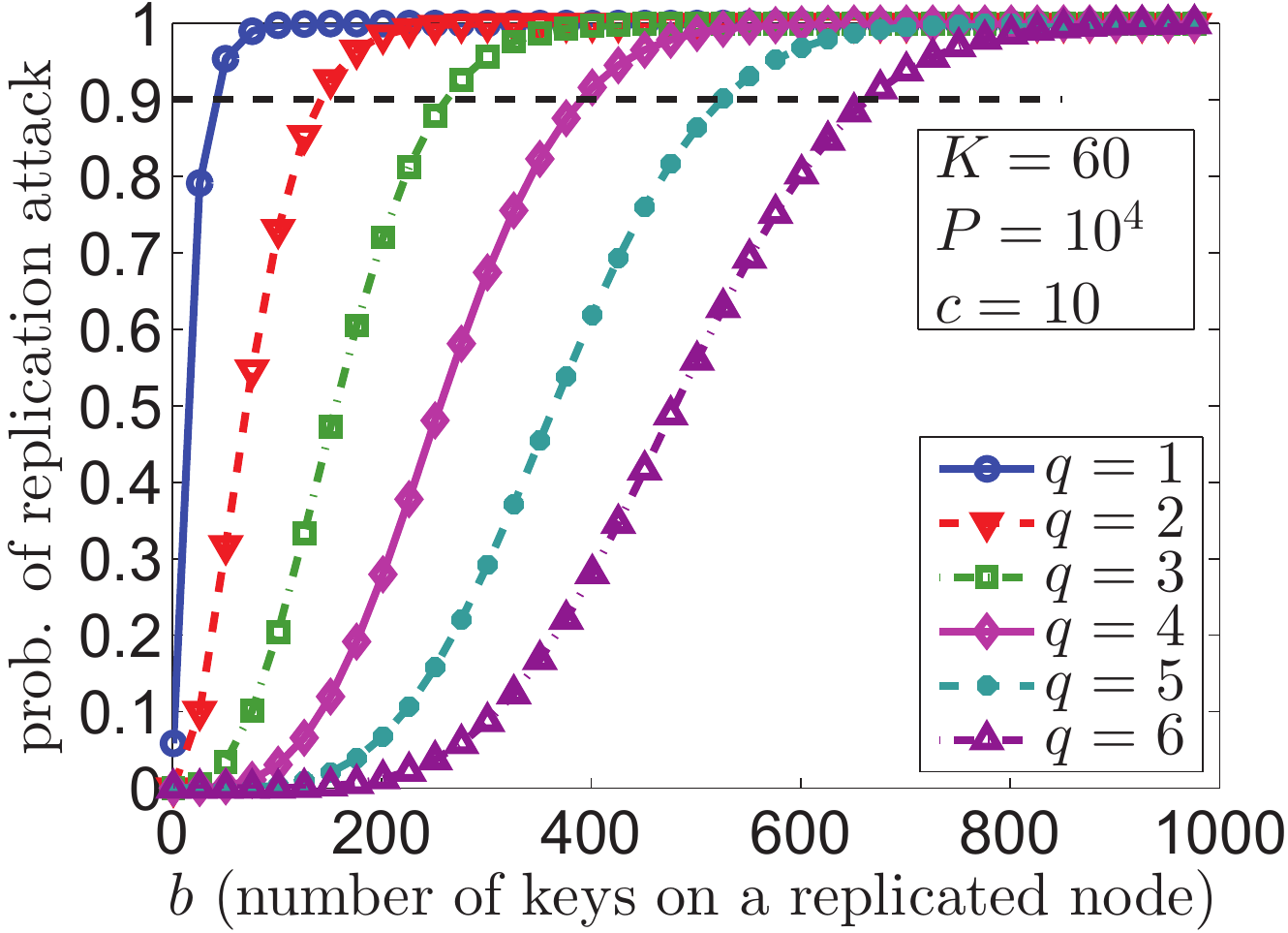}}
\hspace{-2pt}\subfigure[]{\label{fig:replicationq4}\includegraphics[height=0.17\textwidth]{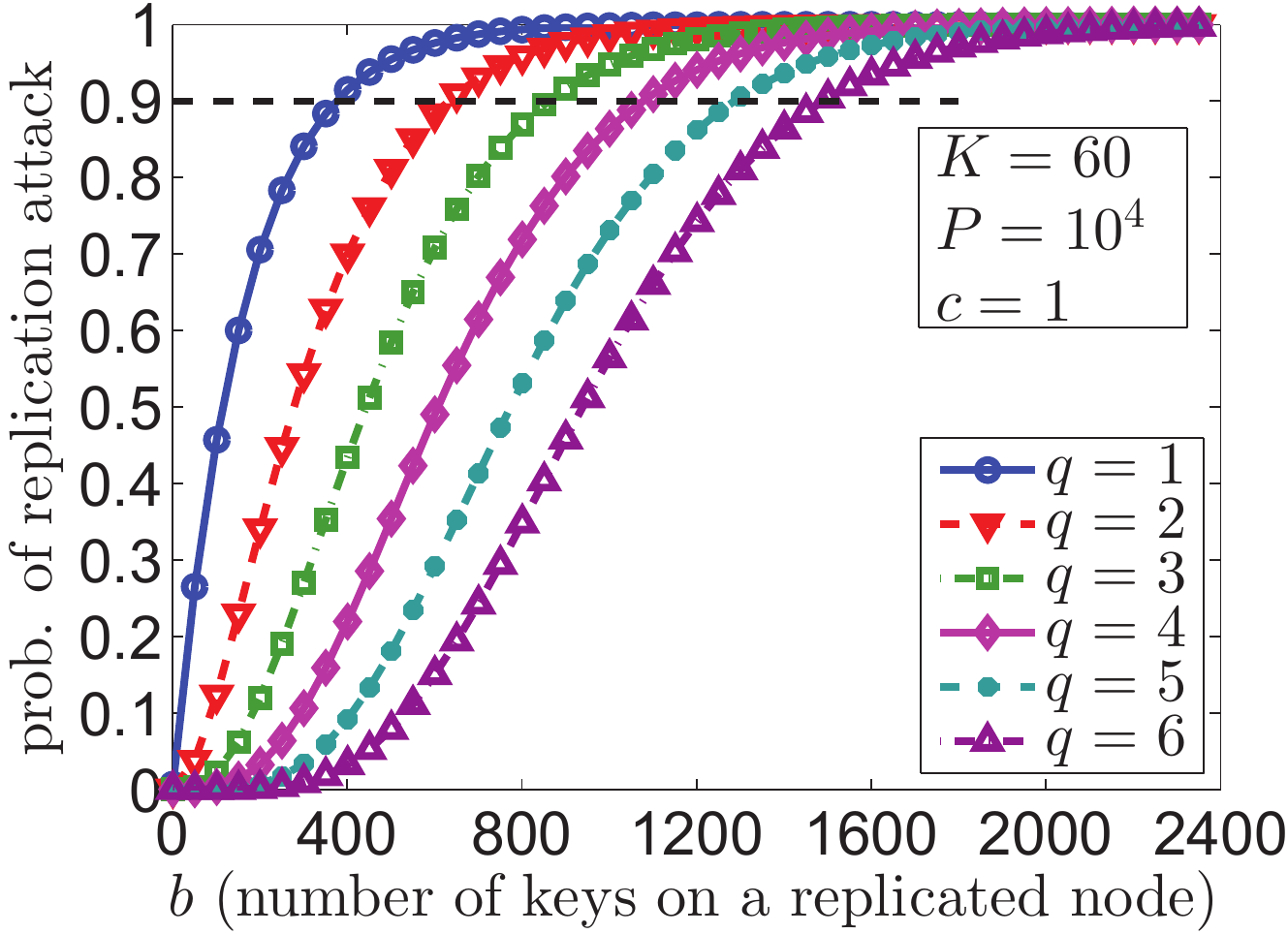}}
\vspace{-9pt}\caption{\mbox{Plots for node replication attacks. We set the density of benign nodes as $d = 1$ in subfigures (a) (c) and (d), while considering $d = 8$ in subfigure (b).\vspace{-10pt}}} \label{fig:replicationq}
\end{figure*}

\section{Quantifying Node-Replication Attacks}
\vspace{-3pt} \label{node-quan}

\subsection{Results}
\vspace{-3pt}

In secure sensor networks, a further attack following node capture is the so-called node-replication attack \cite{parno2005distributed,conti2007randomized,zhu2007efficient,fu2008replication}. The idea is that after node capture, the adversary can deploy   replicas of   compromised nodes by extracting keys from   compromised nodes and then inserting some keys into the memory of replica nodes. The resources quantifying the adversary include the number of replicas deployed, and the cost of each replica. To characterize the cost of each replica, a simple metric is the number of keys inserted into its memory (of course, this may not be precise, but it provides qualitative intuition). The more a replica node cost, the more keys it can store. For a resource-limited adversary, it has to tradeoff between the number of replicas deployed, and the amount of keys kept on each replica. Unfortunately, prior studies on node-replication attacks lack of \emph{either} formal \emph{or} tractable analyses \cite{parno2005distributed,conti2007randomized,zhu2007efficient,fu2008replication} to determine the adversary's optimal strategy. We closes this gap and present the following results:
\begin{itemize}
  \item[(i)] If $c$ denoting the number of replicas decreases by a factor of $\Delta$, then $b$ denoting the amount of keys on each replica should increase by a factor of $\sqrt[q]{\Delta}$ to achieve the same probability of a successful replication attack.
  \item[(ii)] If the adversary's cost function $\textrm{cost}(b,c)$ scales with $b^{p_b} c^{p_c}$ for some $p_b,p_c$, then the optimal $(b,c)$ to maximize the replication attack subject to the resource constraint $b^{p_b} c^{p_c} \leq \textrm{budget}$, depends on how the ratio $p_b/p_c$ compares with $q$. Specifically, the optimal $(b,c)$
\begin{itemize}
  \item  is given by   $b = \lfloor\sqrt[p_b]{\textrm{budget}} \rfloor$ and $c=1$ if $p_b/p_c < q$,
  \item  is given by  $b = 1$ and $c=\lfloor\sqrt[p_c]{\textrm{budget}}\rfloor$ if $p_b/p_c > q$, and
  \item has many choices if $p_b/p_c = q$.
\end{itemize} In practice, when we often have $p_b/p_c < q$ (e.g., $p_b=1$, $p_c=1$ and $q>1$), then $b = \lfloor\sqrt[p_b]{\textrm{budget}} \rfloor$ and $c=1$ become optimal, meaning that the adversary should deploy a small number of high-cost sensors, rather than deploy a large number of low-cost sensors, to maximize the replication attack
under a budget.
\end{itemize}

We now explain  the above results (i) and (ii). Let the network density be $d$ (for benign nodes), so that the adversary can put a replica close to $d$ benign nodes on average. We compute the probability that a benign node shares at least $q$ keys with a replica (this will establish a secure link between them and thus induce a successful replication attack). Note that a benign node has $K_n$ keys selected from the pool of $P_n$ keys, and a replica has $d$ keys out of the same key pool. Then the probability that a benign node shares exactly $i$ keys with a replica is $\frac{\binom{b}{i} \binom{P_n-b}{K_n-i}}{\binom{P_n}{K_n}}$ (the idea is the benign node should have $i$ keys the same as the replica, and $K_n-i$ keys from the $(P_n-b)$-size key set in which the replica does not have a key. Then the probability that a replica and a benign node share at least (resp., less than) $q$ keys is $1-\alpha$ (resp., $\alpha$), where we denote $\sum\limits_{i=0}^{q-1} \frac{\binom{b}{i} \binom{P_n-b}{K_n-i}}{\binom{P_n}{K_n}}$ by $\alpha$ for simplicity. Since there are $c$ replicas deployed, and each replica is neighboring  $d$ benign nodes on average, then the probability of a successful replication attack, defined as the probability that at least one replica and at least one benign node establish a secure link, can be computed as $1-\alpha^{cd}$. Similar to the result (\ref{pssimq}) on Page \pageref{pssimq} for the link set-up probability $p_s$ in the $q$-composite scheme, an asymptotic result of $\alpha$ is $1-\frac{1}{q!}\cdot\frac{b^q{K_n}^q}{{P_n}^q}$, which we prove in the full version \cite{full} due to space limitation (in fact, it is straightforward to see that $\alpha$ is $1-p_s$ if $b$ equals $K_n$). Hence,  the probability of a successful replication attack (denoted by $p_{\textrm{replication}}$ below) asymptotically equals $1-(1-\frac{1}{q!}\cdot\frac{b^q{K_n}^q}{{P_n}^q})^{cd}$, which further becomes $\frac{cd}{q!}\cdot\frac{b^q{K_n}^q}{{P_n}^q}$ since the key pool size $P_n$ is much greater than $b$ and $K_n$. Then we see that $p_{\textrm{replication}}$ scales linearly with $c$, while scaling with $b^q$. This implies result (i); i.e., if $c$ denoting the number of replicas decreases by a factor of $\Delta$, then $b$ denoting the amount of keys on each replica should increase by a factor of $\sqrt[q]{\Delta}$ to achieve the same probability of a successful replication attack. We  further obtain result (ii) by considering that the resource-constrained adversary maximizes $b ^ q c$ to maximize $p_{\textrm{replication}}$, subject to the resource constraint $b^{p_b} c^{p_c} \leq \textrm{budget}$.
\vspace{-12pt}

\subsection{Experiments}
\vspace{-3pt}

We present experiments in Figures \ref{fig:replicationq} (a)--(d) on Page \pageref{fig:replicationq} to confirm the above results on replication attacks. In Figures \ref{fig:replicationq} (a) and (b), we vary $c$ (the number of replicated nodes) given different $q$ and $b$ to plot $p_{\textrm{replication}}$. We consider node density $d=1$ in subfigure (a), and $d=8$ in subfigure (b). We also plot the horizontal line that corresponds to $p_{\textrm{replication}}$ being $0.9$. We take subfigure (a) as an example to validate our result (i) above. In Figure \ref{fig:replicationq1}, after $c$ decreases by a factor of $\Delta$, then $b$ should increase by a factor of $\sqrt[q]{\Delta}$ to achieve the same $p_{\textrm{replication}}$. Let $c(b=100)$ and $c(b=200)$ denote $c$ corresponding to $b=100$ and $b=100$. Then for the upper two lines in Figure \ref{fig:replicationq1}, given $q=1$, we compare ${\frac{c(b=100)}{c(b=200)}}$ with $2$, and the upper part of  Table \ref{replication-table} shows they are actually equal for the data points presented. For the lower two lines in Figure \ref{fig:replicationq1}, given $q=3$, we compare $\sqrt[3]{\frac{c(b=100)}{c(b=200)}}$ with $2$, and the lower part of  Table \ref{replication-table} shows they are quite close for the data points presented. Hence, Table \ref{replication-table} has confirmed our results on replication attacks.

In Figures \ref{fig:replicationq} (c) and (d), we  vary $b$ (number of keys on each replica) given different $q$ to plot $p_{\textrm{replication}}$. We consider $c=10$ in subfigure (c), and $c=1$ in subfigure (d), and set node density $d=1$ in both subfigures. Recall that $p_{\textrm{replication}}$ scales with $\frac{cd}{q!} \big(\frac{bK_n}{P_n}\big)^q$, and thus decreases as $q$ increases for $bK_n$ smaller than $P_n$. This is confirmed in the plots.
\vspace{-7pt}

\begin{table}[H]
\centering
\caption{This table confirms our results on replication attacks.}
\vspace{-7pt}
\label{replication-table}
\begin{tabular}{lllll}
\hline
\multicolumn{5}{|l|}{$q=1$ (the upper two lines in Figure \ref{fig:replicationq1})}                                                                                                              \\ \hline
\multicolumn{1}{|l|}{prob}                                  & \multicolumn{1}{l|}{0.7}  & \multicolumn{1}{l|}{0.9}  & \multicolumn{1}{l|}{0.97} & \multicolumn{1}{l|}{0.99} \\ \hline
\multicolumn{1}{|l|}{$c(b=100)$}                            & \multicolumn{1}{l|}{2}    & \multicolumn{1}{l|}{4}    & \multicolumn{1}{l|}{6}    & \multicolumn{1}{l|}{8}    \\ \hline
\multicolumn{1}{|l|}{$c(b=200)$}                            & \multicolumn{1}{l|}{1}    & \multicolumn{1}{l|}{2}    & \multicolumn{1}{l|}{3}    & \multicolumn{1}{l|}{4}    \\ \hline
\multicolumn{1}{|l|}{$\sqrt[q]{\frac{c(b=100)}{c(b=200)}}$} & \multicolumn{1}{l|}{2}    & \multicolumn{1}{l|}{2}    & \multicolumn{1}{l|}{2}    & \multicolumn{1}{l|}{2}    \\ \hline
                                                            &                           &                           &                           &                           \\[-6pt] \hline
\multicolumn{5}{|l|}{$q=3$ (the lower two lines in Figure \ref{fig:replicationq1})}                                                                                                              \\ \hline
\multicolumn{1}{|l|}{prob}                                  & \multicolumn{1}{l|}{0.3}  & \multicolumn{1}{l|}{0.5}  & \multicolumn{1}{l|}{0.8}  & \multicolumn{1}{l|}{0.9}  \\ \hline
\multicolumn{1}{|l|}{$c(b=100)$}                            & \multicolumn{1}{l|}{17}   & \multicolumn{1}{l|}{43}   & \multicolumn{1}{l|}{75}   & \multicolumn{1}{l|}{104}  \\ \hline
\multicolumn{1}{|l|}{$c(b=200)$}                            & \multicolumn{1}{l|}{3}    & \multicolumn{1}{l|}{6}    & \multicolumn{1}{l|}{10}   & \multicolumn{1}{l|}{12}   \\ \hline
\multicolumn{1}{|l|}{$\sqrt[q]{\frac{c(b=100)}{c(b=200)}}$} & \multicolumn{1}{l|}{1.78} & \multicolumn{1}{l|}{1.93} & \multicolumn{1}{l|}{1.96} & \multicolumn{1}{l|}{2.05} \\ \hline
\end{tabular}
\vspace{-10pt}
\end{table}

\section{Mathematical Details for\\Resilience against Node Capture}\label{sec-derive-pcCapture}
\vspace{-2pt}

We now provide formal details for our results in Section \ref{sec:resi} on the  resilience against node capture.

\subsection{Deriving $p_{\textnormal{compromised}}$} \label{sec-derive-pc}

%

We consider the case where the adversary has captured some random
set of $m$ nodes.
As defined before, $p_{\textnormal{compromised}}$ is the probability
that secure communication
between two non-captured nodes 
 is compromised by the adversary. With $\rho_u$ denoting the
probability that two nodes share exactly $u$ keys in their key rings, for the $q$-composite scheme, Chan
\emph{et al.} \cite{adrian} calculate $p_{\textnormal{compromised}}$ by
\begin{align}
p_{\textnormal{compromised}} & =
\sum_{u=q}^{K_n}\bigg\{\bigg[1-\bigg(1-\frac{K_n}{P_n}\bigg)^m\bigg]^u
\cdot \frac{\rho_u}{p_{s}}\bigg\}. \label{incor}
\end{align}
This result used in many studies
\cite{Chan2008134,Liu2003CCS,YaganThesis,Vu:2010:SWS:1755688.1755703,Yang:2005:TRS:1062689.1062696}
 is unfortunately \emph{incorrect} as formally shown by Yum and
Lee \cite{6170857}. With $A_{\tau}$ denoting the probability that
the $m$ captured nodes have $\tau$ different keys in their key rings
in total, where $K_n \leq \tau \leq \min\{mK_n, P_n\}$, a correct
computation \cite{6170857} of $p_{\textnormal{compromised}}$ is as
follows:
\begin{align}
& p_{\textnormal{compromised}}   = \sum_{\tau=K_n}^{\min\{mK_n, P_n\}}
\Bigg\{  A_{\tau} \cdot \sum_{u=q}^{K_n}\Bigg[
\frac{\binom{\tau}{u}}{\binom{P_n}{u}} \cdot
\frac{\rho_u}{p_{s}}\Bigg]  \Bigg\}, \label{pcompexpr}
\end{align}
where
\begin{align}
 A_{\tau} & =  \binom{P_n}{\tau} \cdot \frac{\binom{\tau}{K_n}^m -
\sum_{\lambda=1}^{\tau-K_n}(-1)^{\lambda+1}\binom{\tau}
{\lambda}\binom{\tau-\lambda}{K_n}^m}{\binom{P_n}{K_n}^m}.
\label{at}
\end{align}
Under a practical condition $P_n \geq 2K_n$, from the result \cite[Proof of Lemma 2]{QcompTech14}, \vspace{-3pt}
it follows that
\begin{align}
\frac{\rho_u}{p_{s}}  & = \frac{1}{u!} \frac{1}{[(K_n-u)!]^2}
\frac{1}{(P_n-2K_n+u)!} \nonumber \\ & \quad \times
\Bigg\{\sum_{u=q}^{K_n}\frac{1}{u!} \frac{1}{[(K_n-u)!]^2} \cdot
 \frac{1}{(P_n-2K_n+u)!}\Bigg\}^{-1} . \label{rhups}
\end{align}
Substituting (\ref{at}) and
(\ref{rhups}) into (\ref{pcompexpr}), we derive the exact
formula of $p_{\textnormal{compromised}}$ through
\begin{align}
& p_{\textnormal{compromised}} \nonumber \\ & =
\Bigg\{\sum_{u=q}^{K_n}\frac{1}{u!} \frac{1}{[(K_n-u)!]^2} \cdot
 \frac{1}{(P_n-2K_n+u)!}\Bigg\}^{-1} \nonumber \\ & ~~ \times
\sum_{\tau=K_n}^{\min\{mK_n, P_n\}} \Bigg\{ \binom{P_n}{\tau}
\sum_{u=q}^{K_n}\Bigg[ \frac{\binom{\tau}{u}}{\binom{P_n}{u}} \cdot
\frac{1}{u!} \frac{1}{[(K_n-u)!]^2}
 \nonumber \\
& \quad \times \hspace{-1pt} \frac{1}{(P_n-2K_n+u)!}\hspace{-1.5pt}
\Bigg] \hspace{-1.5pt} \cdot \hspace{-1.5pt}
\frac{\binom{\tau}{K_n}^m \hspace{-2pt} - \hspace{-2pt}
\sum_{\lambda=1}^{\tau-K_n}\hspace{-1pt}(-1)^{\lambda+1}\hspace{-1pt}\binom{\tau}
{\lambda}\binom{\tau-\lambda}{K_n}^m\hspace{-1pt}}{\binom{P_n}{K_n}^m}
\hspace{-2pt}\Bigg\}.  \label{pcomp_exact}
\end{align}
We see that probability
$p_{\textnormal{compromised}}$ relies on the parameters $K_n,P_n$ and $q$, but does not
directly
depend on $n$. However, (\ref{pcomp_exact}) is too complex to understand how $p_{\textnormal{compromised}}$ varies with respect to the parameters. To this end, we perform an asymptotic analysis in the remainder of this section to study $p_{\textnormal{compromised}}$; i.e., we analyze $p_{\textnormal{compromised}}$ as $n \to \infty$. 

\subsection{Proof of Theorem \ref{thm:pcomp1}} \label{sec:lem:pcomp}


Recall that probability $p_{\textnormal{compromised}}$ is given by
(\ref{pcompexpr}). From $P_n = \omega(mK_n)$ which clearly follows
from condition (\ref{moPnKn}) (i.e., $m =
o\big(\frac{P_n}{K_n}\big)$), it then holds (for all $n$
sufficiently large) that $\min\{mK_n, P_n\} = mK_n$. Thus, we obtain
from (\ref{pcompexpr}) that
\begin{align}
p_{\textnormal{compromised}}   = \sum_{\tau=K_n}^{mK_n} \Bigg\{
A_{\tau} \cdot \sum_{u=q}^{K_n}\Bigg[
\frac{\binom{\tau}{u}}{\binom{P_n}{u}} \cdot
\frac{\rho_u}{p_{s}}\Bigg]  \Bigg\}.\label{pcompexprnew}
\end{align}

With $q \leq u\leq K_n$, we have
\begin{align}
 \binom{\tau}{u}\Bigg/\binom{P_n}{u} =
 \frac{\tau!}{u!(\tau-u)!} \bigg/ \frac{P_n!}{u!(P_n-u)!}
 \leq \frac{\tau^q}{{(P_n-K_n)}^q}. \nonumber
\end{align}
Hence, from $\sum_{u=q}^{K_n}\rho_u=p_{s}$, \f
\begin{align}
\sum_{u=q}^{K_n}\Bigg[ \frac{\binom{\tau}{u}}{\binom{P_n}{u}} \cdot
\frac{\rho_u}{p_{s}}\Bigg]  \leq \sum_{u=q}^{K_n}\Bigg[
\frac{\tau^q}{{(P_n-K_n)}^q} \cdot \frac{\rho_u}{p_{s}}\Bigg]
   = \frac{\tau^q}{{(P_n-K_n)}^q}. \label{sumuq}
\end{align}
Substituting (\ref{sumuq}) and $\sum_{\tau=K_n}^{mK_n}A_{\tau}=1$
into (\ref{pcompexprnew}), we establish
\begin{align}
p_{\textnormal{compromised}}   \hspace{-1pt} \leq \hspace{-1pt} \sum_{\tau=K_n}^{mK_n}\hspace{-2pt} \Bigg[
A_{\tau} \hspace{-2pt} \cdot \hspace{-2pt} \frac{\tau^q}{{(P_n-K_n)}^q} \Bigg]
\hspace{-1pt} \leq \hspace{-1pt} \frac{(mK_n)^q}{{(P_n-K_n)}^q} \label{pcmp_up}  \hspace{-1pt} =\hspace{-1pt}  o(1),
\end{align}
where the last step uses $mK_n = o(P_n-K_n)$, which holds from
condition $m = o\big(\frac{P_n}{K_n}\big)$.

\subsection{Proof of Corollary \ref{cor:pcomp1}}

Corollary \ref{cor:pcomp1} follows directly from Theorem \ref{thm:pcomp1} because $\frac{P_n}{K_n} = \Omega(n)$ and  $m=o(n)$ together imply $m = o\big(\frac{P_n}{K_n}\big)$.

\subsection{Proof of Theorem \ref{thm:pcomp2}} \label{sec:lem:pcomp2}

 From the condition (\ref{moPnKn-sqr}) (i.e., $m = o\big(\sqrt{\frac{P_n}{{K_n}^2}} \hspace{1pt}\big)$), we find for $n$ sufficiently large that
\begin{align}
m < \sqrt{\frac{P_n}{{K_n}^2}} = \frac{\sqrt{P_n}}{K_n} \leq \frac{P_n}{K_n},\label{mexprres}
\end{align}
 so we further obtain $P_n > m K_n$ and hence $\min\{mK_n, P_n\} = mK_n$. Then (\ref{pcompexprnew}) still follows here.

As in (\ref{pcompexprnew}), $p_{\textnormal{compromised}}$ is expressed
as a summation of sequential elements with
$\tau=K_n,K_n+1,\ldots,mK_n$. Clearly, $p_{\textnormal{compromised}}$ is
at least the element with $\tau=mK_n$; i.e.,
\begin{align}
p_{\textnormal{compromised}} & \geq  A_{mK_n} \cdot
\sum_{u=q}^{K_n}\Bigg[ \frac{\binom{mK_n}{u}}{\binom{P_n}{u}} \cdot
\frac{\rho_u}{p_{s}}\Bigg]. \label{a1}
\end{align}
The right hand side  of (\ref{a1}) is also a summation. We only use its element
with $u=q$. Then
\begin{align}
p_{\textnormal{compromised}} & \geq  A_{mK_n} \cdot
\frac{\binom{mK_n}{q}}{\binom{P_n}{q}} \cdot \frac{\rho_q}{p_{s}} .
\label{ptcomp}
\end{align}
The expression of $A_{mK_n}$ can be obtained from (\ref{at}) by setting $\tau$ as $mK_n$, but it is very complex. We note that $A_{mK_n}$ has already been analyzed on Page \pageref{analyzeAmKn} when we interpret (\ref{moPnKn-sqr}). From our analysis therein, $A_{mK_n}$ equals $\frac{\prod_{j=0}^{m-1}\binom{P_n-jK_n}{K_n}}{\left[\binom{P_n}{K_n}\right]^m}$.
Then it holds for all $n$ sufficiently large that
\begin{align}
 A_{mK_n} &
\geq \left[\frac{\binom{P_n-mK_n}{K_n}}{\binom{P_n}{K_n}}\right]^m \geq
\left[\left(\frac{P_n-mK_n}{P_n}\right)^{K_n}\right]^m
\nonumber \\
& = \left(1-\frac{mK_n}{P_n}\right)^{m K_n} \stackrel{(*)}{\geq}  1- \frac{(mK_n)^2}{P_n} , \nonumber
\end{align}
where step (*) holds because
\begin{itemize}
  \item $\frac{mK_n}{P_n} < 1$ holds from (\ref{mexprres}) for all $n$ sufficiently large, and
  \item for $0 \leq x <1$ and positive integer $y$, it holds that $(1-x)^y \geq 1-xy$; see our prior work \cite[Page 20-Fact 2]{ZhaoYaganGligor}.
       \end{itemize}
In view of condition (\ref{moPnKn-sqr})  (i.e., $m = o\big(\sqrt{\frac{P_n}{{K_n}^2}} \hspace{1pt}\big)$), we  then have
\begin{align}
 A_{mK_n} & \geq 1-o(1) . \label{Am}
\end{align}
From condition (\ref{Knomega1}), \f $mK_n = \omega(1)$. Since $q$
does not scale with $n$, then $mK_n = \omega(q)$. Therefore,
\begin{align}
 &\binom{mK_n}{q}\Bigg/\binom{P_n}{q} =
 \frac{(mK_n)!}{q!(mK_n-q)!} \bigg/ \frac{P_n!}{q!(P_n-q)!} \nonumber \\ & \quad
 \geq \frac{(mK_n-q)^q}{{P_n}^q}
  = \frac{(mK_n)^q}{{P_n}^q}  \cdot[1-o(1)].  \label{mK}
\end{align}

Clearly, (\ref{moPnKn-sqr}) implies $\frac{{K_n}^2}{P_n} = o(1)$. Under (\ref{Knomega1}) and
$\frac{{K_n}^2}{P_n} = o(1)$, we use \cite[Lemma 2]{QcompTech14} to derive
\begin{align}
{\rho_q}/{p_{s}} & = 1 - o(1).\label{roqps}
\end{align}
Applying (\ref{Am}) (\ref{mK}) and (\ref{roqps}) to (\ref{ptcomp}),
we establish
\begin{align}
p_{\textnormal{compromised}} & \geq \frac{(mK_n)^q}{{P_n}^q}
\cdot[1-o(1)]. \label{pcmp_down}
\end{align}

Since $\frac{{K_n}^2}{P_n} = o(1)$ implies $P_n = \omega(K_n)$, we know from (\ref{pcmp_up}) that $p_{\textnormal{compromised}}$ is upper bounded by $\big(\frac{mK_n}{P_n} \big)^q \cdot [1 + o(1)]$. This result and (\ref{pcmp_down}) together imply (\ref{ptnc});
i.e.,
\begin{align}
  p_{\textnormal{compromised}}   & \sim
\bigg(\frac{mK_n}{P_n} \bigg)^q. \label{ptnq}
\end{align}

%

 By \cite[Lemma 2]{QcompTech14},
under $\frac{{K_n}^2}{P_n} = o(1)$ and $K_n = \omega(1)$, \h
\begin{align}
p_s  \sim \frac{1}{q!} \bigg( \frac{{K_n}^2}{P_n} \bigg)^{q}.
\label{pssimq}
 \end{align}
 Using (\ref{ptnq}) and (\ref{pssimq}), we obtain $\frac{p_{\textnormal{compromised}}}{p_s} \sim q!
  \big(\frac{m}{K_n}\big)^q.$
  \qeda

\subsection{Proving Corollary \ref{cor:capture_vary_q_with_m}}

From Theorem \ref{thm:pcomp2}, we have $
\frac{p_{\textnormal{compromised}}}{p_s} \sim  q!
  \big(\frac{m}{K_n}\big)^q$. Hence, to minimize $p_{\textnormal{compromised}}$ with respect to $q$ in the asymptotic sense given $p_s$, we will minimize $f(q) =  q! \big(\frac{m}{K_n}\big)^q$.
We derive $f(q+1)/f(q)$ as
\begin{align}
\frac{f(q+1)}{f(q)} & = \frac{m(q+1)}{K_n}. \label{fq1fq}
\end{align}
Given (\ref{fq1fq}), we   obtain the following results.
\begin{itemize}
  \item If $\frac{K_n}{m} < 2$, as $q$ increases, the sequence $f(q)$ always
  increases, as it holds from (\ref{fq1fq}) that\\for $q \geq 1$, then $\frac{f(q+1)}{f(q)} = \frac{m(q+1)}{K_n} \geq \frac{2m}{K_n} > 1$.
\item If $\frac{K_n}{m} = 2$, then $f(1)=f(2)$. For $q \geq 2$, as $q$
  increases, the sequence $f(q)$ increases, since we use (\ref{fq1fq}) to derive
  that\\for $q \geq 2$, then $\frac{f(q+1)}{f(q)}  = \frac{m(q+1)}{K_n} \geq \frac{3m}{K_n} =
\frac{3}{2}.$
\item If $\frac{K_n}{m} > 2$, as $q$ increases, the sequence $f(q)$ first
decreases, then increases, based on (\ref{fq1fq}). In particular,
\begin{itemize}
  \item If $\frac{K_n}{m}$ is not an integer, then\\
  $f(1) > \ldots   >
f(\lfloor\frac{K_n}{m}\rfloor-1)
 > f(\lfloor\frac{K_n}{m}\rfloor)
< f(\lfloor\frac{K_n}{m}\rfloor+1) < \ldots $,
because we obtain from (\ref{fq1fq}) that
\begin{itemize}
  \item[$\mathsmaller{\bullet}$] $f(q+1) < f(q)$ for $q
\leq \big\lfloor\frac{K_n}{m}\big\rfloor -1 < \frac{K_n}{m} - 1$,
and
  \item[$\mathsmaller{\bullet}$] $f(q+1) > f(q)$ for
$q \geq \big\lfloor\frac{K_n}{m}\big\rfloor > \frac{K_n}{m} - 1$.
\end{itemize}
Therefore, $f(q)$ achieves its minimum at $q =
\big\lfloor\frac{K_n}{m}\big\rfloor$.
  \item If $\frac{K_n}{m}$ is an integer, then\\$f(1) > \ldots   > f(\frac{K_n}{m}-1)  =
f(\frac{K_n}{m}) < f(\frac{K_n}{m}+1) < \ldots$,
since we derive from (\ref{fq1fq}) that
\begin{itemize}
  \item[$\mathsmaller{\bullet}$] $f(q+1) < f(q)$ for $q <
\frac{K_n}{m} -1$,
\item[$\mathsmaller{\bullet}$] $f(q+1) = f(q)$ for $q = \frac{K_n}{m} -1$, and
  \item[$\mathsmaller{\bullet}$] $f(q+1) > f(q)$ for $q \geq \frac{K_n}{m}$.
\end{itemize}
In this case, $f(q)$ achieves its minimum at both $q = \frac{K_n}{m} - 1$
and $q = \frac{K_n}{m}$.
\end{itemize}
    \end{itemize}

 To summarize,  an optimal $q$ to minimize $f(q)$ and hence minimize $p_{\textnormal{compromised}}$ is
$q^*=\max\{\big\lfloor\frac{K_n}{m}\big\rfloor, 1\}$ (if $\frac{K_n}{m}$ is an integer greater than $1$, in addition to the above $q^*$, another optimal $q$ is $q = \frac{K_n}{m} - 1$). \qeda

\subsection{Proving Corollary \ref{cor:capture_vary_q_with_pc}}

From Theorem \ref{thm:pcomp2}, we have $
\frac{p_{\textnormal{compromised}}}{p_s} \sim  q!
  \big(\frac{m}{K_n}\big)^q$, which implies $m \sim K_n \times (\frac{p_{\textnormal{compromised}}/p_s}{q!})^{1/q}$. Hence, to maximize $m$ with respect to $q$ in the asymptotic sense given $\frac{p_{\textnormal{compromised}}}{p_s}$ and $K_n$, the optimal $q$ is the solution to maximize $ (\frac{p_{\textnormal{compromised}}/p_s}{q!})^{1/q}$. As noted in Corollary \ref{cor:capture_vary_q_with_pc}, the solution of $q$ is difficult to derive formally, but can given empirically by Table \ref{table-optimal-q-pound} and Figure \ref{figure-optimal-q-pound}.\\ \mbox{~}$~$ \qeda

\section{Conclusion }
\label{sec:Conclusion}

The $q$-composite key predistribution scheme is of interest and
significance as a mechanism to secure communications in wireless
sensor networks. Our work   evaluates the resilience of the
$q$-composite scheme against node capture. We also derive the
critical conditions to guarantee network connectivity with high
probability despite node capture by the adversary, taking into
account of practical transmission constraints. These results provide
design guidelines for secure sensor networks using the $q$-composite
scheme.


\begin{thebibliography}{10}

\bibitem{Alarifi:2006:DSN:1180345.1180359}
A.~Alarifi and W.~Du.
\newblock Diversify sensor nodes to improve resilience against node compromise.
\newblock In {\em ACM Workshop on Security of Ad Hoc and Sensor Networks},
  pages 101--112, 2006.

\bibitem{r1}
S.~R. Blackburn and S.~Gerke.
\newblock Connectivity of the uniform random intersection graph.
\newblock {\em Discrete Mathematics}, 309(16), August 2009.

\bibitem{bloznelis2013}
M.~Bloznelis.
\newblock Degree and clustering coefficient in sparse random intersection
  graphs.
\newblock {\em The Annals of Applied Probability}, 23(3):1254--1289, 2013.

\bibitem{Assortativity}
M.~Bloznelis, J.~Jaworski, and V.~Kurauskas.
\newblock Assortativity and clustering of sparse random intersection graphs.
\newblock {\em Electronic Journal of Probability}, 18(38):1--24, 2013.

\bibitem{Rybarczyk}
M.~Bloznelis, J.~Jaworski, and K.~Rybarczyk.
\newblock Component evolution in a secure wireless sensor network.
\newblock {\em Networks}, 53:19--26, January 2009.

\bibitem{Perfectmatchings}
M.~Bloznelis and T.~{\L}uczak.
\newblock Perfect matchings in random intersection graphs.
\newblock {\em Acta Mathematica Hungarica}, 138(1-2):15--33, 2013.

\bibitem{5717499}
T.~Bonaci, L.~Bushnell, and R.~Poovendran.
\newblock Node capture attacks in wireless sensor networks: A system theoretic
  approach.
\newblock In {\em Proc. IEEE Conference on Decision and Control (CDC)}, pages
  6765--6772, 2010.

\bibitem{adrian}
H.~Chan, A.~Perrig, and D.~Song.
\newblock Random key predistribution schemes for sensor networks.
\newblock In {\em Proc. IEEE Symposium on Security and Privacy}, May 2003.

\bibitem{Chan2008134}
K.~Chan and F.~Fekri.
\newblock A resiliency-connectivity metric in wireless sensor networks with key
  predistribution schemes and node compromise attacks.
\newblock {\em Physical Communication}, 1(2):134 -- 145, 2008.

\bibitem{conti2007randomized}
M.~Conti, R.~Di~Pietro, L.~V. Mancini, and A.~Mei.
\newblock A randomized, efficient, and distributed protocol for the detection
  of node replication attacks in wireless sensor networks.
\newblock In {\em ACM International Symposium on Mobile Ad Hoc Networking and
  Computing}, pages 80--89, 2007.

\bibitem{Conti:2008:EPD:1352533.1352568}
M.~Conti, R.~Di~Pietro, L.~V. Mancini, and A.~Mei.
\newblock Emergent properties: Detection of the node-capture attack in mobile
  wireless sensor networks.
\newblock In {\em Proc. ACM WiSec}, pages 214--219, 2008.

\bibitem{pietro2004connectivity}
R.~Di~Pietro, L.~V. Mancini, A.~Mei, A.~Panconesi, and J.~Radhakrishnan.
\newblock Connectivity properties of secure wireless sensor networks.
\newblock In {\em ACM workshop on Security of Ad hoc and Sensor Networks},
  pages 53--58, 2004.

\bibitem{4198829}
R.~Di~Pietro, L.~V. Mancini, A.~Mei, A.~Panconesi, and J.~Radhakrishnan.
\newblock Sensor networks that are provably resilient.
\newblock In {\em Proc. Securecomm}, pages 1--10, Aug 2006.

\bibitem{DiPietroTissec}
R.~Di~Pietro, L.~V. Mancini, A.~Mei, A.~Panconesi, and J.~Radhakrishnan.
\newblock Redoubtable sensor networks.
\newblock {\em ACM Transactions on Information and Systems Security (TISSEC)},
  11(3):13:1--13:22, 2008.

\bibitem{Du:2005:PKP:1065545.1065548}
W.~Du, J.~Deng, Y.~S. Han, P.~K. Varshney, J.~Katz, and A.~Khalili.
\newblock A pairwise key predistribution scheme for wireless sensor networks.
\newblock {\em ACM Transactions on Information and System Security (TISSEC)},
  8(2):228--258, May 2005.

\bibitem{citeulike:4012374}
P.~Erd\H{o}s and A.~R\'{e}nyi.
\newblock On random graphs, {I}.
\newblock {\em Publicationes Mathematicae (Debrecen)}, 6:290--297, 1959.

\bibitem{virgil}
L.~Eschenauer and V.~Gligor.
\newblock A key-management scheme for distributed sensor networks.
\newblock In {\em Proc. ACM CCS}, 2002.

\bibitem{Fill:2000:RIG:340808.340814}
J.~A. Fill, E.~R. Scheinerman, and K.~B. Singer-Cohen.
\newblock Random intersection graphs when $m = \omega(n)$: An equivalence
  theorem relating the evolution of the ${G}(n, m, p)$ and ${G}(n, p)$ models.
\newblock {\em Random Structures \& Algorithms}, 16(2):156--176, Mar. 2000.

\bibitem{fu2008replication}
H.~Fu, S.~Kawamura, M.~Zhang, and L.~Zhang.
\newblock Replication attack on random key pre-distribution schemes for
  wireless sensor networks.
\newblock {\em Computer Communications}, 31(4):842--857, 2008.

\bibitem{Gupta98criticalpower}
P.~Gupta and P.~R. Kumar.
\newblock Critical power for asymptotic connectivity in wireless networks.
\newblock In {\em Proc. IEEE CDC}, pages 547--566, 1998.

\bibitem{ISIT_RKGRGG}
B.~Krishnan, A.~Ganesh, and D.~Manjunath.
\newblock On connectivity thresholds in superposition of random key graphs on
  random geometric graphs.
\newblock In {\em Proc. IEEE International Symposium on Information Theory
  (ISIT)}, pages 2389--2393, 2013.

\bibitem{Krzywdzi}
K.~Krzywdzi\'{n}ski and K.~Rybarczyk.
\newblock Geometric graphs with randomly deleted edges --- connectivity and
  routing protocols.
\newblock {\em Mathematical Foundations of Computer Science}, 6907:544--555,
  2011.

\bibitem{Liu2003CCS}
D.~Liu and P.~Ning.
\newblock Establishing pairwise keys in distributed sensor networks.
\newblock In {\em Proc. ACM CCS}, pages 52--61, 2003.

\bibitem{parno2005distributed}
B.~Parno, A.~Perrig, and V.~Gligor.
\newblock Distributed detection of node replication attacks in sensor networks.
\newblock In {\em IEEE Symposium on Security and Privacy (S\&P'05)}, pages
  49--63. IEEE, 2005.

\bibitem{citeulike:505396}
M.~Penrose.
\newblock {\em Random Geometric Graphs}.
\newblock {Oxford University Press}, July 2003.

\bibitem{penrose2016connectivity}
M.~D. Penrose et~al.
\newblock Connectivity of soft random geometric graphs.
\newblock {\em The Annals of Applied Probability}, 26(2):986--1028, 2016.

\bibitem{suzuki2006birthday}
K.~Suzuki, D.~Tonien, K.~Kurosawa, and K.~Toyota.
\newblock Birthday paradox for multi-collisions.
\newblock In {\em International Conference on Information Security and
  Cryptology}, pages 29--40. Springer, 2006.

\bibitem{tague2008modeling}
P.~Tague and R.~Poovendran.
\newblock Modeling node capture attacks in wireless sensor networks.
\newblock In {\em Annual Allerton Conference on Communication, Control, and
  Computing}, pages 1221--1224, 2008.

\bibitem{Vu:2010:SWS:1755688.1755703}
T.~M. Vu, R.~Safavi-Naini, and C.~Williamson.
\newblock Securing wireless sensor networks against large-scale node capture
  attacks.
\newblock In {\em Proc. ACM ASIACCS}, ASIACCS '10, pages 112--123, 2010.

\bibitem{Yang:2005:TRS:1062689.1062696}
H.~Yang, F.~Ye, Y.~Yuan, S.~Lu, and W.~Arbaugh.
\newblock Toward resilient security in wireless sensor networks.
\newblock In {\em Proc. ACM MobiHoc}, pages 34--45, 2005.

\bibitem{YaganThesis}
O.~Ya\u{g}an.
\newblock {\em Random Graph Modeling of Key Distribution Schemes in Wireless
  Sensor Networks}.
\newblock PhD thesis, Dept. of ECE, College Park (MD), June 2011.
\newblock Available online at \url{http://hdl.handle.net/1903/11910}.

\bibitem{5383986}
O.~Ya\u{g}an and A.~M. Makowski.
\newblock Random key graphs -- can they be small worlds?
\newblock In {\em Proc. International Conference on Networks and Communications
  (NETCOM)}, pages 313 --318, December 2009.

\bibitem{6170857}
D.~H. Yum and P.~J. Lee.
\newblock Exact formulae for resilience in random key predistribution schemes.
\newblock {\em IEEE Transactions on Wireless Communications}, 11(5):1638--1642,
  May 2012.

\bibitem{full}
J.~Zhao.
\newblock On resilience and connectivity of secure wireless sensor networks
  under node capture attacks (full version).
\newblock 2016.
\newblock Available online at
  \url{https://sites.google.com/site/workofzhao/TIFS-resilience.pdf}

\bibitem{ZhaoYaganGligor}
J.~Zhao, O.~Ya{\u{g}}an, and V.~Gligor.
\newblock $k$-connectivity in random key graphs with unreliable links.
\newblock {\em IEEE Transactions on Information Theory}, 61(7):3810--3836,
  2015.

\bibitem{QcompTech14}
J.~Zhao.
\newblock Topological properties of wireless sensor networks under the
  $q$-composite key predistribution scheme with unreliable links.
\newblock Technical Report CMU-CyLab-14-002, CyLab, Carnegie Mellon University, 2014.


\bibitem{MobiCom14}
J.~Zhao, O.~Ya\u{g}an, and V.~Gligor.
\newblock Connectivity in secure wireless sensor networks under transmission
  constraints.
\newblock In {\em Allerton Conference on Communication, Control, and
  Computing}, 2014.

\bibitem{zhu2007efficient}
B.~Zhu, V.~G.~K. Addada, S.~Setia, S.~Jajodia, and S.~Roy.
\newblock Efficient distributed detection of node replication attacks in sensor
  networks.
\newblock In {\em Computer Security Applications Conference (ACSAC)}, pages
  257--267, 2007.

\end{thebibliography}
\end{document}